\documentclass{aa}  

\newcommand{\msun}{\mathrm{M}_\odot}
\newcommand{\Zsun}{\mathrm{Z}_\odot}

\usepackage{graphicx}
\usepackage{graphicx}
\usepackage{txfonts}
\usepackage{natbib}
\usepackage[colorlinks=true,
    linkcolor=blue,
    filecolor=magenta,      
    urlcolor=blue,
    citecolor=blue]{hyperref}

\usepackage{xcolor}
\usepackage{scrextend}
\usepackage{siunitx}

\newcommand{\orcid}[2][0000-0000-0000-0000]{\href{https://orcid.org/#1}{#2}}
\begin{document}

   \title{A new long gamma-ray burst formation pathway at solar metallicity}
   \subtitle{}

   \author{\orcid[0000-0002-6842-3021]{M. M. Briel}, 
          \inst{1,2}\fnmsep\thanks{E-mail: max.briel@gmail.com}
          \and
          \orcid[0000-0003-1474-1523]{T. Fragos} \inst{1,2}
          \and
          \orcid[0000-0003-4924-7322]{O. S. Salafia} \inst{3,4}
          \and
          \orcid[0000-0001-5876-9259]{G. Ghirlanda} \inst{3,4}
          \and
          \orcid[0000-0002-7464-498X]{E. Zapartas} \inst{5}
          \and
          \orcid[0000-0002-3439-0321]{S. S. Bavera} \inst{1,2} 
          \and
          \orcid[0000-0001-5261-3923]{J. J. Andrews} \inst{6,7}
          \and\\
          \orcid[0000-0001-6692-6410]{S. Gossage} \inst{8,9}
          \and
          \orcid[0000-0003-3684-964X]{K. Kovlakas} \inst{10,11}
          \and
          \orcid[0000-0001-9331-0400]{M. U. Kruckow} \inst{1,2}
          \and
          \orcid[0000-0003-4474-6528]{K. A. Rocha} \inst{8,9,12}
          \and
          \orcid[0000-0003-1749-6295]{P. M. Srivastava} \inst{8,9,13}
          \and
          \orcid[0000-0001-9037-6180]{M. Sun} \inst{8}
          \and
          \orcid[0000-0002-0031-3029]{Z. Xing} \inst{1,2}
          }

   \institute{Département d’Astronomie, Université de Genève, Chemin Pegasi 51, CH-1290 Versoix, Switzerland
        \and
        Gravitational Wave Science Center (GWSC), Université de Genève, CH-1211 Geneva, Switzerland
        \and
        INAF -- Osservatorio Astronomico di Brera, via Emilio Bianchi 46, I-23807 Merate (LC), Italy
        \and
        Istituto Nazionale di Fisica Nucleare -- Sezione di Milano-Bicocca, piazza della Scienza 3, I-20126 Milano (MI), Italy
        \and
        Institute of Astrophysics, Foundation for Research and Technology-Hellas, GR-71110 Heraklion, Greece
        \and
        Department of Physics, University of Florida, 2001 Museum Rd, Gainesville, FL 32611, USA
        \and
        Institute for Fundamental Theory, 2001 Museum Rd, Gainesville, FL 32611, USA
        \and
        Center for Interdisciplinary Exploration and Research in Astrophysics (CIERA), Northwestern University, 1800 Sherman Ave, Evanston, IL 60201, USA
        \and
        NSF-Simons AI Institute for the Sky (SkAI),172 E. Chestnut St., Chicago, IL 60611, USA
        \and
        Institute of Space Sciences (ICE, CSIC), Campus UAB, Carrer de Magrans, 08193 Barcelona, Spain
        \and
        Institut d'Estudis Espacials de Catalunya (IEEC),  Edifici RDIT, Campus UPC, 08860 Castelldefels (Barcelona), Spain
        \and
        Department of Physics and Astronomy, Northwestern University, 2145 Sheridan Road, Evanston, IL 60208, USA
        \and
        Electrical and Computer Engineering, Northwestern University, 2145 Sheridan Road, Evanston, IL 60208, USA
       }

   \date{Received XXX; accepted XXX}
 
  \abstract
   {Long gamma-ray bursts (LGRBs) are generally observed in low-metallicity environments. However, 10 to 20 per cent of LGRBs at redshift $z<2$ are associated with near-solar to super-solar metallicity environments, remaining unexplained by traditional LGRB formation pathways that favour low metallicity progenitors.}
   {In this work, we propose a novel formation channel for LGRBs that is dominant at high metallicities. We explore how a stripped primary star in a binary can be spun up by a second, stable reverse-mass-transfer phase, initiated by the companion star.
   }
   {We use {\tt POSYDON}, a state-of-the-art population synthesis code that incorporates detailed single- and binary-star mode grids, to investigate the metallicity dependence of the stable reverse-mass-transfer LGRB formation channel. We determine the available energy to power an LGRB from the rotational profile and internal structure of a collapsing star, and investigate how the predicted rate density of the proposed channel changes with different star formation histories and criteria for defining a successful LGRB.
   }
   {Stable reverse mass transfer can produce rapidly rotating, stripped stars at collapse. These stars retain enough angular momentum to account for approximately 10–20\% of the observed local LGRB rate density, under a reasonable assumption for the definition of a successful LGRB. However, the local rate density of LGRBs from stable reverse mass transfer can vary significantly, between 1 and 100 Gpc$^{-3}$ yr$^{-1}$, due to strong dependencies on cosmic star formation rate and metallicity evolution, as well as the assumed criteria for successful LGRBs.
   }
   {}
   \keywords{
               }

   \maketitle
%

\section{Introduction}

Gamma-ray bursts (GRBs) are some of the most energetic events in the Universe, lasting between a fraction of a second and several hundreds of seconds. 
Powered by a central engine, jets form and propagate through surrounding material, emitting intense gamma rays.
Their duration has become the defining feature in their classification into short and long GRBs, often linked to their formation pathway.
Several lines of evidence \citep{berger_2014} consistently point to short GRBs ($\lesssim$2 seconds) being produced after mergers of binary neutron stars, in line with the early suggestion by \citet{eichler_1989}. The link has been verified in one case by the joint observations of the gravitational wave event GW170817 \citep{abbott_2017a}, the kilonova AT2017gfo/SSS17a \citep{abbott_2017}, and the gamma-ray burst GRB170817A \citep{abbott_2017b}.
Several long GRBs (LGRBs), on the other hand, have been associated with Type Ic broad-line supernovae \citep[SN Ic-BL;][]{modjaz_2016}, indicating a hydrogen- and helium-poor progenitor massive star. The connection is thought to encompass most LGRBs, even though observations of some LGRBs have shown photometric and spectroscopic optical and near-infrared signatures compatible with kilonova emission \citep{rastinejad_2022, levan_2024}.

Due to their highly-beamed and energetic nature, LGRBs have been observed up to $z \sim 9.4$ \citep[][]{cucchiara_2011} and have been used to estimate the star formation rate at high redshifts.
However, the LGRB rate rises more rapidly over $z \leq 4$ than other star formation estimators \citep{yuksel_2008, kistler_2009, ghirlanda_2022}.
Furthermore, the local population of LGRB hosts ($z \leq 2$) tends towards lower mass hosts compared to the general population of star-forming galaxies \citep{kruhler_2015, vergani_2015, japelj_2016, perley_2016a}.
With more complete and unbiased samples of LGRBs and their host galaxies, \citet{greiner_2015} have shown that between $z=3$ and $z=5$, the LGRB rate traces the UV metrics of the cosmic star formation.
Together with average metallicity measurements of LGRB host galaxies, this has been interpreted as a low-metallicity preference for the formation of LGRBs \citep{levesque_2010a, levesque_2010b, graham_2013, perley_2016a, vergani_2017, palmerio_2019}.
A sharp cut-off of LGRB formation at a sub-solar metallicity can explain these observed phenomena over redshift, although its exact value ranges from $0.3~\Zsun$ to $0.7~\Zsun$ depending on the observational survey \citep{modjaz_2008, salvaterra_2012, graham_2013, perley_2016a, vergani_2017, palmerio_2019} and metallicity calibration used \citep{kewley_2008}.

Despite the preference of LGRBs for sub-solar metallicity host galaxies, GRB 020819 \citep{levesque_2010}, GRB 111005A \citep{michalowski_2018}, and GRB 160804A \citep{heintz_2018} have been observed to have solar or super-solar metallicity explosion sites.
Furthermore, for $z<2.5$, the fraction of near-solar metallicity LGRB host galaxies remains near constant at 10 to 20\% \citep{vergani_2015, palmerio_2019, graham_2023}.
A similar fraction of 20\% of super-solar metallicity LGRB host galaxies has been observed by \citet{kruhler_2015} with the VLT X-SHOOTER emission-line spectroscopy.
However, these fractions might be significantly lower due to observational biases.
Firstly, a metallicity measurement of the host galaxy requires a spectrum, which is easier to obtain for more massive, brighter hosts, which are also generally more metal-rich.
Although the LGRB sample might be complete, the LGRB host sample with measured metallicities is not \citep[for example, see][]{graham_2023}.
Secondly, if a metallicity measurement is done, in most cases it only probes the average line-of-sight metallicity of the LGRB host due to their distance, but their metallicity distribution can be highly non-uniform \citep{metha_2020}.
While the fraction of high-metallicity hosts is, thus, difficult to constrain, the direct detection of the three high-metallicity explosion sites provides evidence that at least some LGRBs originate from metal-rich environments.

The exact criterion for an LGRB to occur from the collapse of a massive star depends on the assumed central engine powering the GRB.
The two best-studied scenarios involve either a highly magnetized, rapidly rotating proto-neutron star (a proto-magnetar) or a spinning black hole (BH) with an accretion disk (the standard `collapsar' scenario).
In the former scenario, the proto-magnetar powers the LGRB through its fast rotation and high magnetic field strength \citep{usov_1992,thompson_1994, thompson_2004, metzger_2011, mazzali_2014}. 
In the collapsar scenario, instead, a BH is formed and sufficient angular momentum is available in the stellar envelope to form an accretion disk around it, allowing jets to form either by the \citet{blandford_1977} mechanism, or because of energy deposition by annihilations of neutrino-antineutrino pairs formed in the inner accretion disk  \citep{eichler_1989, woosley_1993, macfadyen_1999}.
Independent of either central engine scenario, the progenitor star has to (I) reach iron core collapse stripped of hydrogen and helium, (II) contain sufficient angular momentum to either spin up the proto-magnetar or to form an accretion disk, and (III) have a successful and collimated jet breakout through the stellar structure and produce gamma-ray emission \citep{matzner_2003,modjaz_2016, salafia_2020}.
Combined, these three criteria require the stellar progenitor to be a rapidly-rotating, highly-stripped star, where the outer hydrogen and most of the helium envelopes have been removed.
This outcome gives rise to two main formation pathway questions and conditions: (a) How does the stellar core retain or gain sufficient angular momentum at core-collapse? (b) How does the star lose its envelope?

From a single-star perspective, an initially rapidly rotating zero-age main-sequence (ZAMS) star will expand to a red supergiant after finishing core hydrogen burning. The subsequent evolution of its rotation depends heavily on the assumed coupling between the core and outer layers of the star \citep{qin_2019}.
Astroseismology observations of slightly less massive giant stars \citep{fuller_2014, cantiello_2014, denhartogh_2020} indicate a strong internal coupling \citep[][]{eggenberger_2022, moyano_2023}. As the star expands post-main sequence, this results in an efficient redistribution of angular momentum from the core to the outer stellar layers and leads to the formation of slowly spinning BHs at the collapse of the core \citep{qin_2018, fuller_2019}.
Strong stellar winds will further enhance this by removing mass and angular momentum from the outer layers.
For a single star to form a rapidly-rotating highly-stripped star, the metallicity has to be low to minimise mass and angular momentum loss ($Z<0.3Z_\odot$) and the initial rotation needs to be sufficient for rotational mixing to keep the star close to chemical homogeneity \citep[quasi-chemically homogeneous;][]{maeder_1987, meynet_2007}.
As a result of the strong mixing, the star burns nearly all its hydrogen and avoids the giant phase. Without an extended envelope, angular momentum loss from the core is minimised \citep{yoon_2005, yoon_2006, woosley_2006}.
In a sub-solar metallicity environment, the Wolf-Rayet winds are sufficiently weak \citep{vink_2005} for these quasi-chemically homogeneous stars to retain sufficient angular momentum in their core to power a LGRB.

Binary systems provide additional formation pathways for stars to gain angular momentum and be stripped during their evolution instead of requiring the star to be rotating at birth.
On the other hand, processes such as common envelope evolution can remove angular momentum from the star and restrict the formation of LGRBs.
In general, binary formation channels invoke chemically homogenous evolution or late-time tidal spin-up, although several other mechanisms have been proposed throughout the years \citep[see][for an extensive overview]{fryer_2025}.
First, close massive binary systems can undergo mass transfer and spin up the accreting star sufficiently to undergo quasi-chemically homogeneous evolution at metallicities below $Z = 1/3~\Zsun$ \citep{cantiello_2007, yoon_2008, yoon_2010, eldridge_2011, ghodla_2023}, as long as the accretor has not yet formed a well-defined stellar core.
Secondly, in very tight near-equal-mass-ratio contact binaries, tides can spin up both stars, where the chemically homogenous evolution keeps both stars from expanding while forming the LGRB progenitors \citep{marchant_2016, song_2016, mandel_2016}.
Additionally, if the mass ratio is further from unity, just the initially more massive star in a binary can undergo this tidal spin up \citep{demink_2009, marchant_2017}.
Finally, tidal spin-up can also occur after a helium star has formed in tight binaries with a compact object companion \citep{vandenheuvel_2007, detmers_2008, qin_2018}. These four LGRB formation channels are more common at sub-solar metallicities.

The tidal spin-up and the chemically homogeneous evolution formation channels have been investigated using several binary population synthesis codes.
\citet{izzard_2004a} and \citet{chrimes_2020} looked at the effect of spin-orbit coupling in LGRB formation using binary population synthesis, although the exact nature of systems experiencing the tidal effects was not explored.
The latter includes the treatment accretion-induced chemically homogeneous evolution. However, this was done using an approximate prescription \citep[described in][]{eldridge_2011}, which was calibrated on detailed stellar models from \citet{yoon_2006}.
In the context of binary black hole mergers, \citet{bavera_2022b} investigated LGRB formation through chemically homogenous evolution of close binaries and tidal spin-up in BH-helium star binaries. Similarly, \citet{ghodla_2023} showed the effect that tidally and accretion-induced chemically homogenous evolution have on the expected LGRB population.

Several additional scenarios for LGRB formation have been proposed over the years that generate the LGRB through other mechanisms, such as the merger between a helium star and a BH \citep{fryer_1998} or an explosive common envelope ejection \citep{podsiadlowski_2010a}. The contribution of the former to the cosmic LGRB rate has not yet been quantified, while the latter could produce ${\simeq}10^{-6}$ LGRBs yr$^{-1}$ at solar metallicity, but is expected to be more efficient at lower metallicities \citep{podsiadlowski_2010a}.
Finally, a merger between two helium cores has been suggested to generate a rapidly rotating core \citep{ivanova_2003, fryer_2005}, but its population contribution is poorly studied.

The fact that most theoretical formation channels predict an inverse relation between LGRB progenitors and metallicity aligns well with the observational preference for sub-solar metallicity host galaxies.
However, the three (super-)solar metallicity explosion sites and the apparent lack of evolution in the high-metallicity fraction of LGRB host galaxies across different redshifts \citep{graham_2023} suggest additional formation mechanisms might be at play.
In this work, we propose a previously unexplored formation channel through binary interaction that exhibits a positive relationship between metallicity and LGRB progenitors that could explain these effects.
For binaries with an initial mass ratio close to unity, after a phase of stable mass transfer from the primary onto the companion, a second phase of stable mass transfer from the companion onto the now stripped primary star spins it up sufficiently to power an LGRB. 
We refer to this mass transfer from the initially less massive star (secondary) onto the initially more massive star (primary) as ``reverse mass transfer''. We will consistently refer to the initially more massive star as the primary, while to the initially less massive star as the secondary or companion.
We determine the occurrence of this LGRB formation channel using the next-generation binary population synthesis code, {\tt POSYDON}, which uses grids of detailed single- and binary-star models, as described in Section \ref{sec:method}.
In Section \ref{sec:results}, we explain the formation pathway in detail using an example model and show why the efficiency of this LGRB channel has a positive dependence on metallicity.
Furthermore, we explore the available energy to power a jet in potential LGRB progenitors, and how different criteria for defining a successful LGRB affect the predicted rate density from the stable reverse-mass-transfer channel.
We discuss uncertainties in this stable reverse-mass-transfer formation pathway and the rate density evolution in Section \ref{sec:discussion}, while we summarize our conclusions in Section \ref{sec:conclusion}.

\section{Method} \label{sec:method}

We use the public, state-of-the-art binary population synthesis code, {\tt POSYDON}\footnote{ This work uses the {\tt POSYDON} (\href{https://posydon.org/}{posydon.org}) version from commit \href{https://github.com/POSYDON-code/POSYDON/tree/4a8c215}{4a8c215}.} \citep{fragos_2023, andrews_2025}, which employs grids of detailed single- and binary-star models computed with the 1D stellar evolution code {\tt MESA} \citep{paxton_2011, paxton_2013, paxton_2015, paxton_2018, paxton_2019, jermyn_2023} to create synthetic populations of evolved binary systems.
{\tt POSYDON v2} release\footnote{The {\tt POSYDON v2} grids will be available on the \href{https://zenodo.org/communities/posydon/}{{\tt POSYDON} Zenodo community}.} contains grids at eight metallicities: $10^{-4}, 10^{-3}, 10^{-2}, 0.2, 0.45, 0.1, 1, 2 Z_\odot$ with $\Zsun = 0.0142$, which include self-consistent modelling of internal rotation, as well as angular momentum transport in the stellar interior and between the stars and the orbit of the binary.

\subsection{Stellar and binary evolution physics \label{sec:stellar_physics}}

We summarise the essential stellar and binary physics model assumptions in this section -- see \citet{fragos_2023} and \citet{andrews_2025} for more detailed descriptions.
Rotational mixing and angular momentum transport are implemented using the diffusion approximation and a Spruit–Tayler dynamo \citep{spruit_2002} is included, which results in strong coupling between the stellar core and its envelope during the post-main sequence, and an overall efficient angular momentum transport \citep{heger_2005}. 

The \texttt{POSYDON} H-rich+H-rich binary star grids, which start their evolution as two ZAMS stars, combine different formulations for mass transfer depending on the evolutionary state of the stars when Roche lobe overflow (RLOF) occurs \citep[for details, see][]{fragos_2023}. This grid self-consistently treats rotation, tides, and mass transfer within the binary, as the orbital evolution and stellar structure equations of both stars are solved with \texttt{MESA}.
For stars on the main sequence, the \texttt{contact} scheme is used, which keeps the stars inside their Roche lobe \citep{marchant_2016}.
The \texttt{kolb} scheme \citep{kolb_1990} is used for stars undergoing RLOF post-main-sequence to allow for radial expansion past its Roche lobe since the envelope has a low density.
The grids use the correction described in \citet{xing_2024} to allow mass transfer from the companion onto the primary star with the \texttt{kolb} scheme in \texttt{MESA}.
Either star can initiate the mass transfer, and multiple mass-transfer phases can occur in the lifetime of a binary if the interaction remains stable. Stellar rotation enhances the stellar winds to keep rotation below the critical threshold $\omega_s/\omega_\mathrm{s, crit} \leq 1$. The latter is also a mechanism that can limit the accretion efficiency of a stable RLOF mass-transfer phase.

\texttt{POSYDON} considers mass transfer in the H-rich+H-rich grids as unstable either when (i) a donor mass transfer rate of $0.1 ~\msun$ yr$^{-1}$ is reached, or (ii) when mass is lost from the $L_2$ point, or (iii) when both stars fill their Roche lobe while either has evolved off the main sequence\footnote{In the H-rich+H-rich grids, an additional unstable mass transfer condition is applied in later reruns of previously non-converged models. This condition only impacts low-mass accretors (see Section 4.1 in \citet{andrews_2025}) and falls outside the relevant mass range for this work.}
If a contact phase happens while both stars are on the main sequence, this is modelled following \citet{marchant_2016} and can result in a stable overcontact phase.
Binary {\tt MESA} models reaching one of the unstable mass transfer criteria are stopped and evolved using {\tt POSYDON} through a common envelope phase and subsequent detached or interaction phases. If none of the instability conditions are met, the binary evolution is tracked within {\tt MESA} until the end of core carbon burning\footnote{White dwarf and pair-instability progenitors are stopped earlier. See \citet{fragos_2023} and \citet{andrews_2025} for more details on the exact stopping criteria.}, where the final fate of each star as a neutron star or BH is determined. At this point, we determine whether it is an LGRB progenitor.

\subsection{Core collapse and long gamma-ray burst formation \label{sec:cc_physics}}

In the collapsar scenario, the relativistic jet is powered by the accretion of material from the collapsing star onto the newly-formed BH through an accretion disk.
We use the remnant mass prescription from \citet{patton_2020} to determine whether the core collapse produces a BH or a NS, assuming always a prompt collapse for BH formation. We adopt the methodology from \citet{batta_2019} to track the formation and evolution of the accretion disk during the infall without wind feedback.
In their approach, the accretion disk mass is determined by material in the stellar interior whose specific angular momentum exceeds that of the innermost stable circular orbit (ISCO) of the BH. Following \citet{bavera_2020}, we split the stellar structure into shells of mass $m_\mathrm{shell}$. We assume the innermost $2.5~\msun$ of the star to collapse directly to form a BH.
For each consecutive $i$-th shell, the fraction of the shell with sufficient angular momentum to form a disk is given by $m_{\mathrm{disk},i} = m_\mathrm{shell} \cos(\theta_{\mathrm{disk},i})$, where 

\begin{equation}
    \theta_{\mathrm{disk},i} = \arcsin\left( \left[\frac{j_\mathrm{ISCO,i}}{j_{\mathrm{max},i}} \right]^{1/2} \right),
\end{equation}
with $j_\mathrm{ISCO,i}$ being the specific angular momentum of the inner-most stable orbit (ISCO), and $j_{\mathrm{max},i}=\Omega(r_i) r_i^2$ that of material at the equator of the $i$-th mass shell, located at a stellar radius $r_i$. We use equation (5) from \citet{batta_2019} for $j_\mathrm{ISCO}$.
The amount of angular momentum gained by the BH after accreting the $i$-th shell is $\Delta J = m_{\mathrm{disk},i} j_\mathrm{ISCO}$.
We perform the calculation per mass shell to account for the change in BH spin and the ISCO  during the collapse.

We calculate the jet power assuming the Blandford-Znajek mechanism, which powers the GRB by extracting rotational energy from the BH \citep{blandford_1977}. We decompose the total jet luminosity using two efficiency factors, one based on the magnetic field and one on the BH spin, following \citet{gottlieb_2023, gottlieb_2024}:
\begin{equation}
    L_\mathrm{BZ} = \eta_\phi \eta_a \dot{M} c^2~ \mathrm{erg}~\mathrm{s}^{-1},
\end{equation}
where $\dot{M}$ is the accretion rate through the disk, $\eta_a$ is an efficiency factor that depends on the BH spin, and $\eta_\phi$ is an additional efficiency factor that depends on the magnetic field flux at the BH horizon.
The efficiency of extraction depends highly on the state of the disk and requires magnetohydrodynamic simulations to follow the full disk and jet formation, which is beyond the scope of this work, but we will discuss the choices in more detail in Section \ref{sec:discussion}.
We follow the original spin-dependence from \citet{blandford_1977} of $L_\mathrm{BZ} \propto a^2$ and set $\eta_a = a^2$, although see \citet{tchekhovskoy_2010, tchekhovskoy_2012} or \citet{lowell_2024} for a more complex dependence on the BH spin that also depends on the aspect ratio of the accretion disk.
We track how this efficiency changes as the star collapses and the BH spin increases, but we do not model the extraction of angular momentum from the powering of the jet. We sum over the mass of each shell going to the accretion disk to get the total power of the jet available in the collapse:

\begin{equation} \label{eq:E_BZ}
     E_\mathrm{BZ} = \sum_{i} m_{\mathrm{disk},i}\, a_{i-1}^2\, c^2,
\end{equation}
where $a_{i}$ is the BH spin parameter after accreting the $i$-th shell. While this does not track the change of efficiency from the evolving magnetic field flux at the BH horizon, different $\eta_\phi$ values should provide a reasonable approximation for the energy budget. We discuss the choices for the $\eta_\phi$ efficiency in Section~\ref{sec:stable-reverse-MT-rate}.

\subsection{Binary population properties \label{sec:initial_properties}}

We create stellar populations at each metallicity in {\tt POSYDON} following a \citet{kroupa_2001} initial mass function between $7~\msun$ and $120~\msun$, with a flat mass ratio, and log-period distributions. We initiate \num{1000000} binaries per metallicity and evolve them using the {\tt POSYDON} nearest neighbour interpolation scheme.

{\tt POSYDON} implements a nearest neighbour and an initial-final interpolation method across its detailed model grids, which are evolved to either the end of core-collapse or the onset of unstable mass transfer. The initial-final interpolator maps the initial properties of a system to its final state at these endpoints. Additionally, it includes interpolation across compact object formation to include specific properties, such as BH spin \citep[for more details, see][]{fragos_2023}
However, here, we require a stellar profile to perform the collapse, as described in Section \ref{sec:cc_physics}. To obtain this, we use the nearest neighbour scheme, where each system is matched to its closest {\tt MESA} model in the mass-q-period space of the relevant grid. This provides a stellar profile at carbon exhaustion for binaries that have undergone stable mass transfer or no interaction\footnote{An alternative approach would be to use the stellar profile interpolation method described in \citet{teng_2025}. However, this method was developed in parallel with this study, and became part of the {\tt POSYDON} code base only recently.}. We only select systems with stable reverse mass transfer and sufficient angular momentum in the pre-collapse structure to form an accretion disk during the collapse.

\subsection{Star formation and metallicity history}

We combine the binary populations with metallicity-dependent star formation history prescriptions to determine the occurrence rate of stable reverse-mass-transfer LGRB formation over cosmic time, where we assume a binary fraction of 0.7. The star formation rate and metallicity evolution are extracted from the TNG-100 IllustrisTNG simulation \citep{springel_2018, nelson_2018, pillepich_2018a, naiman_2018, marinacci_2018}. This cosmological simulation contains baryonic and dark matter particles of masses $1.4 \times 10^6~\msun$ and $7.5 \times 10^6~\msun$, respectively, in a comoving volume of (75 Mpc/$h$)$^3$.

The IllustrisTNG simulation is based on the moving-mesh code \textsc{arepo} \citep{springel_2010}, which combines Lagrangian particle-based methods and Eulerian mesh-based methods. Because the former methods tend to underestimate the mixing between cells \citep{wiersma_2009}, while the latter tend to overestimate it \citep{sarmento_2017}, the net effect on the metallicity of galaxies in the IllustrisTNG simulation is an active area of research.
For these reasons, in Section \ref{sec:discussion} we implement two additional empirical star formation histories based on \citet{madau_2017} and \citet{neijssel_2019}, for comparison.
The total star formation rate from \citet{neijssel_2019} was calibrated to produce the observed binary black hole distribution\footnote{Under the assumption that all binary black holes come from isolated binary evolution and given the set of assumptions in their specific binary population synthesis model.} with a log-normal metallicity distribution with $\sigma=0.35$, $\mu = \log_{10}(\left<Z\right>) - \sigma^2\frac{\ln(10)}{2} $ with $\left< Z \right> = 0.035 \times 10^{-0.23z}$, and 
\citet{madau_2017} has a higher SFR at high-redshift with $\mu = \ln(\left<Z\right>) - \sigma^2/2$ and  $\left< Z \right> = 10^{0.153 - 0.074 z^{1.34}}~\Zsun$, which spreads out the metallicity-specific SFR compared to the \citet{neijssel_2019} relationship. We follow the parameters from appendix~A in \citet{bavera_2020}, where they set $\sigma=0.5$.
With both these prescriptions, the cosmic metallicity evolution proceeds slower than in the Illustris-TNG simulation and contains significantly more star formation at very high redshift ($z>15$).
Adopting the cosmological parameter values from \citet{aghanim_2020}, we convolve these star formation histories with the {\tt POSYDON} populations to calculate a total stable reverse-mass-transfer LGRB rate.

\section{Results} \label{sec:results}

Since LGRB formation through stable reverse mass transfer is constrained by multiple stellar and binary evolutionary processes, we first examine a representative evolutionary model to outline the key factors at play. We then shift our attention to the LGRB population and how each process limits their formation in the proposed channel.

\subsection{Stable reverse mass transfer towards LGRB formation}

\subsubsection{Example model}

\begin{figure*}
    \centering
    \includegraphics{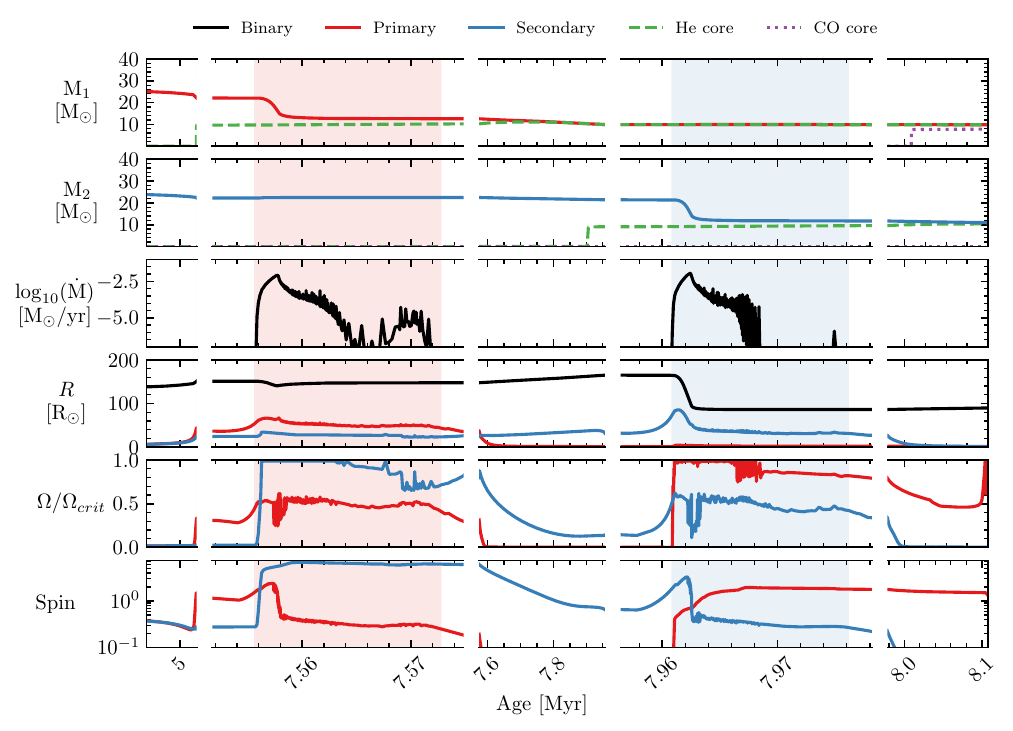}
    \caption{A typical example of the evolution of a potential LGRB progenitor through the stable reverse-mass-transfer channel. The initial properties of the binary are $Z=\Zsun$, $M_\mathrm{1,ZAMS} \approx 25.1~\msun$, $q=0.95$, and $P \approx 26.8$ days. In all panels, red (blue) solid lines refer to the primary (secondary) star. The columns are split into five different evolutionary phases, focusing on the mass transfer phases. The second and fourth columns show the mass transfer and `reverse' mass transfer phases, respectively. The first, third and last columns show the long, nuclear timescale, detached evolution of the system.
    Top two rows: total mass (solid lines), He core mass (dashed lines) and CO core mass (dotted lines). The He (CO) core boundaries are set where the hydrogen (helium) fraction falls below 0.1.
    Third row: mass transfer rate from Roche lobe overflow. 
    The periods of Roche lobe overflow are marked by a red (blue) shading when the donor is the primary (secondary). Fourth row: evolution of stellar radius (red: primary; blue: secondary) and separation of the system (black). Fifth row: surface rotational angular frequency in units of the critical (i.e.\ mass shed) value. Sixth row: the dimensionless spin parameter $cJ/GM^2$ of the star.
    }
    \label{fig:example_model}
\end{figure*}

\begin{figure}
    \centering
    \includegraphics[]{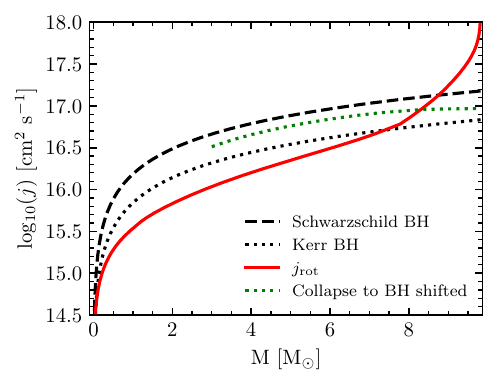}
    \caption{The specific angular momentum distribution of the mass shells of the primary star in the example binary shown in Figure~\ref{fig:example_model}, at core carbon depletion in red. The kink at ${\simeq}7.8~\msun$ is the transition between the carbon-oxygen core and helium shell. It forms an accretion disk of $1.46~\msun$ during the collapse. For comparison, we plot the specific angular momentum of the ISCO, $j_\mathrm{ISCO}$, for a non-rotating Schwarzshild BH (dashed black) and a maximally rotating Kerr BH (dotted black), as a function of BH mass. Additionally, we track the evolution of $j_\mathrm{ISCO}$ for the example primary during the collapse as its mass and spin change (dashed green). It tracks $j_\mathrm{ISCO}$ at the collapse of each mass shell and starts at $3~\msun$ with a BH of $2.5~\msun$ and $0.5~\msun$ neutrino mass loss.
    }
    \label{fig:angular_momentum}
\end{figure}

From the binaries in our population that undergo stable reverse mass transfer, here we take an example binary with sufficient angular momentum at the collapse of the primary star to produce an accretion disk around the BH.
It consists of a primary and secondary star of initial masses $M_1 \approx 25.1~\msun$ and $M_2 \approx 23.8~\msun$, respectively, at solar metallicity, orbiting each other at an initial period of $P \approx 26.8$ days. Figure \ref{fig:example_model} shows the time evolution of several properties of the binary, during a time window that begins close to the end of the main sequence phase. The binary's evolution can be separated into five main phases:
\begin{enumerate}
    \item Initially, the binary evolves detached until the primary reaches core hydrogen depletion at 7.6 Myr. At this stage, the primary has lost ${\simeq}2.97~\msun$ due to the stellar winds. The primary evolves off the main sequence and expands past its Roche lobe radius, thus initiating the first mass transfer phase. 
    \item During the mass transfer phase of ${\sim}16$ kyr, the companion is spun up rapidly, while accreting only ${\simeq}0.27~\msun$. Any additional transferred material (${\simeq}9.49~\msun$) is quickly lost by the rotation-enhanced winds. This initial mass transfer phase removes most of the hydrogen envelope from the primary star before detaching, leaving a hydrogen-rich envelope of ${\simeq}2.43~\msun$ and a surface hydrogen fraction of ${\simeq}0.33$.
    \item The low surface hydrogen fraction signals that Wolf-Rayet winds have kicked in and the remaining hydrogen envelope is quickly removed in the subsequent detached phase until a fully stripped helium star remains.
    \item At ${\sim}8.0$ Myr the secondary, now $21.4~\msun$, expands due to core hydrogen exhaustion and initiates a second mass transfer phase. It loses ${\simeq}9.69~\msun$, while the fully-stripped primary star only accretes ${\simeq}0.04~\msun$ before it reaches critical rotation. When the secondary reaches core helium ignition, it exits the mass transfer phase at a stage similar to the primary with a limited hydrogen envelope of ${\simeq}2.22~\msun$ and a similar surface hydrogen fraction of ${\simeq}0.32$. The properties between the primary and secondary are similar because they start with a mass ratio of $0.95$ but the primary is now rapidly rotating.
    \item The primary continues to evolve for another ${\sim}100$ kyr and remains compact until core carbon exhaustion. It loses only $0.06~\msun$ during this phase and the primary maintains a sufficiently large angular momentum reservoir in the internal structure to form a $1.46~\msun$ accretion disk during the collapse into the BH, as shown in Figure \ref{fig:angular_momentum}. Stellar mass shells with angular momentum above $j_\mathrm{isco}$ form an accretion disk around the BH, as described in Section~\ref{sec:cc_physics}.
    This accretion disk results in a total energy budget of $E_\mathrm{BZ}/\eta_\phi \approx 2.60 \times 10^{54}$ erg. With an $\eta_\phi$ of $10^{-3}$, this system has a typical LGRB energy of ${\sim}10^{51}$ erg and, with the binding energy of the primary only being $\sim 10^{50}$ erg, the jet is likely to breakout successfully.
\end{enumerate}

The essential components of this LGRB formation channel are the occurrence of stable reverse mass transfer and the angular momentum reservoir at collapse.
Since these have complex dependencies on the initial properties of the binary and vary across metallicity, in what follows we first discuss the conditions for the secondary to fill its Roche lobe before the primary star has collapsed at solar metallicity. Then we discuss for what initial condition this interaction is stable and leads to disk formation at collapse. Finally, we investigate why this LGRB formation channel only occurs at near-solar metallicities.

\begin{figure*}
    \centering
    \includegraphics[width=0.942\textwidth]{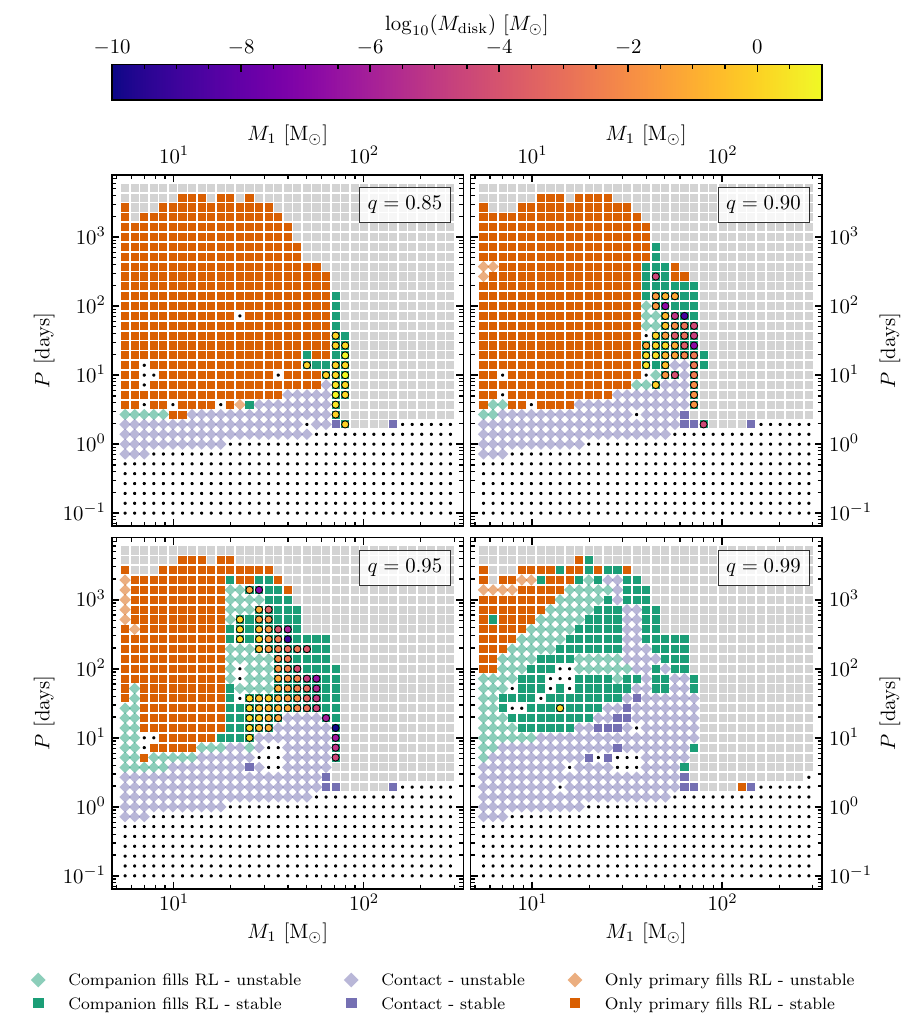}
    \caption{Four grid slices at $\Zsun$ with $q=0.85$, $0.90$, $0.95$, and $0.99$. Orange symbols represent systems where only the primary fills its Roche lobe throughout the evolution of the binary model. Green squares indicate systems where also the secondary fills its Roche lobe, but not at the same time as the primary. Purple symbols indicate systems where both stars fill their Roche lobe at the same time. The type of symbol shows the stability of the interaction: squares indicate stable mass transfer, while binary models reaching instability criteria are indicated with a diamond. Grey squares indicate no mass transfer, while black dots indicate a non-converged model or Roche lobe overflow when the model is initiated.
    For systems where the companion fills its Roche lobe and the interaction remains stable, we indicate if an accretion disk is formed during the collapse of the model. This is indicated by a coloured circle inside the square symbol, where the colour indicates the mass of the accretion disk.}
    \label{fig:Zsun_multi_q}
\end{figure*}

\subsubsection{Dependence on mass ratio, period and donor mass}

In our example model, both stars fill their Roche lobe on the Hertzsprung gap after completing hydrogen burning in their cores. This moment of interaction is common in the proposed LGRB formation channel and requires the main-sequence lifetimes of the binary components to be similar. Otherwise, the primary star will have reached carbon exhaustion before the secondary has left the main sequence.
This evolutionary timescale difference is influenced by the mass of the primary and the binary mass ratio. 
Specifically, the more equal the masses of the stars in the binary, the smaller the lifetime difference.
As a result, many systems with $q$ close to unity have the secondary fill its Roche lobe before the primary completes its evolution.
This effect can be seen in Figure \ref{fig:Zsun_multi_q}, where example grid slices of solar metallicity \texttt{POSYDON} detailed binary models with initial mass ratios of $q=0.85$, $q=0.90$, $q=0.95$, and $q=0.99$ are shown.
The green markers indicate systems with the secondary expanding past its Roche lobe, while the orange markers indicate systems with only the primary star filling its Roche lobe. Systems with both stars filling their Roche lobe at the same time are marked with purple, as contact systems.
As the mass ratio approaches $1$, the green area becomes larger, indicating that more systems have reverse mass transfer.
Additionally, the minimum mass ratio for the secondary to fill its Roche lobe depends on the absolute mass of the donor star.
Massive stars are dominated by radiation pressure, which causes their main-sequence lifetimes to become similar towards higher masses.
This means that binaries with more massive primaries have a broader range of mass ratios where the secondary can initiate a reverse mass transfer phase.
As a result, the highest mass primaries in Figure \ref{fig:Zsun_multi_q}, still have reverse mass transfer at $q=0.85$, while the lower mass systems do not.

In Figure \ref{fig:Zsun_multi_q}, it is important to address that the secondary does fill its Roche lobe in contact systems, but often reaches the $L_2$ overflow criteria for unstable mass transfer or, at longer periods, one of the stars leaves the main sequence while in contact. In general, only when the initial period is long enough can the secondary fill its Roche lobe without the binary reaching an unstable contact state.

A stable mass transfer phase is a requirement to provide sufficient angular momentum to the primary star to produce an accretion disk at collapse. In Figure \ref{fig:Zsun_multi_q} at $q=0.95$, two islands of unstable reverse mass transfer are present in the regime for Roche lobe overflow initiated by the secondary. The higher period island around $10^3$ days is a result of reaching the maximum mass transfer rate, while the lower period island near $10^2$ days is a result of $L_2$ overflow.
While reverse mass transfer occurs more often nearing $q=1$, the grid at $q=0.99$ in Figure \ref{fig:Zsun_multi_q} at solar metallicity shows that this is often unstable.
Many binaries at $q=0.99$ reach a contact phase due to both stars evolving on a similar timescale, which causes many massive donors to reach contact off the main sequence even for large separations.
The island of $L_2$ overflow remains and the diagonal unstable mass transfer is caused by the maximum mass transfer rate.
There is still a large range of stable reverse mass transfer present at $q=0.99$, but not as extensive as at $q=0.95$ and many of the primary masses are too small to form a BH at collapse.

The final limitation in the formation of LGRB through the stable reverse-mass-transfer channel is the angular momentum content of the primary at collapse. 
The circles inside the squares in grid slices in Figure \ref{fig:Zsun_multi_q} show the stable reverse-mass-transfer binaries that form an accretion disk during their collapse.
Their colour indicates the mass of the accretion disk that is formed. 
The $q=0.90$ and $q=0.95$ slices show a gradient in the disk mass with donor mass and period.
In general, wider systems produce a less massive accretion disk than shorter-period binaries, and the widest periods with stable reverse mass transfer do not produce a rapidly rotating star at collapse. The boundary of accretion disk formation moves to shorter periods as the donor mass increases and is a result of hydrogen being present when the reverse mass transfer occurs, which inhibits the transfer of angular momentum into the core of the primary.

We explore the gradient in the accretion disk mass as a function of donor mass and period in Appendix \ref{app:disk_gradient}. In short, at the same mass and mass ratio, the accreted angular momentum and mass during the stable reverse-mass-transfer phase decrease with increasing period due to the mass transfer being shorter and, thus, there being less time to diffuse the angular momentum into deeper layers of the accretor. After the stable reverse-mass-transfer phase, the fraction of angular momentum loss is similar across different initial periods, around 80\%. Thus, short-period systems, which gained more angular momentum through mass transfer, reach core carbon depletion with a larger angular momentum reservoir, sufficient to produce an accretion disk during collapse.

The gradient as a function of donor mass originates from a slightly different physical phenomenon. At the same period and mass ratio with an increasing initial donor mass, more angular momentum is gained by the primary during the reverse-mass-transfer phase. However, the initially more massive primary loses more angular momentum and mass due to stellar winds after the stable reverse mass transfer. A lower total angular momentum amount and the increased total mass at core carbon depletion limit the mass of the formed accretion disk. This effect is visible in Figure \ref{fig:Zsun_multi_q}, where at the same period, an initially more massive star in the regime of stable reverse mass transfer produces a smaller or no accretion disk compared to a lower mass primary.

\subsubsection{Metallicity dependence}
\begin{figure}
    \centering
    \includegraphics[width=0.99\linewidth]{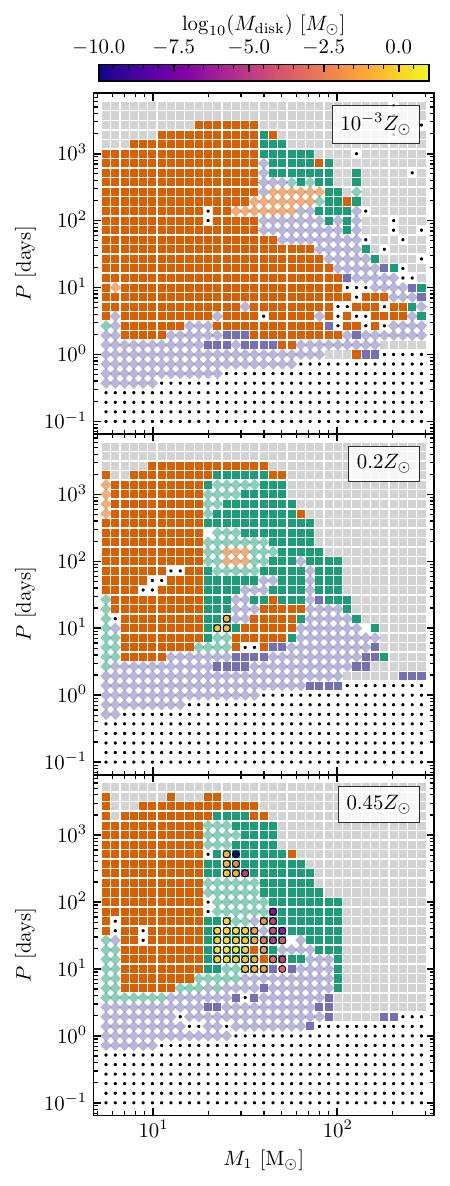}
    \caption{Grids slices at $q=0.95$ at $Z=10^{-3}~\Zsun$, $0.2~\Zsun$, and $0.45~\Zsun$. The same symbol and colours are used as in Figure \ref{fig:Zsun_multi_q}. Less stable reverse mass transfer (green squares) occurs as the metallicity decreases and those that do, do not produce accretion disks upon their collapse.}
    \label{fig:met_q=0.95}
\end{figure}

The criteria for stable reverse mass transfer with sufficient angular momentum at collapse provide a limited parameter space at solar metallicity to potentially produce LGRBs, resulting in a rate of ${\simeq} 10^{-5}$ events yr$^{-1}$ Gpc$^{-3}$.
The efficiency of this formation channel decreases with decreasing metallicity, with the rate going to zero below $0.2~\Zsun$. We attribute this to a combination of effects based on metallicity, which we highlight in Figure~\ref{fig:met_q=0.95}, where we show grid slices of detailed binary models with an initial binary mass ratio of $q=0.95$, for a range of metallicities.

At $Z=0.45~\Zsun$, the region with sufficient angular momentum at collapse in the stable reverse-mass-transfer channel shrinks, as shown in Figure \ref{fig:met_q=0.95}.
Weaker main-sequence winds and partial stripping of the primary star during the first mass transfer phase \citep[][]{gotberg_2017} lead to the formation of a more massive helium core. Additionally, if a hydrogen envelope is still present during the reverse mass transfer, the angular momentum transferred to the primary does not propagate into the core due to the abundance gradient between the core and the hydrogen shell. This results in less angular momentum gain during the reverse mass transfer, and as the primary loses its hydrogen envelope before collapse, it loses the majority of its angular momentum, preventing the formation of an accretion disk. At the same time, the islands of unstable mass transfer from $L_2$ overflow and the maximum $\dot{M}$ are expanding.

Moving to $Z=0.2~\Zsun$, two additional effects are starting to show: short-period systems no longer experience reverse mass transfer, while long-period systems are now experiencing a contact phase, which becomes dominant towards even lower metallicities, as the $Z=10^{-3}~\Zsun$ panel in Figure \ref{fig:met_q=0.95} shows.
In wider binaries, the mass transfer no longer sufficiently strips the primary star for it to reach the Wolf-Rayet wind regime, instead, it finds a thermal equilibrium close to or at its Roche lobe \citep[][]{klencki_2022, ercolino_2024}.
As the companion evolves off its main sequence and also fills its Roche lobe, the system goes into contact, either because the primary is already filling its Roche lobe or because its equilibrium state is near the Roche lobe, such that the radial expansion from mass accretion causes the primary to fill its Roche lobe.
On the other hand, short-period binaries no longer experience reverse mass transfer, because the companion remains compact and hydrogen-burning until the primary reaches carbon exhaustion.
Due to lower stellar wind mass loss, these accretors retain the gained angular momentum from the initial mass transfer phase.
At lower metallicities, the radial expansion of the stars is already less compared to solar metallicity due to the lower opacity, and the increased rotational mixing further limits the radial expansion, which restricts a reverse-mass-transfer phase. The combination of these effects creates a LGRB formation channel that becomes more efficient as metallicity increases, while completely disappearing at metallicities below $0.2~\Zsun$.

\subsection{Stable reverse mass transfer LGRB rate} \label{sec:stable-reverse-MT-rate}

\begin{figure}
    \centering
    \includegraphics[width=\linewidth]{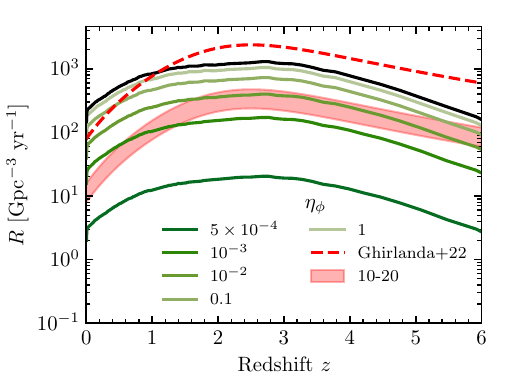}
    \caption{The redshift evolution of the stable reverse-mass-transfer LGRB event rate density for the IllustrisTNG metallicity and star formation history. The total potential population is indicated in black, while the dashed red line shows the intrinsic LGRB rate from \citet{ghirlanda_2022} with the best-fit parameters based on observations. The red shaded area is 10\% to 20\% of this rate. The five green shaded lines indicate the rate density for an energy limit of $10^{51}~\mathrm{erg}$ and different $\eta_\phi$.}
    \label{fig:cosmic_rates}
\end{figure}

The combined requirements of stable reverse mass transfer and the angular momentum reservoir at collapse provide strong constraints on the formation of LGRB progenitors: the reverse-mass-transfer channel only contributes at $Z \geq 0.2~\Zsun$.
As described in Section \ref{sec:method}, we use \texttt{POSYDON} to generate burst stellar populations at eight different metallicities and select systems undergoing stable reverse mass transfer that will form an accretion disk on collapse.
We convolve these populations with the detailed star formation rate and metallicity distribution across cosmic time from the IllustrisTNG simulation in order to compute the cosmic LGRB rate from the stable reverse-mass-transfer channel.
The black line in Figure \ref{fig:cosmic_rates} shows the resulting total cosmic rate of systems that produce an accretion disk on collapse.
This is the maximum rate that can be achieved by reverse-mass-transfer progenitors according to our model, in the presence of a highly efficient jet-launching and powering mechanism, where every system with an accretion disk successfully produces a LGRB. This would correspond to a rate of ${\simeq}100$ events per cubic gigaparsec per year at $z=0$. For comparison, the most recent LGRB rate density estimate based on detailed modelling of the observed population, carefully accounting for selection effects, put the total intrinsic cosmic rate at $z=0$ at $79^{+57}_{-33}$ Gpc$^{-3}$ yr$^{-1}$ \citep{ghirlanda_2022}. 

Given that most observed LGRBs occur in low-metallicity environments, this indicates that some effect must be limiting a large fraction of our stable reverse mass-transfer model progenitors from producing a LGRB. Theoretically, one would expect that such an effect follows from the required energy for the jet to propagate through the stellar envelope and successfully break out of it \citep[e.g.][]{matzner_2003,bromberg_2011,salafia_2020}. The binding energy of the stellar structure at core carbon exhaustion is less than the values of $E_\mathrm{BZ}/\eta_\phi$, when $\eta_\phi=1$. While the actual energy breakout in general depends on the properties of the stellar envelope and also on those of the jet at launch. Here, for simplicity, we model both as a universal threshold on the jet energy. Because we left the efficiency factor $\eta_\phi$ unspecified (see Section~\ref{sec:method}), we set an energy limit for a successful LGRB of $10^{51}~\mathrm{erg}$ based on the typical breakout energy from jet propagation simulation \citep[for example, see][]{harrison_2018}. In what follows, we explore five values for the efficiency, $\eta_\phi$, namely $1$, $0.1$, $10^{-2}$, $10^{-3}$, and $5\times10^{-4}$.

The total rate of disk formation through the stable reverse-mass-transfer channel is comparable to the local rate density of observed LGRBs \citep{ghirlanda_2022}. Given that the inferred fraction of high-metallicity host galaxies is 10\%-20\%, this requires an efficiency that removes 80\%-90\% of events.
The most restrictive efficiency with $\eta_\phi = 5 \times 10^{-4}$ results in a volumetric rate of ${\simeq}2.01$ Gpc$^{-3}$ yr$^{-1}$ at $z=0$; with a total of ${\simeq}15.0$ Gpc$^{-3}$ yr$^{-1}$ events below $z<2.5$. 
With $\eta_\phi = 10^{-3}$, the reverse-mass-transfer channel rate is around 18\% of the latest estimate of the observed cosmic LGRB rate.
This is remarkably close to the observational fraction of solar and super-solar host galaxies of 10-20 per cent.
However, because the normalisation of the observed rate is uncertain due to selection effects, such as the intrinsic energy of the jet, one should be careful to interpret a comparison of our theoretical rates against these observations. These cannot be constrained without considering all formation channels concurrently, which is beyond the scope of this work.
The intrinsic rate density evolution with redshift for different choices in the efficiency is shown in Figure \ref{fig:cosmic_rates}.  We report their rate densities at $z=0$ and averaged within $z=2.5$ in Table~\ref{tab:illustris_local_rates}.

\begin{figure}
    \centering
    \includegraphics{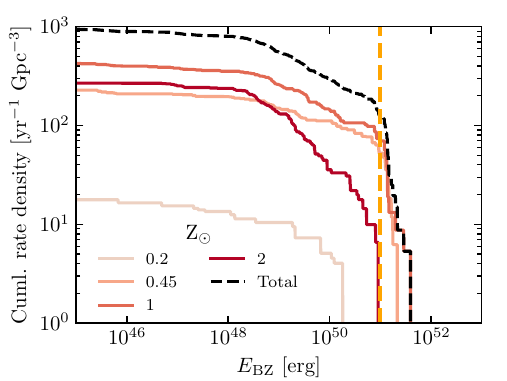}
    \caption{The cumulative rate density ($R[E > E_x]$) within $z<2.5$ from the IllustrisTNG simulation split up per metallicity with $\eta_\phi = 10^{-3}$. The vertical line indicates a typical breakout energy of $10^{51}$ erg.}
    \label{fig:illustrisTNG_energy_budget}
\end{figure}

In Figure \ref{fig:illustrisTNG_energy_budget} we show the cumulative distribution of $E_\mathrm{BZ}$ with $\eta_\phi=10^{-3}$ for all our simulated progenitors within $z<2.5$ (black dashed line) and for progenitors with specific metallicities within the same redshift range (coloured lines). We limit the redshift range to $z<2.5$ to cover most observable LGRBs. Including higher redshift systems reduces the average volumetric rate since fewer events from the stable reverse-mass-transfer channel occur at low metallicity.

Although the progenitor masses at collapse only cover a range from $10$ to $40~\msun$, $E_\mathrm{BZ}$ covers several orders of magnitude from $10^{44}$ to $10^{52}$ erg.
The most energetic events come from the solar metallicity population, although their contribution to the rate changes depending on the star formation history (SFH; see Section \ref{sec:discussion}). The median value of $E_\mathrm{BZ}$ for these models is around ${\simeq}10^{49}~\mathrm{erg}$, but a tail of much higher-energy jets is present, with clustering around $E_\mathrm{BZ}\sim 10^{51}\,\mathrm{erg}$, which causes the steep decrease in the cumulative rate from ${\simeq}100$ Gpc$^{-3}$ yr$^{-1}$ to ${\simeq}10$ Gpc$^{-3}$ yr$^{-1}$ around this energy.

\begin{table} 
\centering
\begin{tabular}{|r|c|c|}
\hline
   & \multicolumn{2}{c|}{Rate[Gpc$^{-3}$ yr$^{-1}$]} \\
\hline
\textbf{$\eta_\phi$} & \textbf{$z=0$ } & \textbf{$z < 2.5$}   \\
\hline
\multicolumn{3}{|c|}{\textbf{IllustrisTNG}} \\
\hline
$5 \times 10^{-4}$ & 2.01 & 15.0 \\
$10^{-3}$ & 15.82 & 126.0 \\
$10^{-2}$ & 39.41 & 297.3 \\
$10^{-1}$ & 78.73 & 549.4 \\
$1$ & 119.62 & 800.0 \\
no cut & 148 & 983 \\
\hline
\multicolumn{3}{|c|}{\textbf{\citet{madau_2017}} } \\
\hline
$5 \times 10^{-4}$ & 0.314 & 3.428 \\
$10^{-3}$ & 1.228 & 14.912 \\
$10^{-2}$ & 14.809 & 191.996 \\
$10^{-1}$ & 29.953 & 369.611 \\
$1$ & 46.785 & 550.654 \\
no cut & 61 & 709 \\
\hline
\multicolumn{3}{|c|}{\textbf{\citet{neijssel_2019}} } \\
\hline
$5 \times 10^{-4}$ & 0.172 & 7.805 \\ 
$10^{-3}$ & 0.428 & 24.105 \\ 
$10^{-2}$ & 11.551 & 259.448 \\ 
$10^{-1}$ & 33.566 & 532.925 \\ 
$1$ & 66.037 & 822.030 \\ 
no cut & 81 & 1049 \\ 
\hline
\end{tabular}
\bigskip
\caption{The cosmic reverse-mass-transfer LGRB rate density from the IllustrisTNG simulation at $z=0$ and the average density in the region $z<2.5$ for an energy limit of $10^{51}~\mathrm{erg}$ with different $\eta_\phi$ values.}
\label{tab:illustris_local_rates}
\end{table}

\section{Discussion} \label{sec:discussion}

While LGRBs have been found in high-metallicity environments, no collapsar formation channel with a positive metallicity relationship has previously been proposed.
We have introduced a new formation channel capable of forming sufficient LGRBs to match observational or theoretical fractional rates in high-metallicity environments.
However, much uncertainty remains in this fraction, as well as in the absolute rate of the total LGRBs and this channel.
Since the main contributors to this uncertainty are the observational biases and hidden LGRB fraction, which would require a population study that includes all possible LGRB formation pathways in order to be explored, we instead focus on the physics which might impact the presence and occurrence of the reverse-mass-transfer LGRB formation channel.

\subsection{Implications of energy limit \label{sec:energy_limit}}

We have implemented a universal energy limit as a proxy for successful jet breakout with different efficiencies, which is equivalent to different energy limits with a constant efficiency, i.e. a $10^{51}~\mathrm{erg}$ limit with a $\eta_\phi = 10^{-3}$ selects the same progenitors as a $10^{54}~\mathrm{erg}$ limit with a $\eta_\phi=1$.
For the stable reverse-mass-transfer channel to produce 10-20\% of the local LGRB rate, a low disk-to-jet-energy conversion or a high jet-breakout-energy threshold is required.
This implies that successful jets are a minority, while choked jets occur an order of magnitude more frequently in the proposed channel. This fraction aligns with \citet{bromberg_2012} and from observed low-luminosity GRBs the inferred intrinsic rate can be up to an order of magnitude higher than their high-luminosity counterpart \citep{liang_2007, guetta_2007, virgili_2009}.
It suggests the existence of a large population of relativistic supernovae with potentially low-luminosity gamma-ray emission caused by these choked jets \citep[see, for example,][]{soderberg_2006, wang_2007, nakar_2012}.

Despite the broad range of jet energies, a high jet-breakout-energy limit or low disk-to-jet-energy conversion leads to a narrow energy range for successful jets, which declines steeply above the minimum energy. Moreover, successful jets possess only slightly more energy than needed for breakout implying the presence of a strong cocoon with comparable energy to the jet.
For a structured jet, this results in a shallow angular profile \citep{salafia_2020}. Thus, the apparent diversity of LGRBs from this channel would primarily be due to viewing angles rather than an actual broad intrinsic energy distribution.

\subsection{Type Ic broad-line and potential LGRB companions}

LGRBs and SN Ic-BL have been associated with one another, but this relationship is not universal; not all Type Ic-BL are accompanied by a LGRB and vice versa \citep{modjaz_2016, japelj_2018}. Type Ic-BL without a LGRBs have distinct spectral features from those with a LGRB \citep{modjaz_2016}, and could indicate a failed jet breakout \citep[see][and references therein]{corsi_2021}. In general, Type Ic-BL are associated more with super-solar host galaxies compared to LGRBs \citep{japelj_2018}, which aligns with the positive metallicity dependence of the stable reverse-mass-transfer LGRB channel. Additionally, if this channel produces the observed high-metallicity population of LGRBs, it produces an order of magnitude more choked jets compared to successful jets (see Section~\ref{sec:energy_limit}). Although the observed relative rate of SN Ic-BL to core-collapse SNe is poorly constrained, based on a single SN Ic-BL observation in the volume-limited Lick Observatory Supernova Search, SN Ic-BL make up 4\% of stripped-envelope supernovae \citep{shivvers_2017}. Combined with the stripped-envelope supernova rate from \citet{frohmaier_2021}, this results in a rough SN Ic-BL rate estimate of ${\sim}300$ Gpc$^{-3}$ yr$^{-1}$.  Assuming that all disk-forming systems produce a jet and $\eta_\phi=10^{-3}$, ${\simeq}86$ choked jets occur per Gpc$^{3}$ per year at $z=0$ from the stable reverse-mass-transfer channel. This is comparable to the observationally inferred rate and could imply that either more dominant LGRB formation channels must have a smaller ratio of choked-to-successful jets to not overproduce the SN Ic-BL or not all failed LGRB accretion disk systems produce a SN Ic-BL.

More puzzling are the three local LGRB events without an associated Type Ic-BL \citep[e.g.][]{fynbo_2006, gal-yam_2006, xu_2009}. GRB 111005A, one of the three LGRB in a high-metallicity environment, is one such LGRB \citep{tanga_2018}. Their origin is poorly understood in the framework of the jet moving outwards through the stellar structure, producing a Type Ic-BL supernova, and could indicate a new LGRB formation mechanism \citep[e.g.][]{gal-yam_2006}. Either way, the near-solar metallicity environment does not have to be an indicator for a unique mechanism, as the stable reverse-mass-transfer channel can produce LGRBs at those metallicities through the standard collapsar scenario.

A special feature of the stable reverse-mass-transfer channel is the presence of a (partially) stripped companion at collapse. Its mass ranges from $10~\msun$ to $50~\msun$ with most potential LGRB progenitors having a companion around $15~\msun$ at collapse. While no clear LGRB in a binary system has been observed \citep[although see][]{ivywang_2022}, the presence of a companion will affect the GRB emission \citep{zou_2021}.
Furthermore, several local Type Ic-BL supernovae have been investigated for companions with limited success. For SN 2002ap, an upper limit $\sim10~\msun$ for a companion has been set \citep{zapartas_2017a}, while a potential companion in SN 2016coj is under debate \citep{terreran_2019}. The stripped nature of the companion could complicate its observability. However, a $15~\msun$ helium star would appear as a Wolf-Rayet star at $\Zsun$ \citep{shenar_2020}. The detection of or upper limits on a companion in nearby LGRB and SN Ic-BL can provide an independent observable on their formation mechanisms.

\subsection{Jet physics}

The relationship between the progenitor star and jet energy remains highly uncertain and general relativistic, radiation, magnetohydrodynamic simulations have shown that the efficiencies depend on the state of the disk \citep{tchekhovskoy_2010}.
A popular approach is the magnetically arrested disk (MAD) state because it produces powerful jets even for slowly rotating BHs and causes rapidly spinning BHs to spin down \citep{jacquemin-ide_2024}.
In this state, the mass accretion rate and spin-down timescale determine the jet power. In such disks $\eta_\phi=1$, while $\eta_a$ can be fitted following \citet{lowell_2024}. This can result in efficiencies above 1 due to the extraction of angular momentum from the BH and results in high-energy jets.
These even occur for slowly spinning BHs as long as the progenitor has sufficient angular momentum to form an accretion disk \citep{fryer_2025}.
However, recent simulations of collapsars show that the accretion rate remains sufficiently high that neutrino cooling keeps the disk thin and the MAD state is not reached when assuming a proto-neutron star is the origin of the BH magnetic field \citep{gottlieb_2024}.

In a non-MAD state, the magnetic field is thought to determine the jet power, which \citet{kawanaka_2013} estimated using the ram pressure at the inner-most stable orbit around the BH following \citet{beckwith_2009}. This provides a $L_\mathrm{BZ} \propto \dot{M}$ relationship, similar to that implemented here and as found by \citet{blandford_1977} and \citet{heger_2003}, although the latter uses a $10^{0.1/(1-a_\mathrm{BH})-0.1}$ spin-dependence.
Although dependent on the progenitor structure, many of the energetic LGRBs have a high accretion rate through their collapse, making the chosen relationship for $L_\mathrm{BZ}$ reasonable. On the other hand, $E\propto a^2$ might require improvement \citep[see, for example:][]{tchekhovskoy_2010,russell_2013}. Furthermore, we do not model the extraction of angular momentum from the BH while it collapses, which will reduce the total extracted energy. On the other hand, we partially account for this with different energy limits.
As a consensus arises on the jet power mechanism in the future, this relationship can be improved.

Besides the potential energy budget to form a jet, an essential component in LGRB formation in the collapsar scenario is the jet breakout through the stellar structure.
The jet breakout depends on the injected energy and active duration of the central engine, and the density profile of the stellar structure.
While the internal stellar density profile at core-carbon exhaustion is available, we do not evolve the final evolutionary phases due to the required computational time, and, thus, do not have a representative profile at collapse available. We verify that the free-fall timescales of the core-carbon exhausted models are similar to observed $T_{90}$ values (See Appendix~\ref{app:free-fall-timescale}).
Furthermore, the injected energy depends on several poorly constrained parameters, such as the injection opening angle, the disk mass to jet energy conversion, and the total jet energy emitted in the gamma-ray.
A full end-to-end pipeline with computationally intensive hydrodynamical simulations is required to accurately model the link between the progenitor model and the jet properties from the collapsar scenario \citep{gottlieb_2024}.

In general, failed jet breakouts are more likely to occur in the low-potential energy models, as these models provide insufficient mass to sustain a successful jet propagation through the stellar envelope. 
In contrast, a high-potential energy model is more likely to overcome the energy requirement for jet breakout and cause an observable LGRB.
The proposed limits show that only the most energetic systems from the stable reverse-mass-transfer channel must be able to lead to LGRBs in order to match the observed fractions of high-metallicity ones at $z<2$, which can be achieved with an $\eta_\phi \lesssim 10^{-3}$.

\subsection{Star formation history}

\begin{figure}
    \centering
    \includegraphics{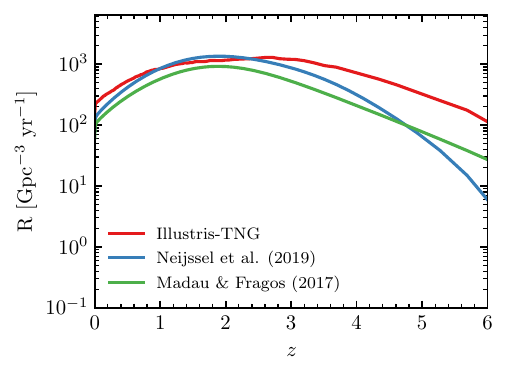}
    \caption{The total potential stable reverse-mass-transfer LGRB rate density over cosmic time from the IllustrisTNG simulation, \citet{madau_2017} with a log-normal metallicity distribution with $\sigma=0.5$, and \citet{neijssel_2019} with a ln-normal distribution with $\sigma=0.39$. No energy requirements are implemented.}
    \label{fig:cosmic_SFH}
\end{figure}

Although the total star formation rate density in the local Universe is relatively well constrained, the metallicity evolution and spread are less understood.
It is the high-metallicity-specific star formation rate that gives rise to uncertainty in the shape of the stable reverse-mass-transfer LGRBs and in the weighting of each model over redshift.
The former can be seen in Figure \ref{fig:cosmic_SFH}, where we compare the LGRB rates over redshift from the IllustrisTNG simulation with the two empirical star formation history prescriptions described in Section \ref{sec:method}. 
The metallicity distribution redistributes the weight between populations of different metallicities but does not alter the shape of the potential LGRB energy distribution per metallicity (see Appendix \ref{app:sfh}).
The ln-normal metallicity distribution from \citet{neijssel_2019} produces a very peaked metallicity-specific star formation rate, reducing the contribution from the reverse-mass-transfer LGRB channel at high redshift to the LGRB rate, while the higher metallicity and wider distribution from \citet{bavera_2020} allow for more near-solar metallicity star formation at higher redshift.

The IllustrisTNG simulation, on the other hand, is rapidly enriched compared to the \citet{neijssel_2019} prescription, which boosts the LGRBs from the stable reverse-mass-transfer channel at high redshift and is similar to the metallicity distribution from \citet{madau_2017}.
Due to its metallicity distribution, the \citet{madau_2017} rate extends high-metallicity star formation to higher redshifts, resulting in even higher rates than the fast-enriching TNG simulation. 
Despite these differences at high redshift, the differences in the local Universe ($z<2.5$) are of a factor of 2.

A stronger effect in the local Universe comes from the energy cuts.
Due to the reweighing of each LGRB progenitor, the cumulative potential energy distribution is significantly different per star formation history, thus, the energy cut can suddenly remove more events.
For example, the most restrictive cut in Figure \ref{fig:illustrisTNG_energy_budget} reduces the stable reverse-mass-transfer LGRB rate to 2.01 LGRB Gpc$^{-3}$ yr$^{-1}$, while the \citet{neijssel_2019} prescription in Figure \ref{fig:cosmic_neijssel} reduces it to 0.17 LGRB Gpc$^{-3}$ yr$^{-1}$.
This is due to the energy distribution at each metallicity, which each SFH rebalances, as can be seen in Appendix~\ref{app:sfh}. As a result, the choice in mean metallicity and metallicity spread has a major impact on the final rate and the limit of the jet energy.

\subsection{Additional caveats}

An essential step in the metallicity-dependence of the stable nature of the reverse mass transfer is the further stripping of the partially stripped primary star by stellar winds.
In most cases, before the first mass transfer, the donor star lies in the metallicity-dependent mass loss regime for OB stars from \citet{vink_2001}. At high metallicities, this already removes significant material from the primary star, such that it reaches the strong Wolf-Rayet winds after the first mass transfer phase ends. These winds quickly remove any remaining hydrogen envelope. Their strength determines if the star becomes fully stripped and the amount of angular momentum remaining after a reverse mass transfer event.
The Wolf-Rayet and the main-sequence massive star mass loss become less with decreasing metallicity, reducing the possibility for the primary to become fully stripped.
At the same time, the mass transfer is less efficient at removing the complete hydrogen envelope at lower metallicities than solar. This is due to the primary star finding an equilibrium state close to its Roche lobe with a significant hydrogen envelope left \cite{klencki_2022}, which is an internal property of the star and not dependent on the stellar wind prescriptions.

If the metallicity dependence for O stars follows a shallower relation of $\dot{M} \sim Z^{0.42}$ as suggested by \citet{vink_2021}, the stable reverse-mass-transfer scenario is extended to lower metallicities.
On the other hand, empirically derived mass loss rates of the Large and Small Magellanic Clouds have indicated stronger metallicity-dependencies for massive stars of $Z^2$ and $ Z^{0.83}$, respectively \citep{mokiem_2007a, ramachandran_2019}.
Either way, the number of stable reverse-mass-transfer systems will increase or decrease depending on the metallicity-dependence of the massive star winds, which will impact the preferred potential energy limit to match the observed high-metallicity LGRB fraction.
Irrespective of the metallicity-dependence of the mass-loss rate, the mass-loss rates for OB stars at solar metallicity remain similar \citep{vink_2021}, thus minimally affecting the contribution of the near-solar metallicity populations.

After the stable reverse-mass-transfer phase, the surface hydrogen fraction of the primary can return to above 0.4, which is the boundary for the Wolf-Rayet winds. As such, the stellar winds for OB stars are applied to this phase of the evolution, which will transition back into the Wolf-Rayet winds as the small hydrogen layer is stripped and the surface hydrogen fraction reaches below 0.4. The transition to the \citet{vink_2001} winds means a lower mass loss in comparison to the \citet{nugis_2000} Wolf-Rayet winds are applied. If the mass loss during this phase were treated as a Wolf-Rayet wind, more mass and angular momentum would be lost in the final phase, which would shrink the number of primaries with sufficient angular momentum to produce an accretion disk. On the other hand, the total mass and angular momentum loss strongly depend on how far the core helium burning has progressed when the stable reverse mass transfer takes place. Generally, for potential LGRB progenitors in the stable reverse-mass-transfer channel, the interaction occurs late in the core helium-burning phase of the primary. Otherwise, the hydrogen envelope is not yet fully stripped, which inhibits the transport of angular momentum into the helium core. With no clear observations of hydrogen-accreted helium stars, the exact treatment of their mass loss is unclear.

The first mass transfer in the binary can result in rejuvenation of the companion, where the efficiency of semi-convection affects the parameter range for which this occurs \citep[e.g.][]{braun_1995}. The {\tt POSYDON} v2 grids use $\alpha_\mathrm{sc}=0.1$ (for more details, see \citet{andrews_2025}), a relatively inefficient semi-convection, which limits the parameter space for rejuvenation. Furthermore, the stable reverse-mass-transfer channel predominantly occurs around $q\approx 0.95$ at $\Zsun$. This implies that the companion is far along the main-sequence when the first mass transfer takes place. This limits the effect of the accreted material on the further evolution and the timing of the reverse mass transfer phase. It is unlikely that more efficient semi-convection can lead to more rejuvenation and less stable reverse mass transfer in this regime.
At low metallicities, the first mass transfer does affect the companions, and they remain compact until the primary reaches carbon-exhaustion. More efficient semi-convection could result in more rejuvenation, while less efficient semi-convection could limit the rejuvenation and allow for a reverse mass transfer before the primary collapses. However, these interactions are likely to be unstable, since the primary will not be fully stripped at the moment of the interaction. As such, the choice of semi-convection parameter should not affect the overall results of this work.

Although we self-consistently calculate the collapse of the progenitor into the BH, we first determine the formation of a BH using the \citet{patton_2020} prescription.
Individual models in this explodability landscape are sensitive to the numerical setup, such as nuclear reaction network and spatial resolution \citep{farmer_2016}. These variations could be indicative of inherent chaos in the final stages of massive star evolution \citep{sukhbold_2018}, but introduce uncertainty into the mapping between the core-carbon depleted structure and compact object formation.
Additionally, this prescription assumes that implosions lead to BH formation, while explosions lead to neutron star formation.

While the majority of LGRB progenitors from the stable reverse-mass-transfer channel are well into the BH formation regime, their parameter space at low masses is limited by neutron star formation, as shown in the right grid slice in Figure \ref{fig:Zsun_multi_q}. Allowing for more BH formation would increase the contribution of this formation channel. However, the additional parameter space is limited, because the timescale difference between the binary components is too large at low masses for reverse mass transfer to occur.
Additionally, the \citet{patton_2020} prescription uses the average central carbon abundance and carbon-oxygen core mass at helium depletion to determine the explodability. Late-stage binary interactions, such as mass transfer after core helium ignition, can alter the stellar structure sufficiently to boost explodability \citep{laplace_2021}. Despite that, the {\tt POSYDON} {\tt MESA} models are evolved until carbon depletion, due to the \citet{patton_2020} prescription, late-stage mass transfer effects are not considered.

\begin{figure}
    \centering
    \includegraphics[]{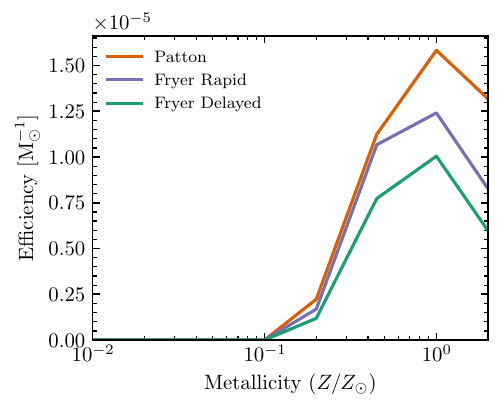}
    \caption{The number of stable reverse-mass-transfer LGRBs per solar mass star formation (efficiency) as a function of metallicity. These metallicity bias functions are shown for the two \citet{fryer_2012} (rapid in purple; delayed in green) and the \citet{patton_2020} (orange) prescriptions.}
    \label{fig:other_SN}
\end{figure}

Switching the remnant mass prescription to the rapid or delayed prescriptions from \citet{fryer_2012} should give an indicator of how different prescriptions for estimating the remnant masses affect our results. These prescriptions use the core properties at carbon depletion to determine BH formation through fallback onto a proto-compact object. Figure \ref{fig:other_SN} shows minimal differences between their metallicity bias functions, which propagate to nearly the same cosmic rates. Slightly less stable reverse-mass-transfer LGRBs occur with the \citet{fryer_2012} prescriptions. The star formation history, metallicity distributions, and energy cut have a significantly stronger impact on this rate than what stars collapse as BHs.

\section{Conclusions} \label{sec:conclusion}

Although LGRBs are more frequently observed in low metallicity environments, 10 to 20\% of LGRB host galaxies at $z<2$ exhibit solar and super-solar metallicities \citep{vergani_2015, palmerio_2019, graham_2023}. Moreover, several LGRB events have been associated with high-metallicity explosion sites \citep{levesque_2010a, michalowski_2018, heintz_2018}.
Traditional binary LGRB formation pathways predominantly have a low-metallicity preference and are rare to non-existent at solar metallicity.
In this work, we proposed a new collapsar LGRB formation pathway through stable reverse mass transfer that is limited to near solar-metallicity stellar binaries, creating a unique formation mechanism which has the opposite dependence on metallicity compared to traditional scenarios. 

In this formation scenario, a partially stripped star is formed by mass transfer after hydrogen depletion (case B). At solar metallicity, it becomes completely stripped by stellar winds from before (OB-type stellar winds) and after the mass transfer (Wolf-Rayet star winds). These wind regimes introduce the metallicity dependence to this LGRB formation channel. The stripped star becomes rapidly rotating when case B mass transfer is initiated by the initially less massive star in the binary system. Mass accretion onto the stripped star carries sufficient angular momentum, which the star is able to maintain till the end of its evolution and form a substantial accretion disk around the proto-BH during collapse.

The existence of this stable reverse-mass-transfer pathway for forming rapidly rotating stripped stars at core-collapse is a robust prediction of population synthesis with {\tt POSYDON}.
However, linking these progenitors to jet breakout and observable LGRBs on a population scale is significantly affected by theoretical uncertainties and observational biases. As such, the absolute rate is highly dependent on the star formation history, metallicity evolution and successful LGRB energy limits.
Nevertheless, sufficient angular momentum, and thus energy, is available in the structure to power jets with LGRB energies around $10^{52}$ erg.
Moreover, the rate is sufficiently high to produce 10 to 20\% of the intrinsic observed LGRB rate from \citet{ghirlanda_2022}. This contribution is consistent with the fraction of LGRBs observed in high-metallicity environments. 

However, the absolute rate of LGRBs from the stable reverse-mass-transfer channel can vary between 1 and 100 LGRB Gpc$^{-3}$ yr$^{-1}$ depending on the star formation rate, metallicity evolution, and energy threshold for a successful LGRB.
To refine the contribution of this high-metallicity LGRB formation pathway, a comprehensive, simultaneous population analysis of all LGRB formation channels is required, which will benefit from a more extensive exploration of the link between rapidly rotating stellar models and successful LGRB observations.

\begin{acknowledgements}

We thank the anonymous reviewer for their insightful comments on the manuscript.
The POSYDON project is supported primarily by two sources: the Swiss National Science Foundation (PI Fragos, project numbers PP00P2\_211006 and CRSII5\_213497) and the Gordon and Betty Moore Foundation (PI Kalogera, grant award GBMF8477). 

MMB was supported by the Boninchi Foundation, the project number CRSII5\_21349, and the Swiss Government Excellence Scholarship.
OSS acknowledges funding by the INAF Finanziamento per la Ricerca Fondamentale grant no. 1.05.23.04.04. 
GG and OSS acknowledge financial support by the INAF Grant ``GRB PrOmpt Emis-
sion Modular Simulator'' (POEMS -- 1.05.23.06.04) and European Union-Next Generation EU, PRIN 2022 RFF M4C21.1 (PEACE -- 202298J7KT).
JJA acknowledges support for Program number (JWST-AR-04369.001-A) provided through a grant from the STScI under NASA contract NAS5-03127.
PMS, SG, KAR, and MS were supported by the project numbers GBMF8477 and GBMF12341.
K.A.R.\ is also supported by the Riedel Family Fellowship and thanks the LSSTC Data Science Fellowship Program, which is funded by LSSTC, NSF Cybertraining Grant No.\ 1829740, the Brinson Foundation, and the Moore Foundation; their participation in the program has benefited this work.
KK is supported by a fellowship program at the Institute of Space Sciences (ICE-CSIC) funded by the program Unidad de Excelencia Mar\'ia de Maeztu CEX2020-001058-M.
Z.X. acknowledges support from the China Scholarship Council (CSC).
EZ acknowledges support from the Hellenic Foundation for Research and Innovation (H.F.R.I.) under the “3rd Call for H.F.R.I. Research Projects to support Post-Doctoral Researchers” (Project No: 7933).

\end{acknowledgements}

%
%

\bibliographystyle{aa}
\bibliography{bibtex}

\begin{thebibliography}{147}
\expandafter\ifx\csname natexlab\endcsname\relax\def\natexlab#1{#1}\fi

\bibitem[{Abbott {et~al.}(2017{\natexlab{a}})Abbott, Abbott, Abbott, Acernese,
  Ackley, Adams, Adams, Addesso, Adhikari, Adya, Affeldt, Afrough, Agarwal,
  Agathos, Agatsuma, Aggarwal, Aguiar, Aiello, Ain, Ajith, Allen, Allen,
  Allocca, Aloy, Altin, Amato, Ananyeva, Anderson, Anderson, Angelova, Antier,
  Appert, Arai, Araya, Areeda, Arnaud, Arun, Ascenzi, Ashton, Ast, Aston,
  Astone, Atallah, Aufmuth, Aulbert, AultONeal, Austin, {Avila-Alvarez}, Babak,
  Bacon, Bader, Bae, Baker, Baldaccini, Ballardin, Ballmer, Banagiri, Barayoga,
  Barclay, Barish, Barker, Barkett, Barone, Barr, Barsotti, Barsuglia, Barta,
  Bartlett, Bartos, Bassiri, Basti, Batch, Bawaj, Bayley, Bazzan, B{\'e}csy,
  Beer, Bejger, Belahcene, Bell, Berger, Bergmann, Bero, Berry, Bersanetti,
  Bertolini, Betzwieser, Bhagwat, Bhandare, Bilenko, Billingsley, Billman,
  Birch, Birney, Birnholtz, Biscans, Biscoveanu, Bisht, Bitossi, Biwer,
  Bizouard, Blackburn, Blackman, Blair, Blair, Blair, Bloemen, Bock, Bode,
  Boer, Bogaert, Bohe, Bondu, Bonilla, Bonnand, Boom, Bork, Boschi, Bose,
  Bossie, Bouffanais, Bozzi, Bradaschia, Brady, Branchesi, Brau, Briant,
  Brillet, Brinkmann, Brisson, Brockill, Broida, Brooks, Brown, Brown, Brunett,
  Buchanan, Buikema, Bulik, Bulten, Buonanno, Buskulic, Buy, Byer, Cabero,
  Cadonati, Cagnoli, Cahillane, Bustillo, Callister, Calloni, Camp, Canepa,
  Canizares, Cannon, Cao, Cao, Capano, Capocasa, Carbognani, Caride, Carney,
  Diaz, Casentini, Caudill, Cavagli{\`a}, Cavalier, Cavalieri, Cella, Cepeda,
  {Cerd{\'a}-Dur{\'a}n}, Cerretani, Cesarini, Chamberlin, Chan, Chao, Charlton,
  Chase, {Chassande-Mottin}, Chatterjee, Chatziioannou, Cheeseboro, Chen, Chen,
  Chen, Cheng, Chia, Chincarini, Chiummo, Chmiel, Cho, Cho, Chow, Christensen,
  Chu, Chua, Chua, Chung, Chung, Ciani, Ciolfi, Cirelli, Cirone, Clara, Clark,
  Clearwater, Cleva, Cocchieri, Coccia, Cohadon, Cohen, Colla, Collette,
  Cominsky, Jr, Conti, Cooper, Corban, Corbitt, {Cordero-Carri{\'o}n}, Corley,
  Cornish, Corsi, Cortese, Costa, Coughlin, Coughlin, Coulon, Countryman,
  Couvares, Covas, Cowan, Coward, Cowart, Coyne, Coyne, Creighton, Creighton,
  Cripe, Crowder, Cullen, Cumming, Cunningham, Cuoco, Canton, D{\'a}lya,
  Danilishin, D'Antonio, Danzmann, Dasgupta, Costa, Dattilo, Dave, Davier,
  Davis, Daw, Day, De, DeBra, Degallaix, Laurentis, Del{\'e}glise, Pozzo,
  Demos, Denker, Dent, Pietri, Dergachev, Rosa, DeRosa, Rossi, DeSalvo,
  de~Varona, Devenson, Dhurandhar, D{\'i}az, Fiore, Giovanni, Girolamo, Lieto,
  Pace, Palma, Renzo, Doctor, Dolique, Donovan, Dooley, Doravari, Dorrington,
  Douglas, {\'A}lvarez, Downes, Drago, Dreissigacker, Driggers, Du, Ducrot,
  Dupej, Dwyer, Edo, Edwards, Effler, Eggenstein, Ehrens, Eichholz, Eikenberry,
  Eisenstein, Essick, Estevez, Etienne, Etzel, Evans, Evans, Factourovich,
  Fafone, Fair, Fairhurst, Fan, Farinon, Farr, Farr, {Fauchon-Jones}, Favata,
  Fays, Fee, Fehrmann, Feicht, Fejer, {Fernandez-Galiana}, Ferrante, Ferreira,
  Ferrini, Fidecaro, Finstad, Fiori, Fiorucci, Fishbach, Fisher, {Fitz-Axen},
  Flaminio, Fletcher, Fong, Font, Forsyth, Forsyth, Fournier, Frasca, Frasconi,
  Frei, Freise, Frey, Frey, Fries, Fritschel, Frolov, Fulda, Fyffe, Gabbard,
  Gadre, Gaebel, Gair, Gammaitoni, Ganija, Gaonkar, {Garcia-Quiros}, Garufi,
  Gateley, Gaudio, Gaur, Gayathri, Gehrels, Gemme, Genin, Gennai, George,
  George, Gergely, Germain, Ghonge, Ghosh, Ghosh, Ghosh, Giaime, Giardina,
  Giazotto, Gill, Glover, Goetz, Goetz, Gomes, Goncharov, Gonz{\'a}lez, Castro,
  Gopakumar, Gorodetsky, Gossan, Gosselin, Gouaty, Grado, Graef, Granata,
  Grant, Gras, Gray, Greco, Green, Gretarsson, Groot, Grote, Grunewald,
  Gruning, Guidi, Guo, Gupta, Gupta, Gushwa, Gustafson, Gustafson, Halim, Hall,
  Hall, Hamilton, Hammond, Haney, Hanke, Hanks, Hanna, Hannam, Hannuksela,
  Hanson, Hardwick, Harms, Harry, Harry, Hart, Haster, Haughian, Healy,
  Heidmann, Heintze, Heitmann, Hello, Hemming, Hendry, Heng, Hennig,
  Heptonstall, Heurs, Hild, Hinderer, Hoak, Hofman, Holt, Holz, Hopkins, Horst,
  Hough, Houston, Howell, Hreibi, Hu, Huerta, Huet, Hughey, Husa, Huttner,
  {Huynh-Dinh}, Indik, Inta, Intini, Isa, Isac, Isi, Iyer, Izumi, Jacqmin,
  Jani, Jaranowski, Jawahar, {Jim{\'e}nez-Forteza}, Johnson,
  {Johnson-McDaniel}, Jones, Jones, Jonker, Ju, Junker, Kalaghatgi, Kalogera,
  Kamai, Kandhasamy, Kang, Kanner, Kapadia, Karki, Karvinen, Kasprzack,
  Kastaun, Katolik, Katsavounidis, Katzman, Kaufer, Kawabe, K{\'e}f{\'e}lian,
  Keitel, Kemball, Kennedy, Kent, Key, Khalili, Khan, Khan, Khan, Khazanov,
  Kijbunchoo, Kim, Kim, Kim, Kim, Kim, Kim, Kimbrell, King, King,
  {Kinley-Hanlon}, Kirchhoff, Kissel, Kleybolte, Klimenko, Knowles, Koch,
  Koehlenbeck, Koley, Kondrashov, Kontos, Korobko, Korth, Kowalska, Kozak,
  Kr{\"a}mer, Kringel, Krishnan, Kr{\'o}lak, Kuehn, Kumar, Kumar, Kumar, Kuo,
  Kutynia, Kwang, Lackey, Lai, Landry, Lang, Lange, Lantz, Lanza,
  {Lartaux-Vollard}, Lasky, Laxen, Lazzarini, Lazzaro, Leaci, Leavey, Lee, Lee,
  Lee, Lee, Lee, Lehmann, Lenon, Leonardi, Leroy, Letendre, Levin, Li, Linker,
  Littenberg, Liu, Lo, Lockerbie, London, Lord, Lorenzini, Loriette, Lormand,
  Losurdo, Lough, Lousto, Lovelace, L{\"u}ck, Lumaca, Lundgren, Lynch, Ma,
  Macas, Macfoy, Machenschalk, MacInnis, Macleod, Hernandez,
  {Maga{\~n}a-Sandoval}, Zertuche, Magee, Majorana, Maksimovic, Man, Mandic,
  Mangano, Mansell, Manske, Mantovani, Marchesoni, Marion, M{\'a}rka,
  M{\'a}rka, Markakis, Markosyan, Markowitz, Maros, Marquina, Martelli,
  Martellini, Martin, Martin, Martynov, Mason, Massera, Masserot, Massinger,
  {Masso-Reid}, Mastrogiovanni, Matas, Matichard, Matone, Mavalvala, Mazumder,
  McCarthy, McClelland, McCormick, McCuller, McGuire, McIntyre, McIver,
  McManus, McNeill, McRae, McWilliams, Meacher, Meadors, Mehmet, Meidam,
  {Mejuto-Villa}, Melatos, Mendell, Mercer, Merilh, Merzougui, Meshkov,
  Messenger, Messick, Metzdorff, Meyers, Miao, Michel, Middleton, Mikhailov,
  Milano, Miller, Miller, Miller, Millhouse, {Milovich-Goff}, Minazzoli,
  Minenkov, Ming, Mishra, Mitra, Mitrofanov, Mitselmakher, Mittleman, Moffa,
  Moggi, Mogushi, Mohan, Mohapatra, Montani, Moore, Moraru, Moreno, Morriss,
  Mours, {Mow-Lowry}, Mueller, Muir, Mukherjee, Mukherjee, Mukherjee, Mukund,
  Mullavey, Munch, Mu{\~n}iz, Muratore, Murray, Napier, Nardecchia,
  Naticchioni, Nayak, Neilson, Nelemans, Nelson, Nery, Neunzert, Nevin,
  Newport, Newton, Ng, Nguyen, Nichols, Nielsen, Nissanke, Nitz, Noack, Nocera,
  Nolting, North, Nuttall, Oberling, O'Dea, Ogin, Oh, Oh, Ohme, Okada, Oliver,
  Oppermann, Oram, O'Reilly, Ormiston, Ortega, O'Shaughnessy, Ossokine,
  Ottaway, Overmier, Owen, Pace, Page, Page, Pai, Pai, Palamos, Palashov,
  Palomba, {Pal-Singh}, Pan, Pan, Pang, Pang, Pankow, Pannarale, Pant,
  Paoletti, Paoli, Papa, Parida, Parker, Pascucci, Pasqualetti, Passaquieti,
  Passuello, Patil, Patricelli, Pearlstone, Pedraza, Pedurand, Pekowsky, Pele,
  Penn, Perez, Perreca, Perri, Pfeiffer, Phelps, Piccinni, Pichot,
  Piergiovanni, Pierro, Pillant, Pinard, Pinto, Pirello, Pitkin, Poe, Poggiani,
  Popolizio, Porter, Post, Powell, Prasad, Pratt, Pratten, Predoi, Prestegard,
  Prijatelj, Principe, Privitera, Prodi, Prokhorov, Puncken, Punturo, Puppo,
  P{\"u}rrer, Qi, Quetschke, Quintero, {Quitzow-James}, Raab, Rabeling,
  Radkins, Raffai, Raja, Rajan, Rajbhandari, Rakhmanov, Ramirez,
  {Ramos-Buades}, Rapagnani, Raymond, Razzano, Read, Regimbau, Rei, Reid,
  Reitze, Ren, Reyes, Ricci, Ricker, Rieger, Riles, Rizzo, Robertson, Robie,
  Robinet, Rocchi, Rolland, Rollins, Roma, Romano, Romel, Romie, Rosi{\'n}ska,
  Ross, Rowan, R{\"u}diger, Ruggi, Rutins, Ryan, Sachdev, Sadecki, Sadeghian,
  Sakellariadou, Salconi, Saleem, Salemi, Samajdar, Sammut, Sampson, Sanchez,
  Sanchez, {Sanchis-Gual}, Sandberg, Sanders, Sassolas, Sathyaprakash, Saulson,
  Sauter, Savage, Sawadsky, Schale, Scheel, Scheuer, Schmidt, Schmidt,
  Schnabel, Schofield, Sch{\"o}nbeck, Schreiber, Schuette, Schulte, Schutz,
  Schwalbe, Scott, Scott, Seidel, Sellers, Sengupta, Sentenac, Sequino,
  Sergeev, Shaddock, Shaffer, Shah, Shahriar, Shaner, Shao, Shapiro, Shawhan,
  Sheperd, Shoemaker, Shoemaker, Siellez, Siemens, Sieniawska, Sigg, Silva,
  Singer, Singh, Singhal, Sintes, Slagmolen, Smith, Smith, Smith, Somala, Son,
  Sonnenberg, Sorazu, Sorrentino, Souradeep, Spencer, Srivastava, Staats,
  Staley, Steinke, Steinlechner, Steinlechner, Steinmeyer, Stevenson, Stone,
  Stops, Strain, Stratta, Strigin, Strunk, Sturani, Stuver, Summerscales, Sun,
  Sunil, Suresh, Sutton, Swinkels, Szczepa{\'n}czyk, Tacca, Tait, Talbot,
  Talukder, Tanner, T{\'a}pai, Taracchini, Tasson, Taylor, Taylor, Tewari,
  Theeg, Thies, Thomas, Thomas, Thomas, Thorne, Thorne, Thrane, Tiwari, Tiwari,
  Tokmakov, Toland, Tonelli, Tornasi, {Torres-Forn{\'e}}, Torrie,
  T{\"o}yr{\"a}, Travasso, Traylor, Trinastic, Tringali, Trozzo, Tsang, Tse,
  Tso, Tsukada, Tsuna, Tuyenbayev, Ueno, Ugolini, Unnikrishnan, Urban, Usman,
  Vahlbruch, Vajente, Valdes, van Bakel, van Beuzekom, van~den Brand, Broeck,
  {Vander-Hyde}, van~der Schaaf, van Heijningen, van Veggel, Vardaro, Varma,
  Vass, Vas{\'u}th, Vecchio, Vedovato, Veitch, Veitch, Venkateswara,
  Venugopalan, Verkindt, Vetrano, Vicer{\'e}, Viets, Vinciguerra, Vine, Vinet,
  Vitale, Vo, Vocca, Vorvick, Vyatchanin, Wade, Wade, Wade, Walet, Walker,
  Wallace, Walsh, Wang, Wang, Wang, Wang, Wang, Ward, Warner, Was, Watchi,
  Weaver, Wei, Weinert, Weinstein, Weiss, Wen, Wessel, We{\ss}els, Westerweck,
  Westphal, Wette, Whelan, Whitcomb, Whiting, Whittle, Wilken, Williams,
  Williams, Williamson, Willis, Willke, Wimmer, Winkler, Wipf, Wittel, Woan,
  Woehler, Wofford, Wong, Worden, Wright, Wu, Wysocki, Xiao, Yamamoto, Yancey,
  Yang, Yap, Yazback, Yu, Yu, Yvert, Zadro{\.z}ny, Zanolin, Zelenova, Zendri,
  Zevin, Zhang, Zhang, Zhang, Zhang, Zhao, Zhou, Zhou, Zhu, Zhu, Zimmerman,
  Zucker, Zweizig, Collaboration, Collaboration), Burns, Veres, Kocevski,
  Racusin, Goldstein, Connaughton, Briggs, Blackburn, Hamburg, Hui, von
  Kienlin, McEnery, Preece, {Wilson-Hodge}, Bissaldi, Cleveland, Gibby, Giles,
  Kippen, McBreen, Meegan, Paciesas, Poolakkil, Roberts, Stanbro, Monitor),
  Savchenko, Ferrigno, Kuulkers, Bazzano, Bozzo, Brandt, Chenevez, Courvoisier,
  Diehl, Domingo, Hanlon, Jourdain, Laurent, Lebrun, Lutovinov, Mereghetti,
  Natalucci, Rodi, Roques, Sunyaev, Ubertini, \& {(Integral)}}]{abbott_2017b}
Abbott, B.~P., Abbott, R., Abbott, T.~D., {et~al.} 2017{\natexlab{a}}, ApJ,
  848, L13

\bibitem[{Abbott {et~al.}(2017{\natexlab{b}})Abbott, Abbott, Abbott, Acernese,
  Ackley, Adams, Adams, Addesso, Adhikari, Adya, Affeldt, Afrough, Agarwal,
  Agathos, Agatsuma, Aggarwal, Aguiar, Aiello, Ain, Ajith, Allen, Allen,
  Allocca, Altin, Amato, Ananyeva, Anderson, Anderson, Angelova, Antier,
  Appert, Arai, Araya, Areeda, Arnaud, Arun, Ascenzi, Ashton, Ast, Aston,
  Astone, Atallah, Aufmuth, Aulbert, AultONeal, Austin, {Avila-Alvarez}, Babak,
  Bacon, Bader, Bae, Bailes, Baker, Baldaccini, Ballardin, Ballmer, Banagiri,
  Barayoga, Barclay, Barish, Barker, Barkett, Barone, Barr, Barsotti,
  Barsuglia, Barta, Barthelmy, Bartlett, Bartos, Bassiri, Basti, Batch, Bawaj,
  Bayley, Bazzan, B{\'e}csy, Beer, Bejger, Belahcene, Bell, Berger, Bergmann,
  Bernuzzi, Bero, Berry, Bersanetti, Bertolini, Betzwieser, Bhagwat, Bhandare,
  Bilenko, Billingsley, Billman, Birch, Birney, Birnholtz, Biscans, Biscoveanu,
  Bisht, Bitossi, Biwer, Bizouard, Blackburn, Blackman, Blair, Blair, Blair,
  Bloemen, Bock, Bode, Boer, Bogaert, Bohe, Bondu, Bonilla, Bonnand, Boom,
  Bork, Boschi, Bose, Bossie, Bouffanais, Bozzi, Bradaschia, Brady, Branchesi,
  Brau, Briant, Brillet, Brinkmann, Brisson, Brockill, Broida, Brooks, Brown,
  Brown, Brunett, Buchanan, Buikema, Bulik, Bulten, Buonanno, Buskulic, Buy,
  Byer, Cabero, Cadonati, Cagnoli, Cahillane, Calder{\'o}n~Bustillo, Callister,
  Calloni, Camp, Canepa, Canizares, Cannon, Cao, Cao, Capano, Capocasa,
  Carbognani, Caride, Carney, Carullo, Casanueva~Diaz, Casentini, Caudill,
  Cavagli{\`a}, Cavalier, Cavalieri, Cella, Cepeda, {Cerd{\'a}-Dur{\'a}n},
  Cerretani, Cesarini, Chamberlin, Chan, Chao, Charlton, Chase,
  {Chassande-Mottin}, Chatterjee, Chatziioannou, Cheeseboro, Chen, Chen, Chen,
  Cheng, Chia, Chincarini, Chiummo, Chmiel, Cho, Cho, Chow, Christensen, Chu,
  Chua, Chua, Chung, Chung, Ciani, Ciolfi, Cirelli, Cirone, Clara, Clark,
  Clearwater, Cleva, Cocchieri, Coccia, Cohadon, Cohen, Colla, Collette,
  Cominsky, Constancio, Conti, Cooper, Corban, Corbitt, {Cordero-Carri{\'o}n},
  Corley, Cornish, Corsi, Cortese, Costa, Coughlin, Coughlin, Coulon,
  Countryman, Couvares, Covas, Cowan, Coward, Cowart, Coyne, Coyne, Creighton,
  Creighton, Cripe, Crowder, Cullen, Cumming, Cunningham, Cuoco, Dal~Canton,
  D{\'a}lya, Danilishin, D'Antonio, Danzmann, Dasgupta, Da~Silva~Costa,
  Dattilo, Dave, Davier, Davis, Daw, Day, De, DeBra, Degallaix, De~Laurentis,
  Del{\'e}glise, Del~Pozzo, Demos, Denker, Dent, De~Pietri, Dergachev, De~Rosa,
  DeRosa, De~Rossi, DeSalvo, {de Varona}, Devenson, Dhurandhar, D{\'i}az,
  Dietrich, Di~Fiore, Di~Giovanni, Di~Girolamo, Di~Lieto, Di~Pace, Di~Palma,
  Di~Renzo, Doctor, Dolique, Donovan, Dooley, Doravari, Dorrington, Douglas,
  Dovale~{\'A}lvarez, Downes, Drago, Dreissigacker, Driggers, Du, Ducrot, Dudi,
  Dupej, Dwyer, Edo, Edwards, Effler, Eggenstein, Ehrens, Eichholz, Eikenberry,
  Eisenstein, Essick, Estevez, Etienne, Etzel, Evans, Evans, Factourovich,
  Fafone, Fair, Fairhurst, Fan, Farinon, Farr, Farr, {Fauchon-Jones}, Favata,
  Fays, Fee, Fehrmann, Feicht, Fejer, {Fernandez-Galiana}, Ferrante, Ferreira,
  Ferrini, Fidecaro, Finstad, Fiori, Fiorucci, Fishbach, Fisher, {Fitz-Axen},
  Flaminio, Fletcher, Fong, Font, Forsyth, Forsyth, Fournier, Frasca, Frasconi,
  Frei, Freise, Frey, Frey, Fries, Fritschel, Frolov, Fulda, Fyffe, Gabbard,
  Gadre, Gaebel, Gair, Gammaitoni, Ganija, Gaonkar, {Garcia-Quiros}, Garufi,
  Gateley, Gaudio, Gaur, Gayathri, Gehrels, Gemme, Genin, Gennai, George,
  George, Gergely, Germain, Ghonge, Ghosh, Ghosh, Ghosh, Giaime, Giardina,
  Giazotto, Gill, Glover, Goetz, Goetz, Gomes, Goncharov, Gonz{\'a}lez,
  Gonzalez~Castro, Gopakumar, Gorodetsky, Gossan, Gosselin, Gouaty, Grado,
  Graef, Granata, Grant, Gras, Gray, Greco, Green, Gretarsson, Groot, Grote,
  Grunewald, Gruning, Guidi, Guo, Gupta, Gupta, Gushwa, Gustafson, Gustafson,
  Halim, Hall, Hall, Hamilton, Hammond, Haney, Hanke, Hanks, Hanna, Hannam,
  Hannuksela, Hanson, Hardwick, Harms, Harry, Harry, Hart, Haster, Haughian,
  Healy, Heidmann, Heintze, Heitmann, Hello, Hemming, Hendry, Heng, Hennig,
  Heptonstall, Heurs, Hild, Hinderer, Ho, Hoak, Hofman, Holt, Holz, Hopkins,
  Horst, Hough, Houston, Howell, Hreibi, Hu, Huerta, Huet, Hughey, Husa,
  Huttner, {Huynh-Dinh}, Indik, Inta, Intini, Isa, Isac, Isi, Iyer, Izumi,
  Jacqmin, Jani, Jaranowski, Jawahar, {Jim{\'e}nez-Forteza}, Johnson,
  {Johnson-McDaniel}, Jones, Jones, Jonker, Ju, Junker, Kalaghatgi, Kalogera,
  Kamai, Kandhasamy, Kang, Kanner, Kapadia, Karki, Karvinen, Kasprzack,
  Kastaun, Katolik, Katsavounidis, Katzman, Kaufer, Kawabe, K{\'e}f{\'e}lian,
  Keitel, Kemball, Kennedy, Kent, Key, Khalili, Khan, Khan, Khan, Khazanov,
  Kijbunchoo, Kim, Kim, Kim, Kim, Kim, Kim, Kimbrell, King, King,
  {Kinley-Hanlon}, Kirchhoff, Kissel, Kleybolte, Klimenko, Knowles, Koch,
  Koehlenbeck, Koley, Kondrashov, Kontos, Korobko, Korth, Kowalska, Kozak,
  Kr{\"a}mer, Kringel, Krishnan, Kr{\'o}lak, Kuehn, Kumar, Kumar, Kumar, Kuo,
  Kutynia, Kwang, Lackey, Lai, Landry, Lang, Lange, Lantz, Lanza, Larson,
  {Lartaux-Vollard}, Lasky, Laxen, Lazzarini, Lazzaro, Leaci, Leavey, Lee, Lee,
  Lee, Lee, Lee, Lehmann, Lenon, Leon, Leonardi, Leroy, Letendre, Levin, Li,
  Linker, Littenberg, Liu, Liu, Lo, Lockerbie, London, Lord, Lorenzini,
  Loriette, Lormand, Losurdo, Lough, Lousto, Lovelace, L{\"u}ck, Lumaca,
  Lundgren, Lynch, Ma, Macas, Macfoy, Machenschalk, MacInnis, Macleod,
  Maga{\~n}a~Hernandez, {Maga{\~n}a-Sandoval}, Maga{\~n}a~Zertuche, Magee,
  Majorana, Maksimovic, Man, Mandic, Mangano, Mansell, Manske, Mantovani,
  Marchesoni, Marion, M{\'a}rka, M{\'a}rka, Markakis, Markosyan, Markowitz,
  Maros, Marquina, Marsh, Martelli, Martellini, Martin, Martin, Martynov, Marx,
  Mason, Massera, Masserot, Massinger, {Masso-Reid}, Mastrogiovanni, Matas,
  Matichard, Matone, Mavalvala, Mazumder, McCarthy, McClelland, McCormick,
  McCuller, McGuire, McIntyre, McIver, McManus, McNeill, McRae, McWilliams,
  Meacher, Meadors, Mehmet, Meidam, {Mejuto-Villa}, Melatos, Mendell, Mercer,
  Merilh, Merzougui, Meshkov, Messenger, Messick, Metzdorff, Meyers, Miao,
  Michel, Middleton, Mikhailov, Milano, Miller, Miller, Miller, Millhouse,
  {Milovich-Goff}, Minazzoli, Minenkov, Ming, Mishra, Mitra, Mitrofanov,
  Mitselmakher, Mittleman, Moffa, Moggi, Mogushi, Mohan, Mohapatra, Molina,
  Montani, Moore, Moraru, Moreno, Morisaki, Morriss, Mours, {Mow-Lowry},
  Mueller, Muir, Mukherjee, Mukherjee, Mukherjee, Mukund, Mullavey, Munch,
  Mu{\~n}iz, Muratore, Murray, Nagar, Napier, Nardecchia, Naticchioni, Nayak,
  Neilson, Nelemans, Nelson, Nery, Neunzert, Nevin, Newport, Newton, Ng,
  Nguyen, Nguyen, Nichols, Nielsen, Nissanke, Nitz, Noack, Nocera, Nolting,
  North, Nuttall, Oberling, O'Dea, Ogin, Oh, Oh, Ohme, Okada, Oliver,
  Oppermann, Oram, O'Reilly, Ormiston, Ortega, O'Shaughnessy, Ossokine,
  Ottaway, Overmier, Owen, Pace, Page, Page, Pai, Pai, Palamos, Palashov,
  Palomba, {Pal-Singh}, Pan, Pan, Pang, Pang, Pankow, Pannarale, Pant,
  Paoletti, Paoli, Papa, Parida, Parker, Pascucci, Pasqualetti, Passaquieti,
  Passuello, Patil, Patricelli, Pearlstone, Pedraza, Pedurand, Pekowsky, Pele,
  Penn, Perez, Perreca, Perri, Pfeiffer, Phelps, Piccinni, Pichot,
  Piergiovanni, Pierro, Pillant, Pinard, Pinto, Pirello, Pitkin, Poe, Poggiani,
  Popolizio, Porter, Post, Powell, Prasad, Pratt, Pratten, Predoi, Prestegard,
  Prijatelj, Principe, Privitera, Prix, Prodi, Prokhorov, Puncken, Punturo,
  Puppo, P{\"u}rrer, Qi, Quetschke, Quintero, {Quitzow-James}, Raab, Rabeling,
  Radkins, Raffai, Raja, Rajan, Rajbhandari, Rakhmanov, Ramirez,
  {Ramos-Buades}, Rapagnani, Raymond, Razzano, Read, Regimbau, Rei, Reid,
  Reitze, Ren, Reyes, Ricci, Ricker, Rieger, Riles, Rizzo, Robertson, Robie,
  Robinet, Rocchi, Rolland, Rollins, Roma, Romano, Romano, Romel, Romie,
  Rosi{\'n}ska, Ross, Rowan, R{\"u}diger, Ruggi, Rutins, Ryan, Sachdev,
  Sadecki, Sadeghian, Sakellariadou, Salconi, Saleem, Salemi, Samajdar, Sammut,
  Sampson, Sanchez, Sanchez, {Sanchis-Gual}, Sandberg, Sanders, Sassolas,
  Sathyaprakash, Saulson, Sauter, Savage, Sawadsky, Schale, Scheel, Scheuer,
  Schmidt, Schmidt, Schnabel, Schofield, Sch{\"o}nbeck, Schreiber, Schuette,
  Schulte, Schutz, Schwalbe, Scott, Scott, Seidel, Sellers, Sengupta, Sentenac,
  Sequino, Sergeev, Shaddock, Shaffer, Shah, Shahriar, Shaner, Shao, Shapiro,
  Shawhan, Sheperd, Shoemaker, Shoemaker, Siellez, Siemens, Sieniawska, Sigg,
  Silva, Singer, Singh, Singhal, Sintes, Slagmolen, Smith, Smith, Smith,
  Somala, Son, Sonnenberg, Sorazu, Sorrentino, Souradeep, Spencer, Srivastava,
  Staats, Staley, Steinke, Steinlechner, Steinlechner, Steinmeyer, Stevenson,
  Stone, Stops, Strain, Stratta, Strigin, Strunk, Sturani, Stuver,
  Summerscales, Sun, Sunil, Suresh, Sutton, Swinkels, Szczepa{\'n}czyk, Tacca,
  Tait, Talbot, Talukder, Tanner, T{\'a}pai, Taracchini, Tasson, Taylor,
  Taylor, Tewari, Theeg, Thies, Thomas, Thomas, Thomas, Thorne, Thorne, Thrane,
  Tiwari, Tiwari, Tokmakov, Toland, Tonelli, Tornasi, {Torres-Forn{\'e}},
  Torrie, T{\"o}yr{\"a}, Travasso, Traylor, Trinastic, Tringali, Trozzo, Tsang,
  Tse, Tso, Tsukada, Tsuna, Tuyenbayev, Ueno, Ugolini, Unnikrishnan, Urban,
  Usman, Vahlbruch, Vajente, Valdes, Vallisneri, {van Bakel}, {van Beuzekom},
  {van den Brand}, Van Den~Broeck, {Vander-Hyde}, {van der Schaaf}, {van
  Heijningen}, {van Veggel}, Vardaro, Varma, Vass, Vas{\'u}th, Vecchio,
  Vedovato, Veitch, Veitch, Venkateswara, Venugopalan, Verkindt, Vetrano,
  Vicer{\'e}, Viets, Vinciguerra, Vine, Vinet, Vitale, Vo, Vocca, Vorvick,
  Vyatchanin, Wade, Wade, Wade, Walet, Walker, Wallace, Walsh, Wang, Wang,
  Wang, Wang, Wang, Ward, Warner, Was, Watchi, Weaver, Wei, Weinert, Weinstein,
  Weiss, Wen, Wessel, We{\ss}els, Westerweck, Westphal, Wette, Whelan,
  Whitcomb, Whiting, Whittle, Wilken, Williams, Williams, Williamson, Willis,
  Willke, Wimmer, Winkler, Wipf, Wittel, Woan, Woehler, Wofford, Wong, Worden,
  Wright, Wu, Wysocki, Xiao, Yamamoto, Yancey, Yang, Yap, Yazback, Yu, Yu,
  Yvert, Zadro{\.z}ny, Zanolin, Zelenova, Zendri, Zevin, Zhang, Zhang, Zhang,
  Zhang, Zhao, Zhou, Zhou, Zhu, Zhu, Zimmerman, Zucker, \&
  Zweizig}]{abbott_2017a}
Abbott, B.~P., Abbott, R., Abbott, T.~D., {et~al.} 2017{\natexlab{b}}, Phys.
  Rev. Lett., 119, 161101

\bibitem[{Abbott {et~al.}(2017{\natexlab{c}})Abbott, Abbott, Abbott, Acernese,
  Ackley, Adams, Adams, Addesso, Adhikari, Adya, Affeldt, Afrough, Agarwal,
  Agathos, Agatsuma, Aggarwal, Aguiar, Aiello, Ain, Ajith, Allen, Allen,
  Allocca, Altin, Amato, Ananyeva, Anderson, Anderson, Angelova, Antier,
  Appert, Arai, Araya, Areeda, Arnaud, Arun, Ascenzi, Ashton, Ast, Aston,
  Astone, Atallah, Aufmuth, Aulbert, AultONeal, Austin, {Avila-Alvarez}, Babak,
  Bacon, Bader, Bae, Baker, Baldaccini, Ballardin, Ballmer, Banagiri, Barayoga,
  Barclay, Barish, Barker, Barkett, Barone, Barr, Barsotti, Barsuglia, Barta,
  Barthelmy, Bartlett, Bartos, Bassiri, Basti, Batch, Bawaj, Bayley, Bazzan,
  B{\'e}csy, Beer, Bejger, Belahcene, Bell, Berger, Bergmann, Bero, Berry,
  Bersanetti, Bertolini, Betzwieser, Bhagwat, Bhandare, Bilenko, Billingsley,
  Billman, Birch, Birney, Birnholtz, Biscans, Biscoveanu, Bisht, Bitossi,
  Biwer, Bizouard, Blackburn, Blackman, Blair, Blair, Blair, Bloemen, Bock,
  Bode, Boer, Bogaert, Bohe, Bondu, Bonilla, Bonnand, Boom, Bork, Boschi, Bose,
  Bossie, Bouffanais, Bozzi, Bradaschia, Brady, Branchesi, Brau, Briant,
  Brillet, Brinkmann, Brisson, Brockill, Broida, Brooks, Brown, Brown, Brunett,
  Buchanan, Buikema, Bulik, Bulten, Buonanno, Buskulic, Buy, Byer, Cabero,
  Cadonati, Cagnoli, Cahillane, Bustillo, Callister, Calloni, Camp, Canepa,
  Canizares, Cannon, Cao, Cao, Capano, Capocasa, Carbognani, Caride, Carney,
  Diaz, Casentini, Caudill, Cavagli{\`a}, Cavalier, Cavalieri, Cella, Cepeda,
  {Cerd{\'a}-Dur{\'a}n}, Cerretani, Cesarini, Chamberlin, Chan, Chao, Charlton,
  Chase, {Chassande-Mottin}, Chatterjee, Chatziioannou, Cheeseboro, Chen, Chen,
  Chen, Cheng, Chia, Chincarini, Chiummo, Chmiel, Cho, Cho, Chow, Christensen,
  Chu, Chua, Chua, Chung, Chung, Ciani, Ciolfi, Cirelli, Cirone, Clara, Clark,
  Clearwater, Cleva, Cocchieri, Coccia, Cohadon, Cohen, Colla, Collette,
  Cominsky, Jr., Conti, Cooper, Corban, Corbitt, {Cordero-Carri{\'o}n}, Corley,
  Cornish, Corsi, Cortese, Costa, Coughlin, Coughlin, Coulon, Countryman,
  Couvares, Covas, Cowan, Coward, Cowart, Coyne, Coyne, Creighton, Creighton,
  Cripe, Crowder, Cullen, Cumming, Cunningham, Cuoco, Canton, D{\'a}lya,
  Danilishin, D'Antonio, Danzmann, Dasgupta, Da~Silva~Costa, Dattilo, Dave,
  Davier, Davis, Daw, Day, De, DeBra, Degallaix, Laurentis, Del{\'e}glise,
  Pozzo, Demos, Denker, Dent, Pietri, Dergachev, Rosa, DeRosa, Rossi, DeSalvo,
  de~Varona, Devenson, Dhurandhar, D{\'i}az, Fiore, Giovanni, Girolamo, Lieto,
  Pace, Palma, Renzo, Doctor, Dolique, Donovan, Dooley, Doravari, Dorrington,
  Douglas, {\'A}lvarez, Downes, Drago, Dreissigacker, Driggers, Du, Ducrot,
  Dupej, Dwyer, Edo, Edwards, Effler, Ehrens, Eichholz, Eikenberry, Eisenstein,
  Essick, Estevez, Etienne, Etzel, Evans, Evans, Factourovich, Fafone, Fair,
  Fairhurst, Fan, Farinon, Farr, Farr, {Fauchon-Jones}, Favata, Fays, Fee,
  Fehrmann, Feicht, Fejer, {Fernandez-Galiana}, Ferrante, Ferreira, Ferrini,
  Fidecaro, Finstad, Fiori, Fiorucci, Fishbach, Fisher, {Fitz-Axen}, Flaminio,
  Fletcher, Fong, Font, Forsyth, Forsyth, Fournier, Frasca, Frasconi, Frei,
  Freise, Frey, Frey, Fries, Fritschel, Frolov, Fulda, Fyffe, Gabbard, Gadre,
  Gaebel, Gair, Gammaitoni, Ganija, Gaonkar, {Garcia-Quiros}, Garufi, Gateley,
  Gaudio, Gaur, Gayathri, Gehrels, Gemme, Genin, Gennai, George, George,
  Gergely, Germain, Ghonge, Ghosh, Ghosh, Ghosh, Giaime, Giardina, Giazotto,
  Gill, Glover, Goetz, Goetz, Gomes, Goncharov, Gonz{\'a}lez, Castro,
  Gopakumar, Gorodetsky, Gossan, Gosselin, Gouaty, Grado, Graef, Granata,
  Grant, Gras, Gray, Greco, Green, Gretarsson, Griswold, Groot, Grote,
  Grunewald, Gruning, Guidi, Guo, Gupta, Gupta, Gushwa, Gustafson, Gustafson,
  Halim, Hall, Hall, Hamilton, Hammond, Haney, Hanke, Hanks, Hanna, Hannam,
  Hannuksela, Hanson, Hardwick, Harms, Harry, Harry, Hart, Haster, Haughian,
  Healy, Heidmann, Heintze, Heitmann, Hello, Hemming, Hendry, Heng, Hennig,
  Heptonstall, Heurs, Hild, Hinderer, Hoak, Hofman, Holt, Holz, Hopkins, Horst,
  Hough, Houston, Howell, Hreibi, Hu, Huerta, Huet, Hughey, Husa, Huttner,
  {Huynh-Dinh}, Indik, Inta, Intini, Isa, Isac, Isi, Iyer, Izumi, Jacqmin,
  Jani, Jaranowski, Jawahar, {Jim{\'e}nez-Forteza}, Johnson, Jones, Jones,
  Jonker, Ju, Junker, Kalaghatgi, Kalogera, Kamai, Kandhasamy, Kang, Kanner,
  Kapadia, Karki, Karvinen, Kasprzack, Katolik, Katsavounidis, Katzman, Kaufer,
  Kawabe, K{\'e}f{\'e}lian, Keitel, Kemball, Kennedy, Kent, Key, Khalili, Khan,
  Khan, Khan, Khazanov, Kijbunchoo, Kim, Kim, Kim, Kim, Kim, Kim, Kimbrell,
  King, King, {Kinley-Hanlon}, Kirchhoff, Kissel, Kleybolte, Klimenko, Knowles,
  Koch, Koehlenbeck, Koley, Kondrashov, Kontos, Korobko, Korth, Kowalska,
  Kozak, Kr{\"a}mer, Kringel, Krishnan, Kr{\'o}lak, Kuehn, Kumar, Kumar, Kumar,
  Kuo, Kutynia, Kwang, Lackey, Lai, Landry, Lang, Lange, Lantz, Lanza, Larson,
  {Lartaux-Vollard}, Lasky, Laxen, Lazzarini, Lazzaro, Leaci, Leavey, Lee, Lee,
  Lee, Lee, Lee, Lehmann, Lenon, Leonardi, Leroy, Letendre, Levin, Li, Linker,
  Littenberg, Liu, Lo, Lockerbie, London, Lord, Lorenzini, Loriette, Lormand,
  Losurdo, Lough, Lousto, Lovelace, L{\"u}ck, Lumaca, Lundgren, Lynch, Ma,
  Macas, Macfoy, Machenschalk, MacInnis, Macleod, Hernandez,
  {Maga{\~n}a-Sandoval}, Zertuche, Magee, Majorana, Maksimovic, Man, Mandic,
  Mangano, Mansell, Manske, Mantovani, Marchesoni, Marion, M{\'a}rka,
  M{\'a}rka, Markakis, Markosyan, Markowitz, Maros, Marquina, Marsh, Martelli,
  Martellini, Martin, Martin, Martynov, Mason, Massera, Masserot, Massinger,
  {Masso-Reid}, Mastrogiovanni, Matas, Matichard, Matone, Mavalvala, Mazumder,
  McCarthy, McClelland, McCormick, McCuller, McGuire, McIntyre, McIver,
  McManus, McNeill, McRae, McWilliams, Meacher, Meadors, Mehmet, Meidam,
  {Mejuto-Villa}, Melatos, Mendell, Mercer, Merilh, Merzougui, Meshkov,
  Messenger, Messick, Metzdorff, Meyers, Miao, Michel, Middleton, Mikhailov,
  Milano, Miller, Miller, Miller, Millhouse, {Milovich-Goff}, Minazzoli,
  Minenkov, Ming, Mishra, Mitra, Mitrofanov, Mitselmakher, Mittleman, Moffa,
  Moggi, Mogushi, Mohan, Mohapatra, Montani, Moore, Moraru, Moreno, Morriss,
  Mours, {Mow-Lowry}, Mueller, Muir, Mukherjee, Mukherjee, Mukherjee, Mukund,
  Mullavey, Munch, Mu{\~n}iz, Muratore, Murray, Napier, Nardecchia,
  Naticchioni, Nayak, Neilson, Nelemans, Nelson, Nery, Neunzert, Nevin,
  Newport, Newton, Ng, Nguyen, Nguyen, Nichols, Nielsen, Nissanke, Nitz, Noack,
  Nocera, Nolting, North, Nuttall, Oberling, O'Dea, Ogin, Oh, Oh, Ohme, Okada,
  Oliver, Oppermann, Oram, O'Reilly, Ormiston, Ortega, O'Shaughnessy, Ossokine,
  Ottaway, Overmier, Owen, Pace, Page, Page, Pai, Pai, Palamos, Palashov,
  Palomba, {Pal-Singh}, Pan, Pan, Pang, Pang, Pankow, Pannarale, Pant,
  Paoletti, Paoli, Papa, Parida, Parker, Pascucci, Pasqualetti, Passaquieti,
  Passuello, Patil, Patricelli, Pearlstone, Pedraza, Pedurand, Pekowsky, Pele,
  Penn, Perez, Perreca, Perri, Pfeiffer, Phelps, Piccinni, Pichot,
  Piergiovanni, Pierro, Pillant, Pinard, Pinto, Pirello, Pitkin, Poe, Poggiani,
  Popolizio, Porter, Post, Powell, Prasad, Pratt, Pratten, Predoi, Prestegard,
  Price, Prijatelj, Principe, Privitera, Prodi, Prokhorov, Puncken, Punturo,
  Puppo, P{\"u}rrer, Qi, Quetschke, Quintero, {Quitzow-James}, Raab, Rabeling,
  Radkins, Raffai, Raja, Rajan, Rajbhandari, Rakhmanov, Ramirez,
  {Ramos-Buades}, Rapagnani, Raymond, Razzano, Read, Regimbau, Rei, Reid,
  Reitze, Ren, Reyes, Ricci, Ricker, Rieger, Riles, Rizzo, Robertson, Robie,
  Robinet, Rocchi, Rolland, Rollins, Roma, Romano, Romel, Romie, Rosi{\'n}ska,
  Ross, Rowan, R{\"u}diger, Ruggi, Rutins, Ryan, Sachdev, Sadecki, Sadeghian,
  Sakellariadou, Salconi, Saleem, Salemi, Samajdar, Sammut, Sampson, Sanchez,
  Sanchez, {Sanchis-Gual}, Sandberg, Sanders, Sassolas, Sathyaprakash, Saulson,
  Sauter, Savage, Sawadsky, Schale, Scheel, Scheuer, Schmidt, Schmidt,
  Schnabel, Schofield, Sch{\"o}nbeck, Schreiber, Schuette, Schulte, Schutz,
  Schwalbe, Scott, Scott, Seidel, Sellers, Sengupta, Sentenac, Sequino,
  Sergeev, Shaddock, Shaffer, Shah, Shahriar, Shaner, Shao, Shapiro, Shawhan,
  Sheperd, Shoemaker, Shoemaker, Siellez, Siemens, Sieniawska, Sigg, Silva,
  Singer, Singh, Singhal, Sintes, Slagmolen, Smith, Smith, Smith, Somala, Son,
  Sonnenberg, Sorazu, Sorrentino, Souradeep, Spencer, Srivastava, Staats,
  Staley, Steinke, Steinlechner, Steinlechner, Steinmeyer, Stevenson, Stone,
  Stops, Strain, Stratta, Strigin, Strunk, Sturani, Stuver, Summerscales, Sun,
  Sunil, Suresh, Sutton, Swinkels, Szczepa{\'n}czyk, Tacca, Tait, Talbot,
  Talukder, Tanner, T{\'a}pai, Taracchini, Tasson, Taylor, Taylor, Tewari,
  Theeg, Thies, Thomas, Thomas, Thomas, Thorne, Thorne, Thrane, Tiwari, Tiwari,
  Tokmakov, Toland, Tonelli, Tornasi, {Torres-Forn{\'e}}, Torrie,
  T{\"o}yr{\"a}, Travasso, Traylor, Trinastic, Tringali, Trozzo, Tsang, Tse,
  Tso, Tsukada, Tsuna, Tuyenbayev, Ueno, Ugolini, Unnikrishnan, Urban, Usman,
  Vahlbruch, Vajente, Valdes, van Bakel, van Beuzekom, {van den Brand}, Van
  Den~Broeck, {Vander-Hyde}, {van der Schaaf}, van Heijningen, van Veggel,
  Vardaro, Varma, Vass, Vas{\'u}th, Vecchio, Vedovato, Veitch, Veitch,
  Venkateswara, Venugopalan, Verkindt, Vetrano, Vicer{\'e}, Viets, Vinciguerra,
  Vine, Vinet, Vitale, Vo, Vocca, Vorvick, Vyatchanin, Wade, Wade, Wade, Walet,
  Walker, Wallace, Walsh, Wang, Wang, Wang, Wang, Wang, Ward, Warner, Was,
  Watchi, Weaver, Wei, Weinert, Weinstein, Weiss, Wen, Wessel, Wessels,
  Westerweck, Westphal, Wette, Whelan, Whitcomb, Whiting, Whittle, Wilken,
  Williams, Williams, Williamson, Willis, Willke, Wimmer, Winkler, Wipf,
  Wittel, Woan, Woehler, Wofford, Wong, Worden, Wright, Wu, Wysocki, Xiao,
  Yamamoto, Yancey, Yang, Yap, Yazback, Yu, Yu, Yvert, Zadro{\.z}ny, Zanolin,
  Zelenova, Zendri, Zevin, Zhang, Zhang, Zhang, Zhang, Zhao, Zhou, Zhou, Zhu,
  Zhu, Zimmerman, Zucker, Zweizig, {LIGO Scientific Collaboration and Virgo
  Collaboration}, {Wilson-Hodge}, Bissaldi, Blackburn, Briggs, Burns,
  Cleveland, Connaughton, Gibby, Giles, Goldstein, Hamburg, Jenke, Hui, Kippen,
  Kocevski, McBreen, Meegan, Paciesas, Poolakkil, Preece, Racusin, Roberts,
  Stanbro, Veres, {von Kienlin}, {Fermi GBM}, Savchenko, Ferrigno, Kuulkers,
  Bazzano, Bozzo, Brandt, Chenevez, Courvoisier, Diehl, Domingo, Hanlon,
  Jourdain, Laurent, Lebrun, Lutovinov, {Martin-Carrillo}, Mereghetti,
  Natalucci, Rodi, Roques, Sunyaev, Ubertini, {Integral}, Aartsen, Ackermann,
  Adams, Aguilar, Ahlers, Ahrens, Samarai, Altmann, Andeen, Anderson, Ansseau,
  Anton, Arg{\"u}elles, Auffenberg, Axani, Bagherpour, Bai, Barron, Barwick,
  Baum, Bay, Beatty, Tjus, Bernardini, Besson, Binder, Bindig, Blaufuss, Blot,
  Bohm, B{\"o}rner, Bos, Bose, B{\"o}ser, Botner, Bourbeau, Bourbeau,
  Bradascio, Braun, Brayeur, Brenzke, Bretz, Bron, {Brostean-Kaiser}, Burgman,
  Carver, Casey, Casier, Cheung, Chirkin, Christov, Clark, Classen, Coenders,
  Collin, Conrad, Cowen, Cross, Day, de~Andr{\'e}, Clercq, DeLaunay, Dembinski,
  Ridder, Desiati, de~Vries, de~Wasseige, de~With, DeYoung,
  {D{\'i}az-V{\'e}lez}, di~Lorenzo, Dujmovic, Dumm, Dunkman, Dvorak, Eberhardt,
  Ehrhardt, Eichmann, Eller, Evenson, Fahey, Fazely, Felde, Filimonov, Finley,
  Flis, Franckowiak, Friedman, Fuchs, Gaisser, Gallagher, Gerhardt, Ghorbani,
  Giang, Glauch, Gl{\"u}senkamp, Goldschmidt, Gonzalez, Grant, Griffith, Haack,
  Hallgren, Halzen, Hanson, Hebecker, Heereman, Helbing, Hellauer, Hickford,
  Hignight, Hill, Hoffman, Hoffmann, {Hokanson-Fasig}, Hoshina, Huang, Huber,
  Hultqvist, H{\"u}nnefeld, In, Ishihara, Jacobi, Japaridze, Jeong, Jero,
  Jones, Kalaczynski, Kang, Kappes, Karg, Karle, Kauer, Keivani, Kelley,
  Kheirandish, Kim, Kim, Kintscher, Kiryluk, Kittler, Klein, Kohnen, Koirala,
  Kolanoski, K{\"o}pke, Kopper, Kopper, Koschinsky, Koskinen, Kowalski, Krings,
  Kroll, Kr{\"u}ckl, Kunnen, Kunwar, Kurahashi, Kuwabara, Kyriacou, Labare,
  Lanfranchi, Larson, Lauber, {Lesiak-Bzdak}, Leuermann, Liu, Lu, L{\"u}nemann,
  Luszczak, Madsen, Maggi, Mahn, Mancina, Maruyama, Mase, Maunu, McNally,
  Meagher, Medici, Meier, Menne, Merino, Meures, Miarecki, Micallef,
  Moment{\'e}, Montaruli, Moore, Moulai, Nahnhauer, Nakarmi, Naumann, Neer,
  Niederhausen, Nowicki, Nygren, Pollmann, Olivas, O'Murchadha, Palczewski,
  Pandya, Pankova, Peiffer, Pepper, {P{\'e}rez de los Heros}, Pieloth, Pinat,
  Price, Przybylski, Raab, R{\"a}del, Rameez, Rawlins, Rea, Reimann,
  Relethford, Relich, Resconi, Rhode, Richman, Robertson, Rongen, Rott, Ruhe,
  Ryckbosch, Rysewyk, S{\"a}lzer, Herrera, Sandrock, Sandroos, Santander,
  Sarkar, Sarkar, Satalecka, Schlunder, Schmidt, Schneider, Schoenen,
  Sch{\"o}neberg, Schumacher, Seckel, Seunarine, Soedingrekso, Soldin, Song,
  Spiczak, Spiering, Stachurska, Stamatikos, Stanev, Stasik, Stettner, Steuer,
  Stezelberger, Stokstad, St{\"o}ssl, Strotjohann, Stuttard, Sullivan,
  Sutherland, Taboada, Tatar, Tenholt, {Ter-Antonyan}, Terliuk, Te{\v s}i{\'c},
  Tilav, Toale, Tobin, Toscano, Tosi, Tselengidou, Tung, Turcati, Turley, Ty,
  Unger, Usner, Vandenbroucke, Driessche, van Eijndhoven, Vanheule, van Santen,
  Vehring, Vogel, Vraeghe, Walck, Wallace, Wallraff, Wandler, Wandkowsky, Waza,
  Weaver, Weiss, Wendt, Werthebach, Whelan, Wiebe, Wiebusch, Wille, Williams,
  Wills, Wolf, Wood, Woolsey, Woschnagg, Xu, Xu, Xu, Yanez, Yodh, Yoshida,
  Yuan, Zoll, {IceCube Collaboration}, Balasubramanian, Mate, Bhalerao,
  Bhattacharya, Vibhute, Dewangan, Rao, Vadawale, {AstroSat Cadmium Zinc
  Telluride Imager Team}, Svinkin, Hurley, Aptekar, Frederiks, Golenetskii,
  Kozlova, Lysenko, Oleynik, Tsvetkova, Ulanov, Cline, {IPN Collaboration}, Li,
  Xiong, Zhang, Lu, Song, Cao, Chang, Chen, Chen, Chen, Chen, Chen, Chen, Cui,
  Cui, Deng, Dong, Du, Fu, Gao, Gao, Gao, Ge, Gu, Guan, Guo, Han, Hu, Huang,
  Huo, Jia, Jiang, Jiang, Jin, Jin, Li, Li, Li, Li, Li, Li, Li, Li, Li, Li, Li,
  Liang, Liao, Liu, Liu, Liu, Liu, Liu, Liu, Liu, Lu, Lu, Luo, Ma, Meng, Nang,
  Nie, Ou, Qu, Sai, Sun, Tan, Tao, Tao, Tuo, Wang, Wang, Wang, Wang, Wang, Wen,
  Wu, Wu, Xiao, Xu, Xu, Yan, Yang, Yang, Yang, Zhang, Zhang, Zhang, Zhang,
  Zhang, Zhang, Zhang, Zhang, Zhang, Zhang, Zhang, Zhang, Zhang, Zhang, Zhang,
  Zhang, Zhang, Zhang, Zhao, Zhao, Zhao, Zheng, Zhu, Zhu, Zou, {The
  Insight-HXMT Collaboration}, Albert, Andr{\'e}, Anghinolfi, Ardid, Aubert,
  Aublin, Avgitas, Baret, {Barrios-Mart{\'i}}, Basa, Belhorma, Bertin, Biagi,
  Bormuth, Bourret, Bouwhuis, Br{\^a}nza{\c s}, Bruijn, Brunner, Busto, Capone,
  Caramete, Carr, Celli, Cherkaoui El~Moursli, Chiarusi, Circella, Coelho,
  Coleiro, Coniglione, Costantini, Coyle, Creusot, D{\'i}az, Deschamps, Bonis,
  Distefano, Palma, Domi, Donzaud, Dornic, Drouhin, Eberl, El~Bojaddaini,
  El~Khayati, Els{\"a}sser, Enzenh{\"o}fer, Ettahiri, Fassi, Felis, Fusco, Gay,
  Giordano, Glotin, Gr{\'e}goire, Ruiz, Graf, Hallmann, van Haren, Heijboer,
  Hello, {Hern{\'a}ndez-Rey}, H{\"o}ssl, Hofest{\"a}dt, Hugon, Illuminati,
  James, de~Jong, Jongen, Kadler, Kalekin, Katz, Kiessling, Kouchner, Kreter,
  Kreykenbohm, Kulikovskiy, Lachaud, Lahmann, Lef{\`e}vre, Leonora, Lotze,
  Loucatos, Marcelin, Margiotta, Marinelli, {Mart{\'i}nez-Mora}, Mele, Melis,
  Michael, Migliozzi, Moussa, Navas, Nezri, Organokov, P{\u a}v{\u a}la{\c s},
  Pellegrino, Perrina, Piattelli, Popa, Pradier, Quinn, Racca, Riccobene,
  {S{\'a}nchez-Losa}, Salda{\~n}a, Salvadori, Samtleben, Sanguineti, Sapienza,
  Sieger, Spurio, Stolarczyk, Taiuti, Tayalati, Trovato, Turpin, T{\"o}nnis,
  Vallage, Elewyck, Versari, Vivolo, Vizzoca, Wilms, Zornoza, Z{\'u}{\~n}iga,
  {ANTARES Collaboration}, Beardmore, Breeveld, Burrows, Cenko, Cusumano,
  D'A{\`i}, {de Pasquale}, Emery, Evans, Giommi, Gronwall, Kennea, Krimm, Kuin,
  Lien, Marshall, Melandri, Nousek, Oates, Osborne, Pagani, Page, Palmer,
  Perri, Siegel, Sbarufatti, Tagliaferri, Tohuvavohu, {The Swift
  Collaboration}, Tavani, Verrecchia, Bulgarelli, Evangelista, Pacciani,
  Feroci, Pittori, Giuliani, Monte, Donnarumma, Argan, Trois, Ursi, Cardillo,
  Piano, Longo, Lucarelli, {Munar-Adrover}, Fuschino, Labanti, Marisaldi,
  Minervini, Fioretti, Parmiggiani, Gianotti, Trifoglio, Persio, Antonelli,
  Barbiellini, Caraveo, Cattaneo, Costa, Colafrancesco, D'Amico, Ferrari,
  Morselli, Paoletti, Picozza, Pilia, Rappoldi, Soffitta, Vercellone, {AGILE
  Team}, Foley, Coulter, Kilpatrick, Drout, Piro, Shappee, Siebert, Simon,
  Ulloa, Kasen, Madore, {Murguia-Berthier}, Pan, Prochaska, {Ramirez-Ruiz},
  Rest, {Rojas-Bravo}, {The 1M2H Team}, Berger, {Soares-Santos}, Annis,
  Alexander, Allam, Balbinot, Blanchard, Brout, Butler, Chornock, Cook,
  Cowperthwaite, Diehl, {Drlica-Wagner}, Drout, Durret, Eftekhari, Finley,
  Fong, Frieman, Fryer, {Garc{\'i}a-Bellido}, Gruendl, Hartley, Herner,
  Kessler, Lin, Lopes, Louren{\c c}o, Margutti, Marshall, Matheson, Medina,
  Metzger, Mu{\~n}oz, Muir, Nicholl, Nugent, Palmese, {Paz-Chinch{\'o}n},
  Quataert, Sako, Sauseda, Schlegel, Scolnic, Secco, Smith, Sobreira, Villar,
  Vivas, Wester, Williams, Yanny, Zenteno, Zhang, Abbott, Banerji, Bechtol,
  {Benoit-L{\'e}vy}, Bertin, Brooks, {Buckley-Geer}, Burke, Capozzi, Rosell,
  Kind, Castander, Crocce, Cunha, D'Andrea, {da Costa}, Davis, DePoy, Desai,
  Dietrich, Eifler, Fernandez, Flaugher, Fosalba, Gaztanaga, Gerdes,
  Giannantonio, Goldstein, Gruen, Gschwend, Gutierrez, Honscheid, James,
  Jeltema, Johnson, Johnson, Kent, Krause, Kron, Kuehn, Lahav, Lima, Maia,
  March, Martini, McMahon, Menanteau, Miller, Miquel, Mohr, Nichol, Ogando,
  Plazas, Romer, Roodman, Rykoff, Sanchez, Scarpine, Schindler, Schubnell,
  {Sevilla-Noarbe}, Sheldon, Smith, Smith, Stebbins, Suchyta, Swanson, Tarle,
  Thomas, Troxel, Tucker, Vikram, Walker, Wechsler, Weller, Carlin, Gill, Li,
  Marriner, Neilsen, {The Dark Energy Camera GW-EM Collaboration and the DES
  Collaboration}, Haislip, Kouprianov, Reichart, Sand, Tartaglia, Valenti,
  Yang, {The DLT40 Collaboration}, Benetti, Brocato, Campana, Cappellaro,
  Covino, D'Avanzo, D'Elia, Getman, Ghirlanda, Ghisellini, Limatola, Nicastro,
  Palazzi, Pian, Piranomonte, Possenti, Rossi, Salafia, Tomasella, Amati,
  Antonelli, Bernardini, Bufano, Capaccioli, Casella, Dadina, Cesare, Paola,
  Giuffrida, Giunta, Israel, Lisi, Maiorano, Mapelli, Masetti, Pescalli,
  Pulone, Salvaterra, Schipani, Spera, Stamerra, Stella, Testa, Turatto,
  Vergani, Aresu, Bachetti, Buffa, Burgay, Buttu, Caria, Carretti, Casasola,
  Castangia, Carboni, Casu, Concu, Corongiu, Deiana, Egron, Fara, Gaudiomonte,
  Gusai, Ladu, Loru, Leurini, Marongiu, Melis, Melis, Migoni, Milia, Navarrini,
  Orlati, Ortu, Palmas, Pellizzoni, Perrodin, Pisanu, Poppi, Righini, Saba,
  Serra, Serrau, Stagni, Surcis, Vacca, Vargiu, Hunt, Jin, Klose, Kouveliotou,
  Mazzali, M{\o}ller, Nava, Piran, Selsing, Vergani, Wiersema, Toma, Higgins,
  Mundell, {di Serego Alighieri}, G{\'o}tz, Gao, Gomboc, Kaper, Kobayashi,
  Kopac, Mao, Starling, Steele, {van der Horst}, {GRAWITA: GRAvitational Wave
  Inaf TeAm}, Acero, Atwood, Baldini, Barbiellini, Bastieri, Berenji,
  Bellazzini, Bissaldi, Blandford, Bloom, Bonino, Bottacini, Bregeon, Buehler,
  Buson, Cameron, Caputo, Caraveo, Cavazzuti, Chekhtman, Cheung, Chiang,
  Ciprini, {Cohen-Tanugi}, Cominsky, Costantin, Cuoco, D'Ammando, de~Palma,
  Digel, Lalla, Mauro, Venere, Dubois, Fegan, Focke, Franckowiak, Fukazawa,
  Funk, Fusco, Gargano, Gasparrini, Giglietto, Giordano, Giroletti, Glanzman,
  Green, Grondin, Guillemot, Guiriec, Harding, Horan, J{\'o}hannesson, Kamae,
  Kensei, Kuss, Mura, Latronico, {Lemoine-Goumard}, Longo, Loparco, Lovellette,
  Lubrano, Magill, Maldera, Manfreda, Mazziotta, McEnery, Meyer, Michelson,
  Mirabal, Monzani, Moretti, Morselli, Moskalenko, Negro, Nuss, Ojha, Omodei,
  Orienti, Orlando, Palatiello, Paliya, Paneque, {Pesce-Rollins}, Piron,
  Porter, Principe, Rain{\`o}, Rando, Razzano, Razzaque, Reimer, Reimer,
  Reposeur, Rochester, Parkinson, Sgr{\`o}, Siskind, Spada, Spandre, Suson,
  Takahashi, Tanaka, Thayer, Thayer, Thompson, Tibaldo, Torres, Torresi, Troja,
  Venters, Vianello, Zaharijas, {The Fermi Large Area Telescope Collaboration},
  Allison, Bannister, Dobie, Kaplan, Lenc, Lynch, Murphy, Sadler, {ATCA:
  Australia Telescope Compact Array}, Hotan, James, Oslowski, Raja, Shannon,
  Whiting, {ASKAP: Australian SKA Pathfinder}, Arcavi, Howell, McCully,
  Hosseinzadeh, Hiramatsu, Poznanski, Barnes, Zaltzman, Vasylyev, Maoz, {Las
  Cumbres Observatory Group}, Cooke, Bailes, Wolf, Deller, Lidman, Wang,
  Gendre, Andreoni, Ackley, Pritchard, Bessell, Chang, M{\"o}ller, Onken,
  Scalzo, {Ridden-Harper}, Sharp, Tucker, Farrell, Elmer, Johnston, Krishnan,
  Keane, Green, Jameson, Hu, Ma, Sun, Wu, Wang, Shang, Hu, Ashley, Yuan, Li,
  Tao, Zhu, Zhang, Suntzeff, Zhou, Yang, Orange, Morris, Cucchiara, Giblin,
  Klotz, Staff, Thierry, Schmidt, {OzGrav, DWF (Deeper, Wider, Faster program),
  AST3, and CAASTRO Collaborations}, Tanvir, Levan, Cano, {de Ugarte-Postigo},
  {Gonz{\'a}lez-Fern{\'a}ndez}, Greiner, Hjorth, Irwin, Kr{\"u}hler, Mandel,
  {Milvang-Jensen}, O'Brien, Rol, Rosetti, Rosswog, Rowlinson, Steeghs,
  Th{\"o}ne, Ulaczyk, Watson, Bruun, Cutter, Figuera~Jaimes, Fujii, Fruchter,
  Gompertz, Jakobsson, Hodosan, J{\`e}rgensen, Kangas, Kann, Rabus,
  Schr{\o}der, Stanway, Wijers, {The VINROUGE Collaboration}, Lipunov,
  Gorbovskoy, Kornilov, Tyurina, Balanutsa, Kuznetsov, Vlasenko, Podesta,
  Lopez, Podesta, Levato, Saffe, Mallamaci, Budnev, Gress, Kuvshinov, Gorbunov,
  Vladimirov, Zimnukhov, Gabovich, Yurkov, Sergienko, Rebolo, {Serra-Ricart},
  Tlatov, Ishmuhametova, {MASTER Collaboration}, Abe, Aoki, Aoki, Asakura,
  Baar, Barway, Bond, Doi, Finet, Fujiyoshi, Furusawa, Honda, Itoh, Kanda,
  Kawabata, Kawabata, Kim, Koshida, Kuroda, Lee, Liu, Matsubayashi, Miyazaki,
  Morihana, Morokuma, Motohara, Murata, Nagai, Nagashima, Nagayama, Nakaoka,
  Nakata, Ohsawa, Ohshima, Ohta, Okita, Saito, Saito, Sako, Sekiguchi, Sumi,
  Tajitsu, Takahashi, Takayama, Tamura, Tanaka, Tanaka, Terai, Tominaga,
  Tristram, Uemura, Utsumi, Yamaguchi, Yasuda, Yoshida, Zenko, {J-Gem}, Adams,
  Anupama, Bally, Barway, Bellm, Blagorodnova, Cannella, Chandra, Chatterjee,
  Clarke, Cobb, Cook, Copperwheat, De, Emery, Feindt, Foster, Fox, Frail,
  Fremling, Frohmaier, Garcia, Ghosh, Giacintucci, Goobar, Gottlieb,
  Grefenstette, Hallinan, Harrison, Heida, Helou, Ho, Horesh, Hotokezaka, Ip,
  Itoh, Jacobs, Jencson, Kasen, Kasliwal, Kassim, Kim, Kiran, Kuin, Kulkarni,
  Kupfer, Lau, Madsen, Mazzali, Miller, Miyasaka, Mooley, Myers, Nakar, Ngeow,
  Nugent, Ofek, Palliyaguru, Pavana, Perley, Peters, Pike, Piran, Qi, Quimby,
  Rana, Rosswog, Rusu, Sadler, Sistine, Sollerman, Xu, Yan, Yatsu, Yu, Zhang,
  Zhao, {GROWTH, JAGWAR, Caltech-NRAO, TTU-NRAO, and NuSTAR Collaborations},
  Chambers, Huber, Schultz, Bulger, Flewelling, Magnier, Lowe, Wainscoat,
  Waters, Willman, {Pan-STARRS}, Ebisawa, Hanyu, Harita, Hashimoto, Hidaka,
  Hori, Ishikawa, Isobe, Iwakiri, Kawai, Kawai, Kawamuro, Kawase, Kitaoka,
  Makishima, Matsuoka, Mihara, Morita, Morita, Nakahira, Nakajima, Nakamura,
  Negoro, Oda, Sakamaki, Sasaki, Serino, Shidatsu, Shimomukai, Sugawara,
  Sugita, Sugizaki, Tachibana, Takao, Tanimoto, Tomida, Tsuboi, Tsunemi, Ueda,
  Ueno, Yamada, Yamaoka, Yamauchi, Yatabe, Yoneyama, Yoshii, {The MAXI Team},
  Coward, Crisp, Macpherson, Andreoni, Laugier, Noysena, Klotz, Gendre,
  Thierry, Turpin, {TZAC Consortium}, Im, Choi, Kim, Yoon, Lim, Lee, Lee, Kim,
  Ko, Joe, Kwon, Kim, Lim, Choi, {KU Collaboration}, Fynbo, Malesani, Xu,
  {Nordic Optical Telescope}, Smartt, Jerkstrand, Kankare, Sim, Fraser,
  Inserra, Maguire, Leloudas, Magee, Shingles, Smith, Young, Kotak, {Gal-Yam},
  Lyman, Homan, Agliozzo, Anderson, Angus, Ashall, Barbarino, Bauer, Berton,
  Botticella, Bulla, Cannizzaro, Cartier, Cikota, Clark, De~Cia, Della~Valle,
  Dennefeld, Dessart, Dimitriadis, {Elias-Rosa}, Firth, Fl{\"o}rs, Frohmaier,
  Galbany, {Gonz{\'a}lez-Gait{\'a}n}, Gromadzki, Guti{\'e}rrez, Hamanowicz,
  Harmanen, Heintz, Hernandez, Hodgkin, Hook, Izzo, James, Jonker, Kerzendorf,
  {Kostrzewa-Rutkowska}, Kromer, Kuncarayakti, Lawrence, Manulis, Mattila,
  McBrien, M{\"u}ller, Nordin, O'Neill, Onori, Palmerio, Pastorello, Patat,
  Pignata, Podsiadlowski, Razza, Reynolds, Roy, Ruiter, Rybicki, Salmon, Pumo,
  Prentice, Seitenzahl, Smith, Sollerman, Sullivan, Szegedi, Taddia,
  Taubenberger, Terreran, Van~Soelen, Vos, Walton, Wright, Wyrzykowski, Yaron,
  {ePESSTO}, Chen, Kr{\"u}hler, Schady, Wiseman, Greiner, Rau, Schweyer, Klose,
  Nicuesa~Guelbenzu, {Grond}, Palliyaguru, {Texas Tech University}, Shara,
  Williams, Vaisanen, Potter, Colmenero, Crawford, Buckley, Mao, {SALT Group},
  D{\'i}az, Macri, Garc{\'i}a~Lambas, {Mendes de Oliveira}, Nilo~Castell{\'o}n,
  Ribeiro, S{\'a}nchez, Schoenell, Abramo, Akras, Alcaniz, Artola, Beroiz,
  Bonoli, Cabral, Camuccio, Chavushyan, Coelho, Colazo, {Costa-Duarte},
  Cuevas~Larenas, Dom{\'i}nguez~Romero, Dultzin, Fern{\'a}ndez, Garc{\'i}a,
  Girardini, Gon{\c c}alves, Gon{\c c}alves, Gurovich, {Jim{\'e}nez-Teja},
  Kanaan, Lares, {Lopes de Oliveira}, {L{\'o}pez-Cruz}, Melia, Molino, Padilla,
  Pe{\~n}uela, Placco, Qui{\~n}ones, Ram{\'i}rez~Rivera, Renzi, Riguccini,
  {R{\'i}os-L{\'o}pez}, Rodriguez, Sampedro, Schneiter, Sodr{\'e}, Starck,
  {Torres-Flores}, Tornatore, Zadro{\.z}ny, Castillo, {TOROS: Transient Robotic
  Observatory of the South Collaboration}, {Castro-Tirado}, Tello, Hu, Zhang,
  Cunniffe, Castell{\'o}n, Hiriart, {Caballero-Garc{\'i}a}, Jel{\'i}nek,
  Kub{\'a}nek, {P{\'e}rez del Pulgar}, Park, Jeong, Castro~Cer{\'o}n, Pandey,
  Yock, Querel, Fan, Wang, {The BOOTES Collaboration}, Beardsley, Brown,
  Crosse, Emrich, Franzen, Gaensler, Horsley, {Johnston-Hollitt}, Kenney,
  Morales, Pallot, Sokolowski, Steele, Tingay, Trott, Walker, Wayth, Williams,
  Wu, {MWA: Murchison Widefield Array}, Yoshida, Sakamoto, Kawakubo, Yamaoka,
  Takahashi, Asaoka, Ozawa, Torii, Shimizu, Tamura, Ishizaki, Cherry,
  Ricciarini, Penacchioni, Marrocchesi, {The CALET Collaboration}, Pozanenko,
  Volnova, Mazaeva, Minaev, Krugov, Kusakin, Reva, Moskvitin, Rumyantsev,
  Inasaridze, Klunko, Tungalag, Schmalz, Burhonov, {IKI-GW Follow-up
  Collaboration}, Abdalla, Abramowski, Aharonian, Benkhali, Ang{\"u}ner,
  Arakawa, Arrieta, Aubert, Backes, Balzer, Barnard, Becherini, Tjus, Berge,
  Bernhard, Bernl{\"o}hr, Blackwell, B{\"o}ttcher, Boisson, Bolmont, Bonnefoy,
  Bordas, Bregeon, Brun, Brun, Bryan, B{\"u}chele, Bulik, Capasso, Caroff,
  Carosi, Casanova, Cerruti, Chakraborty, Chaves, Chen, Chevalier,
  Colafrancesco, Condon, Conrad, Davids, Decock, Deil, Devin, {deWilt}, Dirson,
  {Djannati-Ata{\"i}}, Donath, O'C.~Drury, Dutson, Dyks, Edwards, Egberts,
  Emery, Ernenwein, Eschbach, Farnier, Fegan, Fernandes, Fiasson, Fontaine,
  Funk, F{\"u}ssling, Gabici, Gallant, Garrigoux, Gat{\'e}, Giavitto, Giebels,
  Glawion, Glicenstein, Gottschall, Grondin, Hahn, Haupt, Hawkes, Heinzelmann,
  Henri, Hermann, Hinton, Hofmann, Hoischen, Holch, Holler, Horns, Ivascenko,
  Iwasaki, Jacholkowska, Jamrozy, Jankowsky, Jankowsky, Jingo, Jouvin,
  {Jung-Richardt}, Kastendieck, Katarzy{\'n}ski, Katsuragawa, Kerszberg,
  Khangulyan, Kh{\'e}lifi, King, Klepser, Klochkov, Klu{\'z}niak, Komin,
  Kosack, Krakau, Kraus, Kr{\"u}ger, Laffon, Lamanna, Lau, Lees, Lefaucheur,
  Lemi{\`e}re, {Lemoine-Goumard}, Lenain, Leser, Lohse, Lorentz, Liu, Lypova,
  Malyshev, Marandon, Marcowith, Mariaud, Marx, Maurin, Maxted, Mayer,
  Meintjes, Meyer, Mitchell, Moderski, Mohamed, Mohrmann, Mor{\aa}, Moulin,
  Murach, Nakashima, de~Naurois, Ndiyavala, Niederwanger, Niemiec, Oakes,
  O'Brien, Odaka, Ohm, Ostrowski, Oya, Padovani, Panter, Parsons, Pekeur,
  Pelletier, Perennes, Petrucci, Peyaud, Piel, Pita, Poireau, Poon, Prokhorov,
  Prokoph, P{\"u}hlhofer, Punch, Quirrenbach, Raab, Rauth, Reimer, Reimer,
  Renaud, {de los Reyes}, Rieger, Rinchiuso, Romoli, Rowell, Rudak, Rulten,
  Sahakian, Saito, Sanchez, Santangelo, Sasaki, Schlickeiser, Sch{\"u}ssler,
  Schulz, Schwanke, Schwemmer, {Seglar-Arroyo}, Settimo, Seyffert, Shafi,
  Shilon, Shiningayamwe, Simoni, Sol, Spanier, {Spir-Jacob}, Stawarz,
  Steenkamp, Stegmann, Steppa, Sushch, Takahashi, Tavernet, Tavernier, Taylor,
  Terrier, Tibaldo, Tiziani, Tluczykont, Trichard, Tsirou, Tsuji, Tuffs,
  Uchiyama, {van der Walt}, van Eldik, van Rensburg, van Soelen, Vasileiadis,
  Veh, Venter, Viana, Vincent, Vink, Voisin, V{\"o}lk, Vuillaume, Wadiasingh,
  Wagner, Wagner, Wagner, White, Wierzcholska, Willmann, W{\"o}rnlein, Wouters,
  Yang, Zaborov, Zacharias, Zanin, Zdziarski, Zech, Zefi, Ziegler, Zorn,
  {\.Z}ywucka, {H.E.S.S. Collaboration}, Fender, Broderick, Rowlinson, Wijers,
  Stewart, {ter Veen}, Shulevski, {LOFAR Collaboration}, Kavic, Simonetti,
  League, Tsai, Obenberger, Nathaniel, Taylor, Dowell, Liebling, Estes,
  Lippert, Sharma, Vincent, Farella, {LWA: Long Wavelength Array}, Abeysekara,
  Albert, Alfaro, Alvarez, Arceo, {Arteaga-Vel{\'a}zquez}, Avila~Rojas,
  Ayala~Solares, Barber, Becerra~Gonzalez, Becerril, {Belmont-Moreno}, BenZvi,
  Berley, Bernal, Braun, Brisbois, {Caballero-Mora}, Capistr{\'a}n,
  Carrami{\~n}ana, Casanova, Castillo, Cotti, Cotzomi, {Couti{\~n}o de
  Le{\'o}n}, De~Le{\'o}n, {De la Fuente}, Diaz~Hernandez, Dichiara, Dingus,
  DuVernois, {D{\'i}az-V{\'e}lez}, Ellsworth, Engel, {Enr{\'i}quez-Rivera},
  Fiorino, Fleischhack, Fraija, {Garc{\'i}a-Gonz{\'a}lez}, Garfias, Gerhardt,
  Gonz{\~o}lez~Mu{\~n}oz, Gonz{\'a}lez, Goodman, {Hampel-Arias}, Harding,
  Hernandez, {Hernandez-Almada}, Hona, H{\"u}ntemeyer, Iriarte, {Jardin-Blicq},
  Joshi, Kaufmann, Kieda, Lara, Lauer, Lennarz, Le{\'o}n~Vargas, Linnemann,
  Longinotti, Luis~Raya, {Luna-Garc{\'i}a}, {L{\'o}pez-Coto}, Malone,
  Marinelli, Martinez, {Martinez-Castellanos}, {Mart{\'i}nez-Castro},
  {Mart{\'i}nez-Huerta}, Matthews, {Miranda-Romagnoli}, Moreno, Mostaf{\'a},
  Nellen, Newbold, Nisa, {Noriega-Papaqui}, Pelayo, Pretz,
  {P{\'e}rez-P{\'e}rez}, Ren, Rho, Rivi{\`e}re, {Rosa-Gonz{\'a}lez}, Rosenberg,
  {Ruiz-Velasco}, Salazar, Salesa~Greus, Sandoval, Schneider, Schoorlemmer,
  Sinnis, Smith, Springer, Surajbali, Tibolla, Tollefson, Torres, Ukwatta,
  Weisgarber, Westerhoff, Wisher, Wood, Yapici, Yodh, Younk, Zhou, {\'A}lvarez,
  {HAWC Collaboration}, Aab, Abreu, Aglietta, Albuquerque, Albury, Allekotte,
  Almela, Alvarez~Castillo, {Alvarez-Mu{\~n}iz}, Anastasi, Anchordoqui,
  Andrada, Andringa, Aramo, Arsene, Asorey, Assis, Avila, Badescu, Balaceanu,
  Barbato, Barreira~Luz, Becker, Bellido, Berat, Bertaina, Bertou, Biermann,
  Biteau, Blaess, Blanco, Blazek, Bleve, Boh{\'a}{\v c}ov{\'a}, Bonifazi,
  Borodai, Botti, Brack, Brancus, Bretz, Bridgeman, Briechle, Buchholz, Bueno,
  Buitink, Buscemi, {Caballero-Mora}, Caccianiga, Cancio, Canfora, Caruso,
  Castellina, Catalani, Cataldi, Cazon, Chavez, Chinellato, Chudoba, Clay,
  Cobos~Cerutti, Colalillo, Coleman, Collica, Coluccia, Concei{\c c}{\~a}o,
  Consolati, Contreras, Cooper, Coutu, Covault, Cronin, D'Amico, Daniel, Dasso,
  Daumiller, Dawson, Day, de~Almeida, de~Jong, Mauro, {de Mello Neto}, Mitri,
  de~Oliveira, de~Souza, Debatin, Deligny, D{\'i}az~Castro, Diogo, Dobrigkeit,
  D'Olivo, Dorosti, Dos~Anjos, Dova, Dundovic, Ebr, Engel, Erdmann, Erfani,
  Escobar, Espadanal, Etchegoyen, Falcke, Farmer, Farrar, Fauth, Fazzini,
  Feldbusch, Fenu, Fick, Figueira, Filip{\v c}i{\v c}, Freire, Fujii, Fuster,
  Ga{\"i}or, Garc{\'i}a, Gat{\'e}, Gemmeke, {Gherghel-Lascu}, Ghia, Giaccari,
  Giammarchi, Giller, G{\l}as, Glaser, Golup, G{\'o}mez~Berisso,
  G{\'o}mez~Vitale, Gonz{\'a}lez, Gorgi, Gottowik, Grillo, Grubb, Guarino,
  Guedes, Halliday, Hampel, Hansen, Harari, Harrison, Harvey, Haungs, Hebbeker,
  Heck, Heimann, Herve, Hill, Hojvat, Holt, Homola, H{\"o}randel, Horvath,
  Hrabovsk{\'y}, Huege, Hulsman, Insolia, Isar, Jandt, Johnsen, Josebachuili,
  Jurysek, K{\"a}{\"a}p{\"a}, Kampert, Keilhauer, Kemmerich, Kemp, Kieckhafer,
  Klages, Kleifges, Kleinfeller, Krause, Krohm, Kuempel, Kukec~Mezek, Kunka,
  Kuotb~Awad, Lago, LaHurd, Lang, Lauscher, Legumina, {Leigui de Oliveira},
  {Letessier-Selvon}, {Lhenry-Yvon}, Link, Lo~Presti, Lopes, L{\'o}pez,
  L{\'o}pez~Casado, Lorek, Luce, Lucero, Malacari, Mallamaci, Mandat, Mantsch,
  Mariazzi, Maris, Marsella, Martello, Martinez, Mart{\'i}nez~Bravo,
  Mas{\'i}as~Meza, Mathes, Mathys, Matthews, Matthiae, Mayotte, Mazur, Medina,
  {Medina-Tanco}, Melo, Menshikov, Merenda, Michal, Micheletti, Middendorf,
  Miramonti, Mitrica, Mockler, Mollerach, Montanet, Morello, Morlino,
  M{\"u}ller, M{\"u}ller, Muller, M{\"u}ller, Mussa, Naranjo, Nguyen,
  {Niculescu-Oglinzanu}, Niechciol, Niemietz, Niggemann, Nitz, Nosek, Novotny,
  No{\v z}ka, N{\'u}{\~n}ez, Oikonomou, Olinto, Palatka, Pallotta, Papenbreer,
  Parente, Parra, Paul, Pech, Pedreira, P{\c e}kala, {Pe{\~n}a-Rodriguez},
  Pereira, Perlin, Perrone, Peters, Petrera, Phuntsok, Pierog, Pimenta,
  Pirronello, Platino, Plum, Poh, Porowski, Prado, Privitera, Prouza, Quel,
  Querchfeld, Quinn, {Ramos-Pollan}, Rautenberg, Ravignani, Ridky, Riehn,
  Risse, Ristori, Rizi, {Rodrigues de Carvalho}, Rodriguez~Fernandez,
  Rodriguez~Rojo, Roncoroni, Roth, Roulet, Rovero, Ruehl, Saffi, Saftoiu,
  Salamida, Salazar, Saleh, Salina, S{\'a}nchez, {Sanchez-Lucas}, Santos,
  Santos, Sarazin, Sarmento, {Sarmiento-Cano}, Sato, Schauer, Scherini,
  Schieler, Schimp, Schmidt, Scholten, Schov{\'a}nek, Schr{\"o}der,
  Schr{\"o}der, Schulz, Schumacher, Sciutto, Segreto, Shadkam, Shellard, Sigl,
  Silli, {\v S}m{\'i}da, Snow, Sommers, Sonntag, Soriano, Squartini, Stanca,
  Stani{\v c}, Stasielak, Stassi, Stolpovskiy, Strafella, Streich, Suarez,
  {Suarez-Dur{\'a}n}, Sudholz, Suomij{\"a}rvi, Supanitsky, {\v S}up{\'i}k,
  Swain, Szadkowski, Taboada, Taborda, Timmermans, Todero~Peixoto, Tomankova,
  Tom{\'e}, Torralba~Elipe, Travnicek, Trini, Tueros, Ulrich, Unger, Urban,
  Vald{\'e}s~Galicia, Vali{\~n}o, Valore, van Aar, van Bodegom, {van den Berg},
  van Vliet, Varela, C{\'a}rdenas, V{\'a}zquez, Veberi{\v c}, Ventura,
  Vergara~Quispe, Verzi, Vicha, Villase{\~n}or, Vorobiov, Wahlberg, Wainberg,
  Walz, Watson, Weber, Weindl, Wiede{\'n}ski, Wiencke, Wilczy{\'n}ski, Wirtz,
  Wittkowski, Wundheiler, Yang, Yushkov, Zas, Zavrtanik, Zavrtanik, Zepeda,
  Zimmermann, Ziolkowski, Zong, Zuccarello, {The Pierre Auger Collaboration},
  Kim, Schulze, Bauer, {Corral-Santana}, {de Gregorio-Monsalvo},
  {Gonz{\'a}lez-L{\'o}pez}, Hartmann, {Ishwara-Chandra}, Mart{\'i}n, Mehner,
  Misra, Micha{\l}owski, Resmi, {ALMA Collaboration}, Paragi, Agudo, An,
  Beswick, Casadio, Frey, Jonker, Kettenis, Marcote, Moldon, Szomoru, {van
  Langevelde}, Yang, {Euro VLBI Team}, Cwiek, Cwiok, Czyrkowski, Dabrowski,
  Kasprowicz, Mankiewicz, Nawrocki, Opiela, Piotrowski, Wrochna, Zaremba,
  {\.Z}arnecki, {Pi of the Sky Collaboration}, Haggard, Nynka, Ruan, {The
  Chandra Team at McGill University}, Bland, Booler, Devillepoix, de~Gois,
  Hancock, Howie, Paxman, Sansom, Towner, {DFN: Desert Fireball Network},
  Tonry, Coughlin, Stubbs, Denneau, Heinze, Stalder, Weiland, {Atlas}, Eatough,
  Kramer, Kraus, {High Time Resolution Universe Survey}, Troja, Piro,
  Gonz{\'a}lez, Butler, Fox, Khandrika, Kutyrev, Lee, Ricci, Ryan~Jr.,
  {S{\'a}nchez-Ram{\'i}rez}, Veilleux, Watson, Wieringa, Burgess, van Eerten,
  Fontes, Fryer, Korobkin, Wollaeger, {RIMAS and RATIR}, Camilo, Foley,
  Goedhart, Makhathini, Oozeer, Smirnov, Fender, \& Woudt}]{abbott_2017}
Abbott, B.~P., Abbott, R., Abbott, T.~D., {et~al.} 2017{\natexlab{c}}, ApJ,
  848, L12

\bibitem[{Aghanim {et~al.}(2020)Aghanim, Akrami, Ashdown, Aumont, Baccigalupi,
  Ballardini, Banday, Barreiro, Bartolo, Basak, Battye, Benabed, Bernard,
  Bersanelli, Bielewicz, Bock, Bond, Borrill, Bouchet, Boulanger, Bucher,
  Burigana, Butler, Calabrese, Cardoso, Carron, Challinor, Chiang, Chluba,
  Colombo, Combet, Contreras, Crill, Cuttaia, {de Bernardis}, {de Zotti},
  Delabrouille, Delouis, Di~Valentino, Diego, Dor{\'e}, Douspis, Ducout, Dupac,
  Dusini, Efstathiou, Elsner, En{\ss}lin, Eriksen, Fantaye, Farhang, Fergusson,
  {Fernandez-Cobos}, Finelli, Forastieri, Frailis, Fraisse, Franceschi, Frolov,
  Galeotta, Galli, Ganga, {G{\'e}nova-Santos}, Gerbino, Ghosh,
  {Gonz{\'a}lez-Nuevo}, G{\'o}rski, Gratton, Gruppuso, Gudmundsson, Hamann,
  Handley, Hansen, Herranz, Hildebrandt, Hivon, Huang, Jaffe, Jones, Karakci,
  Keih{\"a}nen, Keskitalo, Kiiveri, Kim, Kisner, Knox, Krachmalnicoff, Kunz,
  {Kurki-Suonio}, Lagache, Lamarre, Lasenby, Lattanzi, Lawrence, Le~Jeune,
  Lemos, Lesgourgues, Levrier, Lewis, Liguori, Lilje, Lilley, Lindholm,
  {L{\'o}pez-Caniego}, Lubin, Ma, {Mac{\'i}as-P{\'e}rez}, Maggio, Maino,
  Mandolesi, Mangilli, {Marcos-Caballero}, Maris, Martin, Martinelli,
  {Mart{\'i}nez-Gonz{\'a}lez}, Matarrese, Mauri, McEwen, Meinhold, Melchiorri,
  Mennella, Migliaccio, Millea, Mitra, {Miville-Desch{\^e}nes}, Molinari,
  Montier, Morgante, Moss, Natoli, {N{\o}rgaard-Nielsen}, Pagano, Paoletti,
  Partridge, Patanchon, Peiris, Perrotta, Pettorino, Piacentini, Polastri,
  Polenta, Puget, Rachen, Reinecke, Remazeilles, Renzi, Rocha, Rosset, Roudier,
  {Rubi{\~n}o-Mart{\'i}n}, {Ruiz-Granados}, Salvati, Sandri, Savelainen, Scott,
  Shellard, Sirignano, Sirri, Spencer, Sunyaev, {Suur-Uski}, Tauber,
  Tavagnacco, Tenti, Toffolatti, Tomasi, Trombetti, Valenziano, Valiviita,
  Van~Tent, Vibert, Vielva, Villa, Vittorio, Wandelt, Wehus, White, White,
  Zacchei, \& Zonca}]{aghanim_2020}
Aghanim, N., Akrami, Y., Ashdown, M., {et~al.} 2020, A\&A, 641, A6

\bibitem[{Andrews {et~al.}(2025)Andrews, Bavera, Briel, Chattaraj, Dotter,
  Fragos, {Gallegos-Garcia}, Gossage, Kalogera, Kasdagli, Katsaggelos, Kimball,
  Kovlakas, Kruckow, Liotine, Misra, Rocha, Souropanis, Srivastava, Sun, Teng,
  Xing, Zapartas, \& Zevin}]{andrews_2025}
Andrews, J.~J., Bavera, S.~S., Briel, M., {et~al.} 2025, Submitted to AAS
  journals [\eprint[arXiv]{2411.02376}]

\bibitem[{Batta \& {Ramirez-Ruiz}(2019)}]{batta_2019}
Batta, A. \& {Ramirez-Ruiz}, E. 2019 [\eprint[arXiv]{1904.04835}]

\bibitem[{Bavera {et~al.}(2020)Bavera, Fragos, Qin, Zapartas, Neijssel, Mandel,
  Batta, Gaebel, Kimball, \& Stevenson}]{bavera_2020}
Bavera, S.~S., Fragos, T., Qin, Y., {et~al.} 2020, A\&A, 635, A97

\bibitem[{Bavera {et~al.}(2022)Bavera, Fragos, Zapartas, {Ramirez-Ruiz},
  Marchant, Kelley, Zevin, Andrews, Coughlin, Dotter, Kovlakas, Misra,
  {Serra-Perez}, Qin, Rocha, {Rom{\'a}n-Garza}, Tran, \& Xing}]{bavera_2022b}
Bavera, S.~S., Fragos, T., Zapartas, E., {et~al.} 2022, A\&A, 657, L8

\bibitem[{Beckwith {et~al.}(2009)Beckwith, Hawley, \& Krolik}]{beckwith_2009}
Beckwith, K., Hawley, J.~F., \& Krolik, J.~H. 2009, ApJ, 707, 428

\bibitem[{Berger(2014)}]{berger_2014}
Berger, E. 2014, ARA\&A, 52, 43

\bibitem[{Blandford \& Znajek(1977)}]{blandford_1977}
Blandford, R.~D. \& Znajek, R.~L. 1977, MNRAS, 179, 433

\bibitem[{Braun \& Langer(1995)}]{braun_1995}
Braun, H. \& Langer, N. 1995, A\&A, 297, 483

\bibitem[{Bromberg {et~al.}(2011)Bromberg, Nakar, Piran, \&
  Sari}]{bromberg_2011}
Bromberg, O., Nakar, E., Piran, T., \& Sari, R. 2011, ApJ, 740, 100

\bibitem[{Bromberg {et~al.}(2012)Bromberg, Nakar, Piran, \&
  Sari}]{bromberg_2012}
Bromberg, O., Nakar, E., Piran, T., \& Sari, R. 2012, ApJ, 749, 110

\bibitem[{Cantiello {et~al.}(2014)Cantiello, Mankovich, Bildsten,
  {Christensen-Dalsgaard}, \& Paxton}]{cantiello_2014}
Cantiello, M., Mankovich, C., Bildsten, L., {Christensen-Dalsgaard}, J., \&
  Paxton, B. 2014, ApJ, 788, 93

\bibitem[{Cantiello {et~al.}(2007)Cantiello, Yoon, Langer, \&
  Livio}]{cantiello_2007}
Cantiello, M., Yoon, S.-C., Langer, N., \& Livio, M. 2007, A\&A, 465, L29

\bibitem[{Chrimes {et~al.}(2020)Chrimes, Stanway, \& Eldridge}]{chrimes_2020}
Chrimes, A.~A., Stanway, E.~R., \& Eldridge, J.~J. 2020, MNRAS, 491, 3479

\bibitem[{Corsi \& Lazzati(2021)}]{corsi_2021}
Corsi, A. \& Lazzati, D. 2021, New Astron. Rev., 92, 101614

\bibitem[{Cucchiara {et~al.}(2011)Cucchiara, Levan, Fox, Tanvir, Ukwatta,
  Berger, Kr{\"u}hler, K{\"u}pc{\"u}~Yolda{\c s}, Wu, Toma, Greiner, Olivares,
  Rowlinson, Amati, Sakamoto, Roth, Stephens, Fritz, Fynbo, Hjorth, Malesani,
  Jakobsson, Wiersema, O'Brien, Soderberg, Foley, Fruchter, Rhoads, Rutledge,
  Schmidt, Dopita, Podsiadlowski, Willingale, Wolf, Kulkarni, \&
  D'Avanzo}]{cucchiara_2011}
Cucchiara, A., Levan, A.~J., Fox, D.~B., {et~al.} 2011, ApJ, 736, 7

\bibitem[{{de Mink} {et~al.}(2009){de Mink}, Cantiello, Langer, Pols, Brott, \&
  Yoon}]{demink_2009}
{de Mink}, S.~E., Cantiello, M., Langer, N., {et~al.} 2009, A\&A, 497, 243

\bibitem[{{den Hartogh} {et~al.}(2020){den Hartogh}, Eggenberger, \&
  Deheuvels}]{denhartogh_2020}
{den Hartogh}, J.~W., Eggenberger, P., \& Deheuvels, S. 2020, A\&A, 634, L16

\bibitem[{Detmers {et~al.}(2008)Detmers, Langer, Podsiadlowski, \&
  Izzard}]{detmers_2008}
Detmers, R.~G., Langer, N., Podsiadlowski, {\relax Ph}., \& Izzard, R.~G. 2008,
  A\&A, 484, 831

\bibitem[{Eggenberger {et~al.}(2022)Eggenberger, Moyano, \& {den
  Hartogh}}]{eggenberger_2022}
Eggenberger, P., Moyano, F.~D., \& {den Hartogh}, J.~W. 2022, A\&A, 664, L16

\bibitem[{Eichler {et~al.}(1989)Eichler, Livio, Piran, \&
  Schramm}]{eichler_1989}
Eichler, D., Livio, M., Piran, T., \& Schramm, D.~N. 1989, Nat, 340, 126

\bibitem[{Eldridge {et~al.}(2011)Eldridge, Langer, \& Tout}]{eldridge_2011}
Eldridge, J.~J., Langer, N., \& Tout, C.~A. 2011, MNRAS, 414, 3501

\bibitem[{Ercolino {et~al.}(2024)Ercolino, Jin, Langer, \&
  Dessart}]{ercolino_2024}
Ercolino, A., Jin, H., Langer, N., \& Dessart, L. 2024, A\&A, 685, A58

\bibitem[{Farmer {et~al.}(2016)Farmer, Fields, Petermann, Dessart, Cantiello,
  Paxton, \& Timmes}]{farmer_2016}
Farmer, R., Fields, C.~E., Petermann, I., {et~al.} 2016, ApJS, 227, 22

\bibitem[{Fragos {et~al.}(2023)Fragos, Andrews, Bavera, Berry, Coughlin,
  Dotter, Giri, Kalogera, Katsaggelos, Kovlakas, Lalvani, Misra, Srivastava,
  Qin, Rocha, {Roman-Garza}, Serra, Stahle, Sun, Teng, Trajcevski, Tran, Xing,
  Zapartas, \& Zevin}]{fragos_2023}
Fragos, T., Andrews, J.~J., Bavera, S.~S., {et~al.} 2023, ApJS, 264, 45

\bibitem[{Frohmaier {et~al.}(2021)Frohmaier, Angus, Vincenzi, Sullivan, Smith,
  Nugent, Cenko, {Gal-Yam}, Kulkarni, Law, \& Quimby}]{frohmaier_2021}
Frohmaier, C., Angus, C.~R., Vincenzi, M., {et~al.} 2021, MNRAS, 500, 5142

\bibitem[{Fryer {et~al.}(2012)Fryer, Belczynski, Wiktorowicz, Dominik,
  Kalogera, \& Holz}]{fryer_2012}
Fryer, C.~L., Belczynski, K., Wiktorowicz, G., {et~al.} 2012, ApJ, 749, 91

\bibitem[{Fryer {et~al.}(2025)Fryer, Burns, Ho, Corsi, Lien, Perley, Vail, \&
  Villar}]{fryer_2025}
Fryer, C.~L., Burns, E., Ho, A. Y.~Q., {et~al.} 2025, ApJ, 986, 185

\bibitem[{Fryer \& Heger(2005)}]{fryer_2005}
Fryer, C.~L. \& Heger, A. 2005, ApJ, 623, 302

\bibitem[{Fryer \& Woosley(1998)}]{fryer_1998}
Fryer, C.~L. \& Woosley, S.~E. 1998, ApJ, 502, L9

\bibitem[{Fuller {et~al.}(2014)Fuller, Lecoanet, Cantiello, \&
  Brown}]{fuller_2014}
Fuller, J., Lecoanet, D., Cantiello, M., \& Brown, B. 2014, ApJ, 796, 17

\bibitem[{Fuller \& Ma(2019)}]{fuller_2019}
Fuller, J. \& Ma, L. 2019, ApJ, 881, L1

\bibitem[{Fynbo {et~al.}(2006)Fynbo, Watson, Th{\"o}ne, Sollerman, Bloom,
  Davis, Hjorth, Jakobsson, J{\o}rgensen, Graham, Fruchter, Bersier, Kewley,
  Cassan, Cer{\'o}n, Foley, Gorosabel, Hinse, Horne, Jensen, Klose, Kocevski,
  Marquette, Perley, {Ramirez-Ruiz}, Stritzinger, Vreeswijk, Wijers, Woller,
  Xu, \& Zub}]{fynbo_2006}
Fynbo, J. P.~U., Watson, D., Th{\"o}ne, C.~C., {et~al.} 2006, Nat, 444, 1047

\bibitem[{{Gal-Yam} {et~al.}(2006){Gal-Yam}, Fox, Price, Ofek, Davis, Leonard,
  Soderberg, Schmidt, Lewis, Peterson, Kulkarni, Berger, Cenko, Sari, Sharon,
  Frail, Moon, Brown, Cucchiara, Harrison, Piran, Persson, McCarthy, Penprase,
  Chevalier, \& MacFadyen}]{gal-yam_2006}
{Gal-Yam}, A., Fox, D.~B., Price, P.~A., {et~al.} 2006, Nat, 444, 1053

\bibitem[{Ghirlanda \& Salvaterra(2022)}]{ghirlanda_2022}
Ghirlanda, G. \& Salvaterra, R. 2022, ApJ, 932, 10

\bibitem[{Ghodla {et~al.}(2023)Ghodla, Eldridge, Stanway, \&
  Stevance}]{ghodla_2023}
Ghodla, S., Eldridge, J.~J., Stanway, E.~R., \& Stevance, H.~F. 2023, MNRAS,
  518, 860

\bibitem[{G{\"o}tberg {et~al.}(2017)G{\"o}tberg, de~Mink, \&
  Groh}]{gotberg_2017}
G{\"o}tberg, Y., de~Mink, S.~E., \& Groh, J.~H. 2017, A\&A, 608, A11

\bibitem[{Gottlieb {et~al.}(2023)Gottlieb, Metzger, Quataert, Issa, Martineau,
  Foucart, Duez, Kidder, Pfeiffer, \& Scheel}]{gottlieb_2023}
Gottlieb, O., Metzger, B.~D., Quataert, E., {et~al.} 2023, ApJ, 958, L33

\bibitem[{Gottlieb {et~al.}(2024)Gottlieb, Renzo, Metzger, Goldberg, \&
  Cantiello}]{gottlieb_2024}
Gottlieb, O., Renzo, M., Metzger, B.~D., Goldberg, J.~A., \& Cantiello, M.
  2024, ApJ, 976, L13

\bibitem[{Graham \& Fruchter(2013)}]{graham_2013}
Graham, J.~F. \& Fruchter, A.~S. 2013, ApJ, 774, 119

\bibitem[{Graham {et~al.}(2023)Graham, Schady, \& Fruchter}]{graham_2023}
Graham, J.~F., Schady, P., \& Fruchter, A.~S. 2023, ApJ, 954, 13

\bibitem[{Greiner {et~al.}(2015)Greiner, Fox, Schady, Kr{\"u}hler, Trenti,
  Cikota, Bolmer, Elliott, Delvaux, Perna, Afonso, Kann, Klose, Savaglio,
  Schmidl, Schweyer, Tanga, \& Varela}]{greiner_2015}
Greiner, J., Fox, D.~B., Schady, P., {et~al.} 2015, ApJ, 809, 76

\bibitem[{Guetta \& Valle(2007)}]{guetta_2007}
Guetta, D. \& Valle, M.~D. 2007, ApJ, 657, L73

\bibitem[{Harrison {et~al.}(2018)Harrison, Gottlieb, \& Nakar}]{harrison_2018}
Harrison, R., Gottlieb, O., \& Nakar, E. 2018, MNRAS, 477, 2128

\bibitem[{Heger {et~al.}(2003)Heger, Fryer, Woosley, Langer, \&
  Hartmann}]{heger_2003}
Heger, A., Fryer, C.~L., Woosley, S.~E., Langer, N., \& Hartmann, D.~H. 2003,
  ApJ, 591, 288

\bibitem[{Heger {et~al.}(2005)Heger, Woosley, \& Spruit}]{heger_2005}
Heger, A., Woosley, S.~E., \& Spruit, H.~C. 2005, ApJ, 626, 350

\bibitem[{Heintz {et~al.}(2018)Heintz, Malesani, Wiersema, Jakobsson, Fynbo,
  Savaglio, Cano, Covino, D'Elia, Gomboc, Hammer, Kaper, {Milvang-Jensen},
  M{\o}ller, Piranomonte, Selsing, Rhodin, Tanvir, Th{\"o}ne, {de Ugarte
  Postigo}, Vergani, \& Watson}]{heintz_2018}
Heintz, K.~E., Malesani, D., Wiersema, K., {et~al.} 2018, MNRAS, 474, 2738

\bibitem[{Ivanova \& Podsiadlowski(2003)}]{ivanova_2003}
Ivanova, N. \& Podsiadlowski, P. 2003, in From {{Twilight To Highlight Physics
  Of Supernovae}}, ed. W.~Hillebrandt \& B.~Leibundgut (Berlin, Heidelberg:
  Springer), 19--22

\bibitem[{Ivy~Wang {et~al.}(2022)Ivy~Wang, Zhang, \& Lei}]{ivywang_2022}
Ivy~Wang, X., Zhang, B.-B., \& Lei, W.-H. 2022, ApJ, 931, L2

\bibitem[{Izzard {et~al.}(2004)Izzard, {Ramirez-Ruiz}, \& Tout}]{izzard_2004a}
Izzard, R.~G., {Ramirez-Ruiz}, E., \& Tout, C.~A. 2004, MNRAS, 348, 1215

\bibitem[{{Jacquemin-Ide} {et~al.}(2024){Jacquemin-Ide}, Gottlieb, Lowell, \&
  Tchekhovskoy}]{jacquemin-ide_2024}
{Jacquemin-Ide}, J., Gottlieb, O., Lowell, B., \& Tchekhovskoy, A. 2024, ApJ,
  961, 212

\bibitem[{Japelj {et~al.}(2016)Japelj, Vergani, Salvaterra, D'Avanzo, Mannucci,
  {Fernandez-Soto}, Boissier, Hunt, Atek, {Rodr{\'i}guez-Mu{\~n}oz}, Scodeggio,
  Cristiani, Le~Floc'h, Flores, Gallego, Ghirlanda, Gomboc, Hammer, Perley,
  Pescalli, Petitjean, Puech, Rafelski, \& Tagliaferri}]{japelj_2016}
Japelj, J., Vergani, S.~D., Salvaterra, R., {et~al.} 2016, A\&A, 590, A129

\bibitem[{Japelj {et~al.}(2018)Japelj, Vergani, Salvaterra, Renzo, Zapartas,
  {de Mink}, Kaper, \& Zibetti}]{japelj_2018}
Japelj, J., Vergani, S.~D., Salvaterra, R., {et~al.} 2018, A\&A, 617, A105

\bibitem[{Jermyn {et~al.}(2023)Jermyn, Bauer, Schwab, Farmer, Ball, Bellinger,
  Dotter, Joyce, Marchant, Mombarg, Wolf, Sunny~Wong, Cinquegrana, Farrell,
  Smolec, Thoul, Cantiello, Herwig, Toloza, Bildsten, Townsend, \&
  Timmes}]{jermyn_2023}
Jermyn, A.~S., Bauer, E.~B., Schwab, J., {et~al.} 2023, ApJS, 265, 15

\bibitem[{Kawanaka {et~al.}(2013)Kawanaka, Piran, \& Krolik}]{kawanaka_2013}
Kawanaka, N., Piran, T., \& Krolik, J.~H. 2013, ApJ, 766, 31

\bibitem[{Kewley \& Ellison(2008)}]{kewley_2008}
Kewley, L.~J. \& Ellison, S.~L. 2008, ApJ, 681, 1183

\bibitem[{Kistler {et~al.}(2009)Kistler, Y{\"u}ksel, Beacom, Hopkins, \&
  Wyithe}]{kistler_2009}
Kistler, M.~D., Y{\"u}ksel, H., Beacom, J.~F., Hopkins, A.~M., \& Wyithe, J.
  S.~B. 2009, ApJ, 705, L104

\bibitem[{Klencki {et~al.}(2022)Klencki, Istrate, Nelemans, \&
  Pols}]{klencki_2022}
Klencki, J., Istrate, A., Nelemans, G., \& Pols, O. 2022, A\&A, 662, A56

\bibitem[{Kolb \& Ritter(1990)}]{kolb_1990}
Kolb, U. \& Ritter, H. 1990, A\&A, 236, 385

\bibitem[{Kroupa(2001)}]{kroupa_2001}
Kroupa, P. 2001, MNRAS, 322, 231

\bibitem[{Kr{\"u}hler {et~al.}(2015)Kr{\"u}hler, Malesani, Fynbo, Hartoog,
  Hjorth, Jakobsson, Perley, Rossi, Schady, Schulze, Tanvir, Vergani, Wiersema,
  Afonso, Bolmer, Cano, Covino, D'Elia, {de Ugarte Postigo}, Filgas, Friis,
  Graham, Greiner, Goldoni, Gomboc, Hammer, Japelj, Kann, Kaper, Klose, Levan,
  Leloudas, {Milvang-Jensen}, Nicuesa~Guelbenzu, Palazzi, Pian, Piranomonte,
  {S{\'a}nchez-Ram{\'i}rez}, Savaglio, Selsing, Tagliaferri, Vreeswijk, Watson,
  \& Xu}]{kruhler_2015}
Kr{\"u}hler, T., Malesani, D., Fynbo, J. P.~U., {et~al.} 2015, A\&A, 581, A125

\bibitem[{Laplace {et~al.}(2021)Laplace, Justham, Renzo, G{\"o}tberg, Farmer,
  Vartanyan, \& {de Mink}}]{laplace_2021}
Laplace, E., Justham, S., Renzo, M., {et~al.} 2021, A\&A, 656, A58

\bibitem[{Levan {et~al.}(2024)Levan, Gompertz, Salafia, Bulla, Burns,
  Hotokezaka, Izzo, Lamb, Malesani, Oates, Ravasio, Rouco~Escorial, Schneider,
  Sarin, Schulze, Tanvir, Ackley, Anderson, Brammer, Christensen, Dhillon,
  Evans, Fausnaugh, Fong, Fruchter, Fryer, Fynbo, Gaspari, Heintz, Hjorth,
  Kennea, Kennedy, Laskar, Leloudas, Mandel, {Martin-Carrillo}, Metzger,
  Nicholl, Nugent, Palmerio, Pugliese, Rastinejad, Rhodes, Rossi, Saccardi,
  Smartt, Stevance, Tohuvavohu, {van der Horst}, Vergani, Watson, Barclay,
  Bhirombhakdi, Breedt, Breeveld, Brown, Campana, Chrimes, D'Avanzo, D'Elia,
  De~Pasquale, Dyer, Galloway, Garbutt, Green, Hartmann, Jakobsson, Kerry,
  Kouveliotou, Langeroodi, Le~Floc'h, Leung, Littlefair, Munday, O'Brien,
  Parsons, Pelisoli, Sahman, Salvaterra, Sbarufatti, Steeghs, Tagliaferri,
  Th{\"o}ne, {de Ugarte Postigo}, \& Kann}]{levan_2024}
Levan, A.~J., Gompertz, B.~P., Salafia, O.~S., {et~al.} 2024, Nat, 626, 737

\bibitem[{Levesque {et~al.}(2010{\natexlab{a}})Levesque, Berger, Kewley, \&
  Bagley}]{levesque_2010b}
Levesque, E.~M., Berger, E., Kewley, L.~J., \& Bagley, M.~M.
  2010{\natexlab{a}}, AJ, 139, 694

\bibitem[{Levesque {et~al.}(2010{\natexlab{b}})Levesque, Kewley, Berger, \&
  Zahid}]{levesque_2010a}
Levesque, E.~M., Kewley, L.~J., Berger, E., \& Zahid, H.~J. 2010{\natexlab{b}},
  AJ, 140, 1557

\bibitem[{Levesque {et~al.}(2010{\natexlab{c}})Levesque, Kewley, Graham, \&
  Fruchter}]{levesque_2010}
Levesque, E.~M., Kewley, L.~J., Graham, J.~F., \& Fruchter, A.~S.
  2010{\natexlab{c}}, ApJ, 712, L26

\bibitem[{Liang {et~al.}(2007)Liang, Zhang, Virgili, \& Dai}]{liang_2007}
Liang, E., Zhang, B., Virgili, F., \& Dai, Z.~G. 2007, ApJ, 662, 1111

\bibitem[{Lowell {et~al.}(2024)Lowell, {Jacquemin-Ide}, Tchekhovskoy, \&
  Duncan}]{lowell_2024}
Lowell, B., {Jacquemin-Ide}, J., Tchekhovskoy, A., \& Duncan, A. 2024, ApJ,
  960, 82

\bibitem[{MacFadyen \& Woosley(1999)}]{macfadyen_1999}
MacFadyen, A.~I. \& Woosley, S.~E. 1999, ApJ, 524, 262

\bibitem[{Madau \& Fragos(2017)}]{madau_2017}
Madau, P. \& Fragos, T. 2017, ApJ, 840, 39

\bibitem[{Maeder(1987)}]{maeder_1987}
Maeder, A. 1987, A\&A, 178, 159

\bibitem[{Mandel \& {de~Mink}(2016)}]{mandel_2016}
Mandel, I. \& {de~Mink}, S.~E. 2016, MNRAS, 458, 2634

\bibitem[{Marchant {et~al.}(2017)Marchant, Langer, Podsiadlowski, Tauris,
  de~Mink, Mandel, \& Moriya}]{marchant_2017}
Marchant, P., Langer, N., Podsiadlowski, P., {et~al.} 2017, A\&A, 604, A55

\bibitem[{Marchant {et~al.}(2016)Marchant, Langer, Podsiadlowski, Tauris, \&
  Moriya}]{marchant_2016}
Marchant, P., Langer, N., Podsiadlowski, P., Tauris, T.~M., \& Moriya, T.~J.
  2016, A\&A, 588, A50

\bibitem[{Marinacci {et~al.}(2018)Marinacci, Vogelsberger, Pakmor, Torrey,
  Springel, Hernquist, Nelson, Weinberger, Pillepich, Naiman, \&
  Genel}]{marinacci_2018}
Marinacci, F., Vogelsberger, M., Pakmor, R., {et~al.} 2018, MNRAS, 480, 5113

\bibitem[{Matzner(2003)}]{matzner_2003}
Matzner, C.~D. 2003, MNRAS, 345, 575

\bibitem[{Mazzali {et~al.}(2014)Mazzali, McFadyen, Woosley, Pian, \&
  Tanaka}]{mazzali_2014}
Mazzali, P.~A., McFadyen, A.~I., Woosley, S.~E., Pian, E., \& Tanaka, M. 2014,
  MNRAS, 443, 67

\bibitem[{Metha \& Trenti(2020)}]{metha_2020}
Metha, B. \& Trenti, M. 2020, MNRAS, 495, 266

\bibitem[{Metzger {et~al.}(2011)Metzger, Giannios, Thompson, Bucciantini, \&
  Quataert}]{metzger_2011}
Metzger, B.~D., Giannios, D., Thompson, T.~A., Bucciantini, N., \& Quataert, E.
  2011, MNRAS, 413, 2031

\bibitem[{Meynet \& Maeder(2007)}]{meynet_2007}
Meynet, G. \& Maeder, A. 2007, A\&A, 464, L11

\bibitem[{Micha{\l}owskI {et~al.}(2018)Micha{\l}owskI, Xu, Stevens, Levan,
  Yang, Paragi, Kamble, Tsai, Dannerbauer, van~der Horst, Shao, Crosby,
  Gentile, Stanway, Wiersema, Fynbo, Tanvir, Kamphuis, Garrett, \&
  Bartczak}]{michalowski_2018}
Micha{\l}owskI, M.~J., Xu, D., Stevens, J., {et~al.} 2018, A\&A, 616, A169

\bibitem[{Modjaz {et~al.}(2008)Modjaz, Kewley, Kirshner, Stanek, Challis,
  Garnavich, Greene, Kelly, \& Prieto}]{modjaz_2008}
Modjaz, M., Kewley, L., Kirshner, R.~P., {et~al.} 2008, AJ, 135, 1136

\bibitem[{Modjaz {et~al.}(2016)Modjaz, Liu, Bianco, \& Graur}]{modjaz_2016}
Modjaz, M., Liu, Y.~Q., Bianco, F.~B., \& Graur, O. 2016, ApJ, 832, 108

\bibitem[{Mokiem {et~al.}(2007)Mokiem, {de Koter}, Vink, Puls, Evans, Smartt,
  Crowther, Herrero, Langer, Lennon, Najarro, \& Villamariz}]{mokiem_2007a}
Mokiem, M.~R., {de Koter}, A., Vink, J.~S., {et~al.} 2007, A\&A, 473, 603

\bibitem[{Moyano {et~al.}(2023)Moyano, Eggenberger, Mosser, \&
  Spada}]{moyano_2023}
Moyano, F.~D., Eggenberger, P., Mosser, B., \& Spada, F. 2023, A\&A, 673, A110

\bibitem[{Naiman {et~al.}(2018)Naiman, Pillepich, Springel, {Ramirez-Ruiz},
  Torrey, Vogelsberger, Pakmor, Nelson, Marinacci, Hernquist, Weinberger, \&
  Genel}]{naiman_2018}
Naiman, J.~P., Pillepich, A., Springel, V., {et~al.} 2018, MNRAS, 477, 1206

\bibitem[{Nakar \& Sari(2012)}]{nakar_2012}
Nakar, E. \& Sari, R. 2012, ApJ, 747, 88

\bibitem[{Neijssel {et~al.}(2019)Neijssel, {Vigna-G{\'o}mez}, Stevenson,
  Barrett, Gaebel, Broekgaarden, {de Mink}, Sz{\'e}csi, Vinciguerra, \&
  Mandel}]{neijssel_2019}
Neijssel, C.~J., {Vigna-G{\'o}mez}, A., Stevenson, S., {et~al.} 2019, MNRAS,
  490, 3740

\bibitem[{Nelson {et~al.}(2018)Nelson, Pillepich, Springel, Weinberger,
  Hernquist, Pakmor, Genel, Torrey, Vogelsberger, Kauffmann, Marinacci, \&
  Naiman}]{nelson_2018}
Nelson, D., Pillepich, A., Springel, V., {et~al.} 2018, MNRAS, 475, 624

\bibitem[{Nugis \& Lamers(2000)}]{nugis_2000}
Nugis, T. \& Lamers, H. J. G. L.~M. 2000, A\&A, 360, 227

\bibitem[{Palmerio {et~al.}(2019)Palmerio, Vergani, Salvaterra, Sanders,
  Japelj, {Vidal-Garc{\'i}a}, D'Avanzo, Corre, Perley, Shapley, Boissier,
  Greiner, Le~Floc'h, \& Wiseman}]{palmerio_2019}
Palmerio, J.~T., Vergani, S.~D., Salvaterra, R., {et~al.} 2019, A\&A, 623, A26

\bibitem[{Patton \& Sukhbold(2020)}]{patton_2020}
Patton, R.~A. \& Sukhbold, T. 2020, MNRAS, 499, 2803

\bibitem[{Paxton {et~al.}(2011)Paxton, Bildsten, Dotter, Herwig, Lesaffre, \&
  Timmes}]{paxton_2011}
Paxton, B., Bildsten, L., Dotter, A., {et~al.} 2011, ApJS, 192, 3

\bibitem[{Paxton {et~al.}(2013)Paxton, Cantiello, Arras, Bildsten, Brown,
  Dotter, Mankovich, Montgomery, Stello, Timmes, \& Townsend}]{paxton_2013}
Paxton, B., Cantiello, M., Arras, P., {et~al.} 2013, ApJS, 208, 4

\bibitem[{Paxton {et~al.}(2015)Paxton, Marchant, Schwab, Bauer, Bildsten,
  Cantiello, Dessart, Farmer, Hu, Langer, Townsend, Townsley, \&
  Timmes}]{paxton_2015}
Paxton, B., Marchant, P., Schwab, J., {et~al.} 2015, ApJS, 220, 15

\bibitem[{Paxton {et~al.}(2018)Paxton, Schwab, Bauer, Bildsten, Blinnikov,
  Duffell, Farmer, Goldberg, Marchant, Sorokina, Thoul, Townsend, \&
  Timmes}]{paxton_2018}
Paxton, B., Schwab, J., Bauer, E.~B., {et~al.} 2018, ApJS, 234, 34

\bibitem[{Paxton {et~al.}(2019)Paxton, Smolec, Schwab, Gautschy, Bildsten,
  Cantiello, Dotter, Farmer, Goldberg, Jermyn, Kanbur, Marchant, Thoul,
  Townsend, Wolf, Zhang, \& Timmes}]{paxton_2019}
Paxton, B., Smolec, R., Schwab, J., {et~al.} 2019, ApJS, 243, 10

\bibitem[{Perley {et~al.}(2016)Perley, Tanvir, Hjorth, Laskar, Berger, Chary,
  Postigo, Fynbo, Kr{\"u}hler, Levan, Micha{\l}owski, \&
  Schulze}]{perley_2016a}
Perley, D.~A., Tanvir, N.~R., Hjorth, J., {et~al.} 2016, ApJ, 817, 8

\bibitem[{Pillepich {et~al.}(2018)Pillepich, Nelson, Hernquist, Springel,
  Pakmor, Torrey, Weinberger, Genel, Naiman, Marinacci, \&
  Vogelsberger}]{pillepich_2018a}
Pillepich, A., Nelson, D., Hernquist, L., {et~al.} 2018, MNRAS, 475, 648

\bibitem[{Podsiadlowski {et~al.}(2010)Podsiadlowski, Ivanova, Justham, \&
  Rappaport}]{podsiadlowski_2010a}
Podsiadlowski, P., Ivanova, N., Justham, S., \& Rappaport, S. 2010, MNRAS, 406,
  840

\bibitem[{Qin {et~al.}(2018)Qin, Fragos, Meynet, Andrews, S{\o}rensen, \&
  Song}]{qin_2018}
Qin, Y., Fragos, T., Meynet, G., {et~al.} 2018, A\&A, 616, A28

\bibitem[{Qin {et~al.}(2019)Qin, Marchant, Fragos, Meynet, \&
  Kalogera}]{qin_2019}
Qin, Y., Marchant, P., Fragos, T., Meynet, G., \& Kalogera, V. 2019, ApJ, 870,
  L18

\bibitem[{Ramachandran {et~al.}(2019)Ramachandran, Hamann, Oskinova, Gallagher,
  Hainich, Shenar, Sander, Todt, \& Fulmer}]{ramachandran_2019}
Ramachandran, V., Hamann, W.~R., Oskinova, L.~M., {et~al.} 2019, A\&A, 625,
  A104

\bibitem[{Rastinejad {et~al.}(2022)Rastinejad, Gompertz, Levan, Fong, Nicholl,
  Lamb, Malesani, Nugent, Oates, Tanvir, {de Ugarte Postigo}, Kilpatrick,
  Moore, Metzger, Ravasio, Rossi, Schroeder, Jencson, Sand, Smith,
  Fern{\'a}ndez, Berger, Blanchard, Chornock, Cobb, De~Pasquale, Fynbo, Izzo,
  Kann, Laskar, Marini, Paterson, Escorial, Sears, \&
  Th{\"o}ne}]{rastinejad_2022}
Rastinejad, J.~C., Gompertz, B.~P., Levan, A.~J., {et~al.} 2022, Nat, 612, 223

\bibitem[{Russell {et~al.}(2013)Russell, Gallo, \& Fender}]{russell_2013}
Russell, D.~M., Gallo, E., \& Fender, R.~P. 2013, MNRAS, 431, 405

\bibitem[{Salafia {et~al.}(2020)Salafia, Barbieri, Ascenzi, \&
  Toffano}]{salafia_2020}
Salafia, O.~S., Barbieri, C., Ascenzi, S., \& Toffano, M. 2020, A\&A, 636, A105

\bibitem[{Salvaterra {et~al.}(2012)Salvaterra, Campana, Vergani, Covino,
  D'Avanzo, Fugazza, Ghirlanda, Ghisellini, Melandri, Nava, Sbarufatti, Flores,
  Piranomonte, \& Tagliaferri}]{salvaterra_2012}
Salvaterra, R., Campana, S., Vergani, S.~D., {et~al.} 2012, ApJ, 749, 68

\bibitem[{Sarmento {et~al.}(2017)Sarmento, Scannapieco, \& Pan}]{sarmento_2017}
Sarmento, R., Scannapieco, E., \& Pan, L. 2017, ApJ, 834, 23

\bibitem[{Shenar {et~al.}(2020)Shenar, Gilkis, Vink, Sana, \&
  Sander}]{shenar_2020}
Shenar, T., Gilkis, A., Vink, J.~S., Sana, H., \& Sander, A. A.~C. 2020, A\&A,
  634, A79

\bibitem[{Shivvers {et~al.}(2017)Shivvers, Modjaz, Zheng, Liu, Filippenko,
  Silverman, Matheson, Pastorello, Graur, Foley, Chornock, Smith, Leaman, \&
  Benetti}]{shivvers_2017}
Shivvers, I., Modjaz, M., Zheng, W., {et~al.} 2017, PASP, 129, 054201

\bibitem[{Soderberg {et~al.}(2006)Soderberg, Kulkarni, Nakar, Berger, Cameron,
  Fox, Frail, {Gal-Yam}, Sari, Cenko, Kasliwal, Chevalier, Piran, Price,
  Schmidt, Pooley, Moon, Penprase, Ofek, Rau, Gehrels, Nousek, Burrows,
  Persson, \& McCarthy}]{soderberg_2006}
Soderberg, A.~M., Kulkarni, S.~R., Nakar, E., {et~al.} 2006, Nat, 442, 1014

\bibitem[{Song {et~al.}(2016)Song, Meynet, Maeder, Ekstr{\"o}m, \&
  Eggenberger}]{song_2016}
Song, H.~F., Meynet, G., Maeder, A., Ekstr{\"o}m, S., \& Eggenberger, P. 2016,
  A\&A, 585, A120

\bibitem[{Springel(2010)}]{springel_2010}
Springel, V. 2010, MNRAS, 401, 791

\bibitem[{Springel {et~al.}(2018)Springel, Pakmor, Pillepich, Weinberger,
  Nelson, Hernquist, Vogelsberger, Genel, Torrey, Marinacci, \&
  Naiman}]{springel_2018}
Springel, V., Pakmor, R., Pillepich, A., {et~al.} 2018, MNRAS, 475, 676

\bibitem[{Spruit(2002)}]{spruit_2002}
Spruit, H.~C. 2002, A\&A, 381, 923

\bibitem[{Sukhbold {et~al.}(2018)Sukhbold, Woosley, \& Heger}]{sukhbold_2018}
Sukhbold, T., Woosley, S.~E., \& Heger, A. 2018, ApJ, 860, 93

\bibitem[{Tanga {et~al.}(2018)Tanga, Kr{\"u}hler, Schady, Klose, Graham,
  Greiner, Kann, \& Nardini}]{tanga_2018}
Tanga, M., Kr{\"u}hler, T., Schady, P., {et~al.} 2018, A\&A, 615, A136

\bibitem[{Tchekhovskoy {et~al.}(2012)Tchekhovskoy, McKinney, \&
  Narayan}]{tchekhovskoy_2012}
Tchekhovskoy, A., McKinney, J.~C., \& Narayan, R. 2012, J. Phys. Conf. Ser.,
  372, 12040

\bibitem[{Tchekhovskoy {et~al.}(2010)Tchekhovskoy, Narayan, \&
  McKinney}]{tchekhovskoy_2010}
Tchekhovskoy, A., Narayan, R., \& McKinney, J.~C. 2010, ApJ, 711, 50

\bibitem[{Teng {et~al.}(2025)Teng, Demir, Doctor, Srivastava, Lalvani,
  Kalogera, Katsaggelos, Andrews, Bavera, Briel, Gossage, Kovlakas, Kruckow,
  Rocha, Sun, Xing, \& Zapartas}]{teng_2025}
Teng, E., Demir, U., Doctor, Z., {et~al.} 2025, A\&C, 51, 100935

\bibitem[{Terreran {et~al.}(2019)Terreran, Margutti, Bersier, Brimacombe,
  Caprioli, Challis, Chornock, Coppejans, Dong, Guidorzi, Hurley, Kirshner,
  Migliori, Milisavljevic, Palmer, Prieto, Tomasella, Marchant, Pastorello,
  Shappee, Stanek, Stritzinger, Benetti, Chen, DeMarchi, {Elias-Rosa}, Gall,
  Harmanen, \& Mattila}]{terreran_2019}
Terreran, G., Margutti, R., Bersier, D., {et~al.} 2019, ApJ, 883, 147

\bibitem[{Thompson(1994)}]{thompson_1994}
Thompson, C. 1994, MNRAS, 270, 480

\bibitem[{Thompson {et~al.}(2004)Thompson, Chang, \& Quataert}]{thompson_2004}
Thompson, T.~A., Chang, P., \& Quataert, E. 2004, ApJ, 611, 380

\bibitem[{Usov(1992)}]{usov_1992}
Usov, V.~V. 1992, Nat, 357, 472

\bibitem[{{van den Heuvel} \& Yoon(2007)}]{vandenheuvel_2007}
{van den Heuvel}, E. P.~J. \& Yoon, S.~C. 2007, Astrophys. Space Sci., 311, 177

\bibitem[{Vergani {et~al.}(2017)Vergani, Palmerio, Salvaterra, Japelj,
  Mannucci, Perley, D'Avanzo, Kr{\"u}hler, Puech, Boissier, Campana, Covino,
  Hunt, Petitjean, \& Tagliaferri}]{vergani_2017}
Vergani, S.~D., Palmerio, J., Salvaterra, R., {et~al.} 2017, A\&A, 599, A120

\bibitem[{Vergani {et~al.}(2015)Vergani, Salvaterra, Japelj, Le~Floc'h,
  D'Avanzo, {Fernandez-Soto}, Kr{\"u}hler, Melandri, Boissier, Covino, Puech,
  Greiner, Hunt, Perley, Petitjean, Vinci, Hammer, Levan, Mannucci, Campana,
  Flores, Gomboc, \& Tagliaferri}]{vergani_2015}
Vergani, S.~D., Salvaterra, R., Japelj, J., {et~al.} 2015, A\&A, 581, A102

\bibitem[{Vink \& De~Koter(2005)}]{vink_2005}
Vink, J.~S. \& De~Koter, A. 2005, A\&A, 442, 587

\bibitem[{Vink {et~al.}(2001)Vink, {de Koter}, \& Lamers}]{vink_2001}
Vink, J.~S., {de Koter}, A., \& Lamers, H. J. G. L.~M. 2001, A\&A, 369, 574

\bibitem[{Vink \& Sander(2021)}]{vink_2021}
Vink, J.~S. \& Sander, A. A.~C. 2021, MNRAS, 504, 2051

\bibitem[{Virgili {et~al.}(2009)Virgili, Liang, \& Zhang}]{virgili_2009}
Virgili, F.~J., Liang, E.-W., \& Zhang, B. 2009, MNRAS, 392, 91

\bibitem[{Wang {et~al.}(2007)Wang, Li, Waxman, \& M{\'e}sz{\'a}ros}]{wang_2007}
Wang, X.-Y., Li, Z., Waxman, E., \& M{\'e}sz{\'a}ros, P. 2007, ApJ, 664, 1026

\bibitem[{Wiersma {et~al.}(2009)Wiersma, Schaye, Theuns, Dalla~Vecchia, \&
  Tornatore}]{wiersma_2009}
Wiersma, R. P.~C., Schaye, J., Theuns, T., Dalla~Vecchia, C., \& Tornatore, L.
  2009, MNRAS, 399, 574

\bibitem[{Woosley(1993)}]{woosley_1993}
Woosley, S.~E. 1993, ApJ, 405, 273

\bibitem[{Woosley \& Heger(2006)}]{woosley_2006}
Woosley, S.~E. \& Heger, A. 2006, ApJ, 637, 914

\bibitem[{Xing {et~al.}(2024)Xing, Bavera, Fragos, Kruckow, {Rom{\'a}n-Garza},
  Andrews, Dotter, Kovlakas, Misra, Srivastava, Rocha, Sun, \&
  Zapartas}]{xing_2024}
Xing, Z., Bavera, S.~S., Fragos, T., {et~al.} 2024, A\&A, 683, A144

\bibitem[{Xu {et~al.}(2009)Xu, Starling, Fynbo, Sollerman, Yost, Watson, Foley,
  O'Brien, \& Hjorth}]{xu_2009}
Xu, D., Starling, R. L.~C., Fynbo, J. P.~U., {et~al.} 2009, ApJ, 696, 971

\bibitem[{Yoon {et~al.}(2008)Yoon, Cantiello, \& Langer}]{yoon_2008}
Yoon, S.~C., Cantiello, M., \& Langer, N. 2008, 990, 225

\bibitem[{Yoon \& Langer(2005)}]{yoon_2005}
Yoon, S.~C. \& Langer, N. 2005, A\&A, 443, 643

\bibitem[{Yoon {et~al.}(2006)Yoon, Langer, \& Norman}]{yoon_2006}
Yoon, S.-C., Langer, N., \& Norman, C. 2006, A\&A, 460, 199

\bibitem[{Yoon {et~al.}(2010)Yoon, Woosley, \& Langer}]{yoon_2010}
Yoon, S.-C., Woosley, S.~E., \& Langer, N. 2010, ApJ, 725, 940

\bibitem[{Y{\"u}ksel {et~al.}(2008)Y{\"u}ksel, Kistler, Beacom, \&
  Hopkins}]{yuksel_2008}
Y{\"u}ksel, H., Kistler, M.~D., Beacom, J.~F., \& Hopkins, A.~M. 2008, ApJ,
  683, L5

\bibitem[{Zapartas {et~al.}(2017)Zapartas, {de Mink}, Van~Dyk, Fox, Smith,
  Bostroem, {de Koter}, Filippenko, Izzard, Kelly, Neijssel, Renzo, \&
  Ryder}]{zapartas_2017a}
Zapartas, E., {de Mink}, S.~E., Van~Dyk, S.~D., {et~al.} 2017, ApJ, 842, 125

\bibitem[{Zou {et~al.}(2021)Zou, Zhang, Huang, \& Zhao}]{zou_2021}
Zou, Z.-C., Zhang, B.-B., Huang, Y.-F., \& Zhao, X.-H. 2021, ApJ, 921, 2

\end{thebibliography}

\begin{appendix}

\section{Accretion disk dependence on initial primary mass and period} \label{app:disk_gradient}

Within the stable reverse-mass-transfer region, the mass of the accretion disk at collapse depends on the initial primary mass and period. As the initial period or donor mass increase, the accretion disk mass decreases. As such, the lowest mass and shortest period systems within the stable reverse-mass-transfer region, in general, have the most massive accretion disk when collapsing.
In this section, we will first cover the initial period dependence and then the initial primary mass dependence. We pick representative slices of the stable reverse-mass-transfer region to show the evolution of different stable reverse-mass-transfer binaries.

\subsection{Initial period dependence}

At $\Zsun$, we select systems with $M_\mathrm{ZAMS,1} \approx 39.87$ and $q=0.95$, which are undergoing stable reverse mass transfer. These systems are illustrated in Figure \ref{fig:Zsun_multi_q} and exhibit a correlation between decreasing disk mass and increasing initial period. Only the widest two systems that form an accretion disk at collapse are reversed in disk mass.

\begin{figure}
    \centering
    \includegraphics[width=\linewidth]{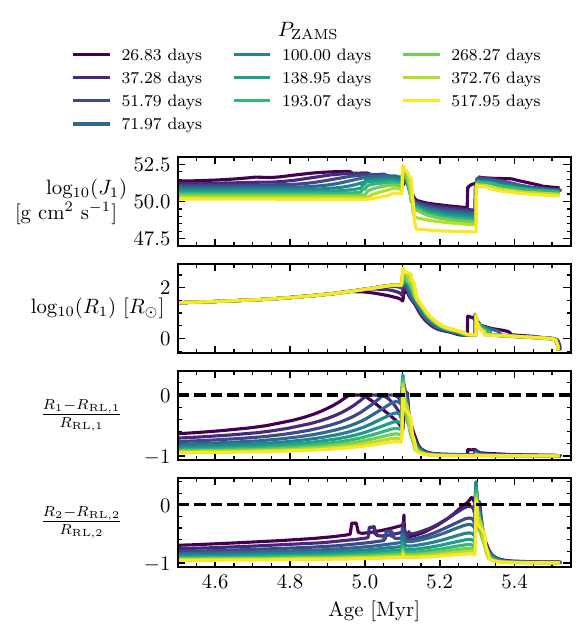}
    \caption{The evolution of stable reverse-mass-transfer systems with $M_\mathrm{ZAMS,1} \approx 39.87$ and $q=0.95$ and different initial periods, including the first mass transfer and reverse phase. The rows from top to bottom show: the total angular momentum in the primary star, the radius of the primary star, and finally the relative Roche lobe filling factor of the primary and secondary. The colour map indicates the initial period at the zero-age main-sequence, with blue indicating the tightest and yellow the widest systems. The horizontal dashed black lines indicate when the star fills its Roche lobe.}
    \label{fig:app-example-period-gradient}
\end{figure}

In Figure \ref{fig:app-example-period-gradient}, we show the evolution of these models focusing on the mass transfer phases. As expected, the shortest period systems (dark blue) fill their Roche lobe already on the main-sequence, as shown in the third row. They shrink back inside their Roche lobe until core-hydrogen depletion is reached. At this stage, all primaries expand, and for the wider systems (green and yellow), this is their first mass transfer phase. The hydrogen envelopes of the primaries in long-period binaries experience less stripping compared to those in short-period systems; consequently, they exhibit a greater radial expansion after the mass transfer. The $P_\mathrm{ZAMS}=37.28~\mathrm{days}$ binary exits the first mass transfer with $4.39~\msun$ of hydrogen envelope left and a radius of ${\sim}84 R_\odot$, while the binary with $P_\mathrm{ZAMS}=517.95~\mathrm{days}$ has $8.27~\msun$ of hydrogen envelope left and a radius of ${\sim}546 R_\odot$. This remains until the remaining hydrogen is lost through stellar winds, which also removes all the angular momentum gained during the first mass transfer phase.

\begin{figure}
    \centering
    \includegraphics[width=\linewidth]{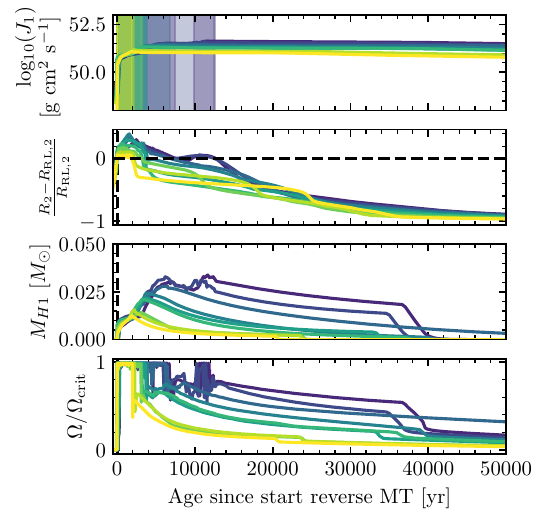}
    \caption{The stable-reverse-mass-transfer phase for the same models as in Figure \ref{fig:app-example-period-gradient} with the same colour map, except for $P=26.83~\mathrm{days}$ which is not included. The models are aligned when starting the stable reverse-mass-transfer phase. The rows from top to bottom show the total angular momentum in the primary, the relative Roche lobe overflow factor of the secondary, the total hydrogen in the primary, and the fraction of critical rotation at the surface. In the top figure, the shaded areas indicate the reverse mass transfer.}
    \label{fig:app-example-period-zoom}
\end{figure}

With a mass ratio of $q=0.95$, the secondary evolves off the main-sequence soon after. In the shortest period system, $P_\mathrm{ZAMS}\approx26.83~\mathrm{days}$, the secondary fills its Roche lobe while still on the main sequence, leading to a relatively prolonged and stable mass transfer phase. This results in the transfer of more mass and angular momentum compared to wider-orbit systems, which typically experience their initial reverse interaction upon core hydrogen depletion. In Figure \ref{fig:app-example-period-zoom}, we zoom in on the stable reverse mass transfer and exclude this short-period binary for clarity.

\begin{figure}
    \centering
    \includegraphics[width=\linewidth]{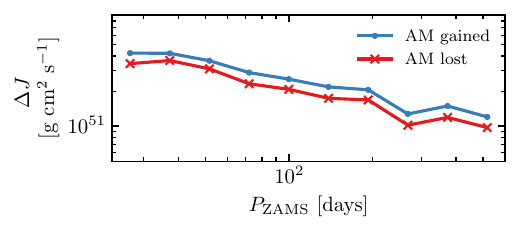}
    \caption{The amount of angular momentum (AM) gained during the stable reverse-mass-transfer phase (blue; dots) and the amount of AM lost between the end of the mass transfer and core carbon depletion (red, cross) for the models in Figure \ref{fig:app-example-period-zoom}. The shortest-period systems gain the most angular momentum, but also lose more angular momentum due to their large radial extension compared to long-period binaries. The fractional loss is approximately 80\% of the AM gained at each initial period.}
    \label{fig:app-example-period-AM-gain}
\end{figure}

In the top row of Figure \ref{fig:app-example-period-zoom}, we show the total angular momentum in the primary star and the duration of the stable reverse-mass-transfer phase as the shaded region. The shortest period systems (dark blue; ${\sim}12000~\mathrm{yr}$) have a 6 times longer mass transfer than the widest periods (yellow; ${\sim}5000~\mathrm{yr}$). Over time, angular momentum at the surface is able to diffuse inwards to spin up deeper layers in the star. For a longer duration mass transfer, the ongoing mass transfer will replenish the angular momentum of the outer layers of the accretor. As a result, the short-period systems are able to accrete more mass and angular momentum, as the total hydrogen in the third row in Figure \ref{fig:app-example-period-zoom} shows, despite the surface being near critical rotation for the majority of the mass transfer. The amount of AM per accreted unit mass is constant around $\sim 7 $ g cm$^{2}$ s$^{-1}$ $\msun$. In Figure \ref{fig:app-example-period-AM-gain}, we show the total amount of angular momentum gained during the stable reverse-mass-transfer phase in blue. The total angular momentum gain in short-period binaries is higher than in long-period systems, but the angular momentum loss in the later evolutionary phases is higher for short-period binaries. Independent of the initial period, approximately 80\% of the accreted angular momentum is lost between the stable reverse mass transfer and core carbon depletion.

With the evolutionary state of the primary being similar for different initial periods, the main difference is the duration of the stable reverse mass transfer leading to more mass accretion and angular momentum gain. A long-period binary has a shorter stable reverse-mass-transfer phase, and gains less angular momentum and mass. With a similar fraction of angular momentum lost across different initial periods, this puts most of the long-period primaries outside the regime for disk formation during collapse, while short-period primaries retain sufficient angular momentum to produce an accretion disk, despite the higher absolute angular momentum loss.

\subsection{Initial primary mass dependence}

\begin{figure}
    \centering
    \includegraphics[width=\linewidth]{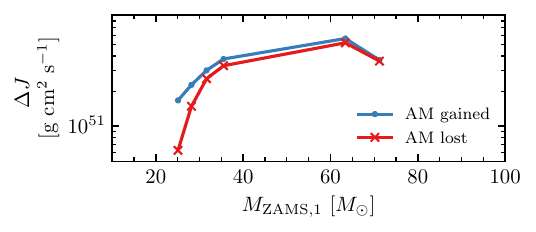}
    \caption{The angular momentum gain (blue) and loss (red) during the reverse mass transfer and post-reverse mass transfer, respectively, as a function of initial donor mass for $P_\mathrm{ZAMS}=19.31~\mathrm{days}$ and $q=0.95$ at $Z=\Zsun$. Only the binaries undergoing stable reverse mass transfer at this period are shown.}
    \label{fig:app-example-mass-AM-gain}
\end{figure}

The dependence on the initial primary mass is more straightforward compared to the period dependence. At the same period and mass ratio, a more massive primary gains more mass and angular momentum during the stable reverse-mass-transfer phase compared to a lower mass companion, as shown as the blue line in Figure \ref{fig:app-example-mass-AM-gain} for $P_\mathrm{ZAMS}=19.31~\mathrm{days}$ and $q=0.95$ at $Z=\Zsun$. At the same time, the interaction occurs earlier in the evolution of the primary due to its larger radius. As such, the duration between the reverse mass transfer and core carbon depletion is shorter for lower mass primaries compared to a higher mass primary, despite the longer evolutionary timescale. Together with a smaller radius and lower mass loss through stellar winds, this leads to less angular momentum loss in lower-mass primaries, as the red line in Figure \ref{fig:app-example-mass-AM-gain} shows. In conclusion, as the initial mass of the primary star increases in a binary, the angular momentum loss before core carbon depletion increases, which decreases the mass of an accretion disk at collapse.

\section{Star formation histories} \label{app:sfh}

\begin{figure}
    \centering
    \includegraphics{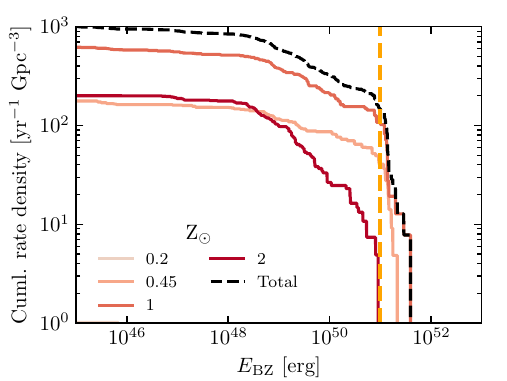}
    \caption{Similar to Figure \ref{fig:illustrisTNG_energy_budget}, but the cumulative rate density from the \citet{neijssel_2019} star formation rate.}
    \label{fig:neijssel_energy_budget}
\end{figure}

\begin{figure}
    \centering
    \includegraphics{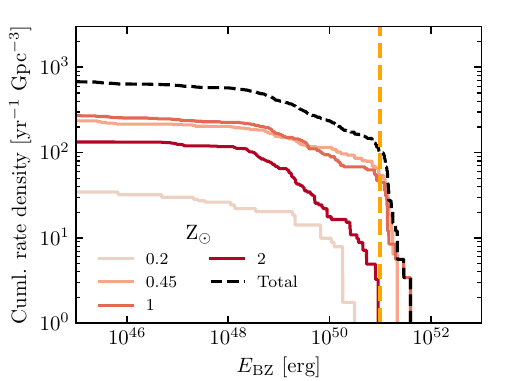}
    \caption{Similar to Figure \ref{fig:illustrisTNG_energy_budget}, but the cumulative rate density from the \citet{madau_2017} star formation rate.}
    \label{fig:madau_energy_budget}
\end{figure}

\begin{figure}
    \centering
    \includegraphics{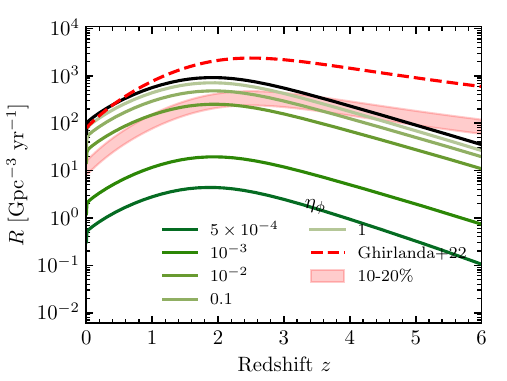}
    \caption{The stable reverse-mass-transfer LGRB rate density from the \citet{madau_2017} star formation history.}
    \label{fig:cosmic_MF}
\end{figure}

\begin{figure}
    \centering
    \includegraphics{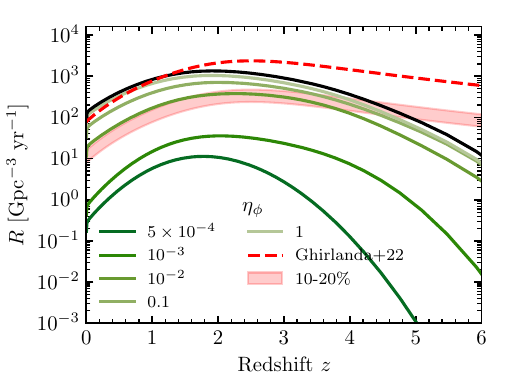}
    \caption{The stable reverse-mass-transfer LGRB rate density from the \citet{neijssel_2019} star formation history.}
    \label{fig:cosmic_neijssel}
\end{figure}

Since the star formation rate density and metallicity evolution impact the cosmic rate density of stable reverse-mass-transfer collapsar LGRB significantly, we perform the same potential energy budget calculations as discussed in Section \ref{sec:method} on the \citet{madau_2017} and \citet{neijssel_2019} star formation rate prescriptions.
Their cumulative energy distributions are shown in Figures \ref{fig:neijssel_energy_budget} and \ref{fig:madau_energy_budget}.
Because they originate from the same underlying {\tt POSYDON} stellar populations, the shape of each metallicity distribution is similar, but the star formation rate of each metallicity is redistributed. E.g. $\Zsun$ events have a total cumulative rate of ${\simeq}290$ Gpc$^{-3}$ yr$^{-1}$ in the \citet{madau_2017} population, but ${\simeq}600$ Gpc$^{-3}$ yr$^{-1}$ in the \citet{neijssel_2019} population. This is an effect of the latter prescription having a narrower metallicity distribution compared to \citet{madau_2017}, which results in more high-metallicity star formation in the local Universe.
Figures \ref{fig:cosmic_neijssel} and \ref{fig:cosmic_MF} show the decrease of the rate density from the energy limits. This effect is similar to that from the Illustris TNG simulation as shown in Figure \ref{fig:cosmic_rates}.

\section{Free-fall collapse timescale} \label{app:free-fall-timescale}

\begin{figure}
    \centering
    \includegraphics[width=\columnwidth]{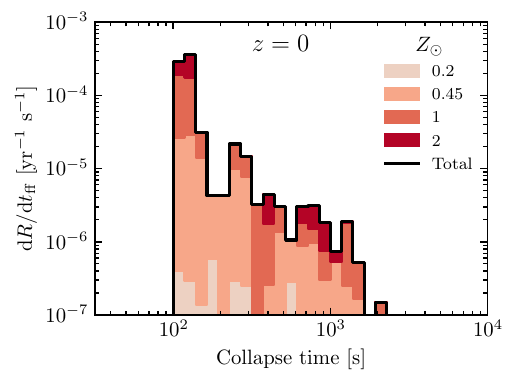}
    \caption{The free-fall collapse timescales of the stable reverse-mass-transfer progenitors at carbon depletion. Each model is weighted by its contribution at $z=0$.  The red shading areas are stacked histograms for each metallicity with the solid black line indicating the total differential rate.}
    \label{fig:free-fall-collapse}
\end{figure}

For each of the primary stars undergoing collapse through the stable reverse-mass-transfer channel, we calculate their free-fall timescale using 

\begin{equation}
 t_\mathrm{ff} = \sqrt{\frac{3 \pi}{32 G \rho}}, 
\end{equation}
where $G$ is the gravitational constant and $\rho$ is the average density of the star at collapse. We use the stellar radius and mass at carbon depletion to calculate $\rho$. 
In Figure \ref{fig:free-fall-collapse}, the duration of each model is weighted by its contribution to the event rate at $z=0$. The free-fall collapse timescale of the stable reverse-mass-transfer LGRB progenitors lies between ${\simeq}108$ and ${\simeq}2204$ seconds with the majority of durations around ${\simeq}120$ seconds.

Observations use the $T_{90}$ duration to quantify the duration of the event, which is defined as the duration between the emission of 5\% and 95\% of the total measured counts of the LGRB. For LGRBs these peak around ${\sim}50$ seconds \cite[see, for example][]{ghirlanda_2022}, but extend up to ${\sim}1000$ seconds. 

The free-fall collapse durations of our models are at the higher end of the observed $T_{90}$ distribution, since we do not account for energies below the noise level nor do we consider only 90\% of the counts. Additionally, the time for the accretion disk to form and the breakout time of the jet are not included in our calculation.

These effects would shorten the perceived duration of the collapse compared to the free-fall duration, although the infall via the disk can extend the duration for the fastest rotating models \citep[see, for example][]{fryer_2025}.
In general, the free-fall collapse durations of our LGRBs progenitors are of the same order of magnitude as the observed $T_{90}$ values for LGRBs.

\end{appendix}

\end{document}